# OCAMS: The OSIRIS-REx Camera Suite

B. Rizk[1], C. Drouet d'Aubigny[1], D. Golish[1], C. Fellows[1], C. Merrill[3], P. Smith[1], M.S. Walker[7], J.E. Hendershot[8], J. Hancock[2], S.H. Bailey[1,3], D. DellaGiustina[1], D. Lauretta[1], R. Tanner[1], M. Williams[1], K. Harshman[1], M. Fitzgibbon[1], W. Verts[4], J. Chen[5], T. Connors[3], D. Hamara[1], A. Dowd[6], A. Lowman[4], M. Dubin[4], R. Burt[2], M. Whiteley[2], M. Watson[2], T. McMahon[3], M. Ward[3], D. Booher[9], M. Read[1], B. Williams[3], M. Hunten[1], E. Little[9], T. Saltzman[1], D. Alfred[3], S. O'Dougherty[4], M. Walthall[9], K. Kenagy[3], S. Peterson[1], B. Crowther[2,11], M.L. Perry[1], C. See[1], S. Selznick[1], C. Sauve[3], M. Beiser[9], W. Black[4], R.N. Pfisterer[12], A. Lancaster[9], S. Oliver[3], C. Oquest[1], D. Crowley[1], C. Morgan[1], C. Castle[10], R. Dominguez[3], M. Sullivan[3]

[1]*Lunar and Planetary Laboratory, University of Arizona, Tucson, AZ, USA*

(bashar@LPL.arizona.edu)

[2]*Space Dynamics Laboratory, Utah State University Foundation, Logan, UT, USA*

[3]*Steward Observatory, University of Arizona, Tucson, AZ, USA*

[4]*College of Optical Sciences, University of Arizona, Tucson, AZ, USA*

[5]*Baja Technology LLC, Tucson, AZ, USA*

[6]*Lithe Technology LLC, Tucson, AZ, USA*

[7]*Goddard Space Flight Center, Greenbelt, MD, USA*

[8]*Ball Aerospace, Greenbelt, MD, USA*

[9]*Raytheon Missile Systems, Tucson, AZ, USA*

[10]*Hofstadter Analytical, Tucson, AZ, USA*

[11]*Synopsys Optical Solutions Group, Pasadena, CA, USA*

[12]*Photon Engineering, L.L.C., Tucson, AZ, USA*

**Abstract** The requirements-driven OSIRIS-REx Camera Suite (OCAMS) acquires images essential to collecting a sample from the surface of Bennu. During proximity operations, these images document the presence of satellites and plumes, record spin state, enable an accurate digital terrain model of the asteroid's shape, and identify any surface hazards. They confirm the presence of sampleable regolith on the surface, observe the sampling event itself, and image the sample head in order to verify its readiness to be stowed. They document Bennu's history as an example of early solar system material, as a microgravity body with a planetesimal size-scale, and as a carbonaceous object. OCAMS is fitted with three cameras. The MapCam records point-source color images on approach to the asteroid in order to connect Bennu's ground-based point-source observational record to later higher-resolution surface spectral imaging. The SamCam documents the sample site before, during, and after it is disturbed by the sample mechanism. The PolyCam, using its focus mechanism, observes the sample site at sub-centimeter resolutions, revealing surface texture and morphology. While their imaging requirements divide naturally between the three cameras, they preserve a strong degree of functional overlap. OCAMS and the other spacecraft instruments allow the OSIRIS-REx mission to collect a sample from a microgravity body on the same visit during which it was first optically acquired from long range, a useful capability as humanity explores near-Earth, Main-Belt and Jupiter Trojan asteroids.

## Acronyms

| | |
|---|---|
| AAM | Asteroid Approach Maneuver |
| APID | Application Process Identification |
| A/D, ADC | Analog-to-Digital Conversion |
| ATLO | Assembly, Test, and Launch Operations |
| C&DH | Command and Data Handling |
| CCD | Charge-Coupled Device |
| CCM | Camera Control Module |
| CDS | Correlated Double Sampling |
| CM | Configuration Management |
| CPU | Computer Processing Unit |
| CTE | Charge Transfer Efficiency |
| CTE | Coefficient of Thermal Expansion |
| DA | Detector Assembly |
| DN | Data Number |
| DPU | Data Processing Unit |
| DRM | Design Reference Mission |
| DTCI | Data Telemetry Command Interface |
| DTM | Digital Terrain Model |
| ECAS | Eight-Color Asteroid Survey |
| EPER | Extended-Pixel Edge Response |
| EQM | Engineering Qualification Model |
| F/ | F-Stop or F-number |
| FITS | Flexible Image Transport System |
| FM | Flight Model |
| FOV | Field of View |
| FPGA | Field-Programmable Gate Array |
| FRED | Fred Optical Engineering Software |
| FSW | Flight Software |
| FWHM | Full-Width-at-Half-Max |
| GNC | Guidance Navigation and Control |
| GSFC | Goddard Space Flight Center |
| HGA | High Gain Antenna |
| IFOV | Instantaneous Field of View |
| IP | Intellectual Property |
| ISO | International Standards Organization |
| LED | Light-Emitting Diode |
| LGA | Low Gain Antenna |
| LIDAR | Light Detection and Ranging |
| LPL | Lunar and Planetary Laboratory |
| LSF | Line-Spread Function |
| LVDS | Low-Voltage Differential Signaling |
| LVPS | Low-Voltage Power Supply |
| MapCam | Mapping Camera |
| MDR | Minimum Detectable Radiance |
| MGA | Medium Gain Antenna |
| MHD | Motor/Heater Interface/Driver |
| MTF | Modulation Transfer Function |
| NASA | National Aeronautics and Space Administration |
| NFPO | New Frontiers Program Office |
| NFT | Natural Feature Tracking |
| NIR | Near-Infrared |
| NIST | National Institute of Standards & Technology |
| OCAMS | OSIRIS-REx Camera Suite |
| OLA | OSIRIS-REx Laser Altimeter |
| OSIRIS-REx | Origins, Spectral Interpretation, Resource Identification, Security–Regolith Explorer |
| OST | Optical Support Tube |
| OTES | OSIRIS-REx Thermal Emission Spectrometer |
| OVIRS | OSIRIS-REx Visible and InfraRed Spectrometer |
| Pan | Panchromatic |
| PAPL | Project-Approved Parts List |
| PCB | Parts Control Board |
| PIL | Parts Identification List |
| PSNIT | Point Source Normalized Irradiance Transmittance |
| PolyCam | Polyfunctional Camera |
| PRT | Platinum Resistance Thermometer |
| QTH | Quartz-Tungsten-Halogen |
| REXIS | Regolith X-ray Imaging Spectrometer |
| RMS | Root-Mean-Square |
| ROI | Region of Interest |
| RTS | Random Telegraph Signal |
| RWA | Reaction Wheel Assembly |
| SamCam | Sample Acquisition Verification Camera |
| S/C | Spacecraft |
| S/N, SNR | Signal-to-Noise Ratio |
| SDRAM | Synchronous Dynamic Random Access Memory |
| SPI | Serial Peripheral Interface |
| SRC | Sample Return Capsule |
| TAG | Touch And Go |
| TAGCAMS | Touch And Go Camera System |
| TAGSAM | Touch And Go Sample Acquisition Maneuver |
| Ti 6Al-4V | Titanium Alloy (Titanium, 6% Aluminum, 4% Vanadium) |
| TID | Total Ionization Dose |
| TVAC | Thermal Vacuum |
| UA | University of Arizona |
| UUT | Unit-Under-Test |
| UV | Ultraviolet |
| VML | Virtual Machine Language |

# Contents

# 1    Introduction

The sample-return mission of the Origins, Spectral Interpretation, Resource Identification and Security–Regolith Explorer (OSIRIS-REx) must thoroughly characterize the near-Earth asteroid 101955 Bennu before being able to acquire a sample from a scientifically interesting location on its surface with minimal risk, either to the spacecraft or to mission success (Ajluni et al. 2015; Beshore et al. 2015; Lauretta 2015, 2016; Bierhaus et al. 2017; Lauretta et al. 2017). In addition, the mission team will map Bennu's global properties, characterize its geologic and dynamical history, document the morphology and chemistry of the sample site, and determine Bennu's spin, surface area, and thermal emission. To this end, the spacecraft includes within its instrument complement the OSIRIS-REx Thermal Emission Spectrometer (OTES), which determines Bennu's mineralogical and thermo-physical properties (Christensen et al. 2017); the OSIRIS-REx Visible and InfraRed Spectrometer (OVIRS), which provides surface maps of interesting materials such as carbonates, silicates, sulfates, oxides, adsorbed water, and a wide range of organic species (Reuter et al. 2017); the OSIRIS-REx Laser Altimeter (OLA), which produces a comprehensive topographic mapping of Bennu's surface (Daly et al. 2017); the Radio Science experiment, which uses radiometric tracking data to estimate significant components of the gravity field (McMahon et al. 2017); the Regolith X-ray Imaging Spectrometer (REXIS), which measures the abundances of key elements (Allen et al. 2013; Masterson et al. 2017); and finally the OSIRIS-REx Camera Suite (OCAMS), whose three imagers visually record Bennu's near-asteroid environment and surface at a sufficiently high resolution as to be able to document the presence of satellites and plumes, to record spin state, to enable an accurate digital terrain mode of its shape (Gaskell et al. 2008), and to identify any surface hazards, described here. The spacecraft and science teams use these instruments to identify the presence of sampleable regolith on the surface, identify features useful to guide the spacecraft's trajectory to the surface, document the sampling process itself, and confirm the existence of sample material inside the sampling head after the sampling event. Complemented by the Touch and Go Camera System (TAGCAMS), another three-camera suite, whose imagers facilitate navigation to the target asteroid and confirm stowage of the sample (Bos et al. 2017), OCAMS enables the OSIRIS-REx mission to accomplish a feat rare in planetary exploration: to collect a sample from a microgravity body on the same visit during which it first acquires it visually at long range (Berry et al. 2013, 2015; McMahon et al. 2014; Antreasian et al. 2016; Clark et al. 2016; Getzandanner et al. 2016; Hamilton et al. 2016; Mario and Debrunner 2016; Scheeres et al. 2016; Dworkin et al. 2017; Hesar et al. 2017).

OCAMS functions primarily as a mission-critical scientific instrument on this journey. Its three imagers satisfy competing optical and radiometric requirements to carry out their primary task: acquiring images during each mission phase that inform the operations of subsequent mission phases and provide key scientific data about the asteroid. These phases include navigation and approach to the asteroid, various proximity-operations campaigns, and eventually, acquisition of the sample (Lauretta et al. 2017). OCAMS' requirements were first established during the generation of the OSIRIS-REx Design Reference Mission (DRM) (Mink et al. 2014; Stevens et al. 2017). The DRM was in turn influenced by OCAMS' design, capabilities, and functional redundancy. This feedback process was repeated until a mature DRM and comprehensive set of imaging requirements for OCAMS emerged. These requirements divided naturally between the three imagers while preserving a strong degree of functional overlap (Smith et al. 2013; Merrill and Williams, 2016).

In addition to informing mission operations, the camera suite also gathers information scientifically interesting in its own right. For example the (Mapping Camera) MapCam records color images of Bennu as a point source in wavelength bands related to the Eight-Color Asteroid Survey (ECAS) to connect Bennu's ground-based point-source observational record to later surface spectral imaging at higher resolutions (Tedesco et al. 1982; Clark et al. 2011). It also records color-filter images of Bennu's surface that allow a high-resolution classification of its surface, contributing to the ranking of sample sites by scientific interest. The Sample Acquisition Verification Camera (SamCam) documents the sample site before, during, and after it is disturbed by the sample mechanism and images the sample head with millimeter resolution post-sampling in order to verify the process. The Polyfunctional Camera (PolyCam) is capable of detecting Bennu from several million kilometers away; its focusing mechanism allows it to image the surface of the asteroid to ranges below 200 m at sub-centimeter resolutions.

Development of the camera suite was a collaboration between the University of Arizona's (UA) Lunar and Planetary Laboratory and several other institutions, including the UA's College of Optical Sciences (PolyCam optical train), Utah State University's Space Dynamics Laboratory (detector read-out modules), Baja Electronics (electronics controller design), UA's Steward Observatory (systems, mission assurance, configuration management, as well as mechanical engineering support), and Teledyne DALSA's Custom Division (detector). Goddard Space Flight Center was responsible for providing NASA management and technical oversight and worked closely with Lockheed Martin Space Systems and the UA to define the interface between instruments and spacecraft.

## 2      Instrument Design

OCAMS supports OSIRIS-REx's methodical approach to the surface of Bennu (Mink et al. 2014). Data gathered during a year of approach-phase and proximity operations identify hazards in the asteroid proximity environment, characterize the asteroid's gravity (through the determination of its shape), rotation, and surface states, and systematically reduce mission risks. Step by step, the shape of the body is determined, and safety-, deliverability-, and science-value maps are developed (Nolan et al. 2013; Lauretta et al. 2017). The presence, within the prospective sample site, of material capable of being ingested by the sample head is confirmed during close-range reconnaissance passes. Images acquired during sampling campaign rehearsals verify the correct execution of the various steps. Finally, the process culminates in verifying the choreographed steps of the sampling event and acquiring post-sampling images to confirm its success or failure. At each phase, and after each campaign, the results of the imaging are examined, analyzed, and compared to expectations to guide OSIRIS-REx's path toward sample acquisition.

To support these operations, the OCAMS optical systems consist of two refractive optical systems of medium and low resolution (MapCam and SamCam, respectively) and one high-resolution reflecting system with a two-lens field corrector, one of which is used within a focusing mechanism (PolyCam). In several aspects, they share a heritage with the Surface Stereo Imager and Robotic Arm Camera on the Phoenix mission to Mars, the previous Surface Stereo Imagers on the Mars Pathfinder and Mars Polar Lander missions, and the three imagers on the Descent Imager/Spectral Radiometer on Cassini/Huygens (Smith et al. 1997; Keller et al. 2001;

Tomasko et al. 2002; Lemmon et al. 2008). To accomplish the mission's requirement to breach nine orders-of-magnitude variation in range to target, the three imagers span a factor of 25 in resolution and field of view (FOV) (Table 1), allowing mission planners a primary and a backup camera to image at any resolution or range short of the sample-acquisition event (Figure 1). The approximate focal ranges for nominal operations are indicated and correspond to PolyCam (180 m to infinity), MapCam (100 m to infinity), and SamCam (3 m to infinity). Resolution is defined here as a 3-pixel subtension on the focal plane. The optical layouts of all three systems are shown in Figures 2–4 and are described in more detail later in this document. All refractive elements are constructed from the radiation-hardened glasses BK7G18, F2G12 and K5G20, including the PolyCam field-correcting doublet, which is made from Schott BK7G18 and F2G12 glasses.

[Insert Figure 1]

[Insert Table 1]

[Insert Figure 2]

[Insert Figure 3]

[Insert Figure 4]

One feature of the camera suite that addresses the mission's variation in range to target is the addition of limited (MapCam, SamCam) or extensive (PolyCam) refocus capability, shown in Figure 1 as the blue and green dots along the SamCam and MapCam curves, respectively, and the continuation of the magenta PolyCam curve from the right edge of the figure down to 200 m. Each imager's refocus capability enables it to operate at additional ranges in support of mission objectives and is achieved by altering the optical train. In all three cameras the position of the detector is set during alignment and never changed afterward.

The PolyCam's refocusing capability is enabled by actuating the first of its two refracting field corrector lenses over a 5.7-mm operational range to focus the camera from infinity to ~180 m using a continuously variable mechanism enabled by four gears (Figures 5–6). The system is driven by a motor identical to those that drive the MapCam and SamCam filter wheels. The focus is graduated in steps that correspond to one-third of a shutter gear rotation, 180 motor steps (Table 2). The full mechanical range includes 40 different shutter rotations, while the operational range has 25 different optically useful shutter rotations.

[Insert Figure 5]

[Insert Figure 6]

[Insert Table 2]

In the MapCam, refocus is achieved by rotating filters of different thicknesses into the optical path between the focusing optics and the detector. In the SamCam, the filter wheel is positioned in front of the focusing optics; it uses a diopter lens in its filter wheel to refocus. These enhancements allow focused imaging at 30-m and 2-m ranges for the MapCam and SamCam, respectively (Figures 7–8). For the MapCam, four medium-width color filters are added that allow high-resolution and high-signal spectral characterizations of the asteroid both as a point source and as an extended object. For the SamCam, two additional copies of the same Panchromatic filter are available in case the primary Pan filter becomes obscured during the initial sampling attempt and additional attempts are required.

[Insert Figure 7]

[Insert Figure 8]

To reduce complexity, all three cameras possess identical focal planes, a 1k×1k frame-transfer Charge-Coupled Device (CCD) array (Figure 9). The arrays are used without their cover glass in order to reduce optical ghosting. They are controlled by a common Camera Control Module (CCM) (Figure 10). The dual-redundant CCM includes primary and redundant power supplies, motor/heater drivers, and controller boards, which interface both to the spacecraft and their detector assemblies. The CCM operates all mechanisms, heaters, and light-emitting diodes (LEDs) used for indexing and calibration. It also gathers housekeeping temperatures, voltages, currents, and software/hardware states.

[Insert Figure 9]

[Insert Figure 10]

The three OCAMS imagers are mounted on the –X, +Y quadrant of the OSIRIS-REx science deck, as can be seen in Figure 11. Their coordinates are given in Table 3. The MapCam is mounted closest to the science deck's edge; the SamCam is positioned close to the SRC and tilted so as to view the deployed Touch And Go Sample Acquisition Maneuver (TAGSAM) arm. The PolyCam is close to OTES.

[Insert Figure 11]

[Insert Table 3]

In the following, we describe each of the three imagers in greater detail.

## 2.1 PolyCam

The OSIRIS-REx Polyfunctional Camera, PolyCam, is a 20-cm-wide aperture, F/3.15 Ritchey-Chretien telescope. At infinite range, it possesses a focal length of 629-mm and a plate scale of 13.5 µrad/pixel. At 200-m range its focal length is 610-mm and its plate scale is 13.9 µrad/pixel. Its field of view is 0.8° wide (Table 1). A view emphasizing the sharply angled sugar scoop baffle is shown in Figure 12.

[Insert Figure 12]

The optical design depicted in Figure 4 combines several elements:

1. The slightly hyperbolic primary ($K = -1.271$) and the strongly hyperbolic secondary mirror ($K = -10.414$) are constructed from Zerodur, the low coefficient of thermal expansion (CTE) lithium-aluminosilicate glass ceramic produced by Schott AG. The back surface of the primary mirror has a double arch structure mounted to the 6061-T6 aluminum base housing using bi-pod flexures made of Invar 36, another low CTE material. The bottoms of the bipod flexures are bolted to the base housing, while the tops are bolted to three pucks, also made of Invar 36, bonded to the mirror. Both Invar 36 parts were heat-treated after rough machining and again after final machining.

2. The optical support tube (OST), also made of Invar 36, maintains the primary and secondary mirrors at a precise distance from each other to tight tolerances over the operational temperature range of the camera. The OST is scalloped at its bottom-edge mounting surface; 12 tongues provide compliance at the mounting interface between the dissimilar materials of Invar 36 and Al. Annular stray light control vanes and a secondary mirror hub with a four-vane spider, also made of Invar 36, are laser welded to the OST.

3. A focusing mechanism, constructed inside a 6061-T6 aluminum case, is built around a two-element radiation-hardened field corrector (the first element of which is actuated by the focus mechanism) (Figures 5 and 6). The focus mechanism also rotates a calibration/solar-blocking shutter in and out of the optical axis.

4. A stray light control sugar scoop baffle with its own annular stray light control vanes is bolted to the top of the OST above the secondary mirror spider. Primary and secondary mirror stray light control baffles, made of Ultem 2300, are bolted through

5. The common OCAMS detector assembly package is accommodated within the base housing. Each assembly consists of three flex-connected circuit boards folded into a compact installed configuration. The design is similar in layout and concept to the MapCam and SamCam detector assemblies.

The PolyCam's focus mechanism accommodates range-to-target changes during mission imaging campaigns. In early concepts, refocusing the PolyCam used glass plates, mounted on a heritage filter wheel, which allowed a small number of finite ranges to be in focus. That concept would have permitted the telescope to focus at only a few object ranges and significantly constrained spacecraft operations. As a result, the refocusing plates were replaced by the focus mechanism concept, which greatly lowered operations risk.

The focus mechanism is capable of 8.3 mm of total travel, 2.6 mm more than its operational range of 5.7 mm. As the mechanism motor moves lens 1 through its full range of travel to achieve focus, it also rotates a sun-blocking shutter in and out of the optical path 40 times. This makes a shutter available within a few seconds throughout the travel interval, a fact that is exploited by the flight software which always moves to the nearest shutter location when safing the camera. The sun-blocking shutter is shaped like a circular wedge with an angular width of 85° that blocks the beam ~23% of the time. As with all OCAMS mechanisms, its operation is controlled directly by the spacecraft's flight software via the CCM by issuing a dynamically alterable command (Table 4). This flexibility is useful when the range to surface changes frequently (e.g., during the reconnaissance phase). The focus position can be adjusted dynamically based on ranging measurements provided to the spacecraft flight software (FSW) by the Guidance Navigation and Control (GNC) subsystem. For simplicity, each of the PolyCam's 40 accessible shutter rotations is divided into three valid focus positions, corresponding to shutter positions 120° apart. A single table defines and controls the focus mechanism's operation. A portion of the table is displayed in Table 2.

[Insert Table 4]

The PolyCam supports mission operations at several different ranges to the asteroid, each with its own resolution and sensitivity requirements. At infinity focus the PolyCam is required to image an object of magnitude 12, the anticipated brightness of the asteroid Bennu during the Approach phase. As the spacecraft nears the surface of the asteroid, major campaigns involving the PolyCam occur at ranges of 3.2 km (Detailed Survey Baseball Diamond), 750 m (Orbital B Site-Specific Surveys), and 225 m (Reconnaissance). During the Reconnaissance phase, the PolyCam images prospective sample sites and resolves objects 2 cm in diameter. This dimension corresponds to that of the maximum-sized regolith unit that can be successfully ingested by TAGSAM. A detail from an image simulated to reproduce the expected illumination, albedo, and geometry for this observation is shown in Figure 13. Additionally, the PolyCam backs up the MapCam in its mapping, shape determination, and optical navigation functions.

[Insert Figure 13]

Thermally, PolyCam is biased cold using a passive design. A radiator pointed to space radiates the heat produced by its detector electronics; four survival heaters controlled by the spacecraft flight software on two different circuits (CCD, focus mechanism, read-out electronics, and secondary mirror) keep critical components of the PolyCam warm when not in use. Four redundant operational heaters controlled by the OCAMS flight software (CCD, focus mechanism, base/primary mirror, and focus motor) are available to heat the optics, electronics, and mechanisms for operations. PolyCam is thermally decoupled from the spacecraft to maximize the effect of on-camera radiating surfaces and heaters. The resulting predicted flight operating temperatures are listed in Table 5 for the various subsystems.

[Insert Table 5]

Flexures are used at several locations throughout the camera to connect parts of the camera made from materials with dissimilar CTE: (1) Deck mounting flexures (Ti 6Al-4V) mount the 6061-T6 aluminum base housing to the carbon fiber spacecraft science deck. (2) Three Invar-36 bi-pod flexures connect the Zerodur primary mirror to the base housing. (3) The scalloped edge of the Invar 36 OST provides compliance to the base housing. In addition to providing compliance between materials with mismatched CTE, Z-shaped flexures are also used to provide compression and keep refractive optics in place. The wide range of thermal environments where the PolyCam must perform requires it to be athermalized. Separation between optics, the most important of which is the primary–secondary separation distance, is maintained largely by the use of near-zero-expansion materials like Zerodur and Invar 36. The judicious use of materials with nonzero CTE in strategic places such as the secondary mirror helps provides further athermalization.

### 2.1.1 Focus Mechanism Operation

The PolyCam focus mechanism uses a stepper motor to power a system of four gears that rotates a threaded helical hub holding lens 1 against a system of grooved rollers (Figures 5 and 6). As lens 1 moves closer to the secondary mirror and away from lens 2, the refractive power of the doublet is reduced and the focal length of the PolyCam increases. The object distance which results in best focus at a fixed image plane moves toward infinity. While the motor drives the lens 1 hub it also drives a wedge-shaped 6061-T6 aluminum shutter which blocks external light from the detector over a shading envelope approximately 85° wide. The side of the shutter facing the detector is grit-blasted to provide a matte finish that is used as a quasi-Lambertian reflector for a pair of green LEDs to provide an internal flat field. This internal calibration source can be used to monitor changes in the detector spatial responsivity due to aging or radiation damage. The linear travel range of this threaded hub corresponds to an operating focus range from 180 m to slightly beyond infinity. The focus mechanism leaves about 1.3 mm of margin (corresponding to about seven shutter rotations) at either end to accommodate temperature and workmanship-related resets of the PolyCam focus induced by launch vibration. The ability to focus on targets beyond infinity allows a simple through-focus test to very precisely determine best focus. A schematic diagram depicting the functional relationship between the four gears of the mechanism is shown in Figure 14. The curve describing the correspondence between motor step position and in-focus range is given in Figure 15.

[Insert Figure 14]
[Insert Figure 15]

The wedge-shaped shutter subtends an 85°angle, blocking just under one-fourth of a full rotation. This conveniently allows three valid imaging focus positions, 120° apart, to subdivide each shutter rotation. A focus table has been constructed that parses the full operating range into 90 valid positions (an excerpt is shown in Table 2). It is loaded into flight software and used to enable the automatic refocusing of the PolyCam in a stepped fashion as the range to target changes.

## 2.2 MapCam

The MapCam is a 125-mm F/3.3, five-element, radiation-hardened refractive system based on a telephoto lens design (Figs. 2 and 16). It possesses a 68-µrad/pixel plate scale, a 4° field of view, and incorporates a filter wheel. Its best focus ranges from 125 m to infinity (Table 1). It adds one refocusing filter that adjusts focus to a 30-m range (Fig. 1).

[Insert Figure 16]

A medium-resolution imager, MapCam's optical design is realized by:

1. A fore-optics positive-negative achromatic doublet and a rear-optics triplet consisting of two positive elements and one negative element (Figure 2)
2. An eight-filter wheel with one Pan filter thickness optimized for operation at ranges between 125 m and infinity and another optimized for focus at ranges between 25 and 35 m (Figure 7)
3. Two sun-safe blocking plates, one of which can also serve as an on-board flat-field tracing calibration
4. Four 60- to 100-nm-wide color filters with passbands based on the Eight-Color Asteroid Survey filter passbands (Tedesco et al. 1982); the thicknesses of the filters are optimized to partially compensate for chromatic aberrations of the MapCam optics over their spectral band and to provide best focus for operation at ranges between 500 m and infinity

The MapCam's mechanical design matches the thermal expansion of three types of the radiation-hardened lens material (N-BK7G18, N-F2G12, and N-K5G20) to a titanium Ti6Al4V lens cell and lens flexures. The base, containing the detector read-out assembly, is constructed of 6061-T6 Al. The deck flexures are Ti6Al4V.

The MapCam's two Pan filters transmit between 500 and 800 nm (Figure 17), the same wavelength range as PolyCam and SamCam. One is optimized to achieve a focus from 125 m to infinity, and the other is optimized to achieve a focus near 30 m. The four color filters are aligned with several filters in the Eight-Color Asteroid Survey (ECAS) filter set, except for the blue filter. This filter is shifted toward longer wavelengths from the ECAS blue filter in order to improve optical and radiometric performance and minimize aging effects due to radiation. The MapCam backs up the PolyCam for asteroid acquisition and sub-cm imaging by reducing the ranges of such observations from 2,000,000 to 500,000 km and from 200 to 30 m, respectively. To carry out the latter observation, the MapCam uses its Pan 30-m filter.

[Insert Figure 17]

MapCam's parameters allow it to survey Bennu from a safe and convenient stand-off distance. The whole surface—northern and southern hemispheres—can be mapped at less than 1-meter resolution by a campaign consuming less than 9 hours while Bennu spins through two rotation periods.

The transmission curves of the four color filters (B, V, W, X) are shown in Figure 17, where they are compared with the spectral behavior of the solar spectrum, the OCAMS detector's quantum efficiency, the Pan filter, and the spectrum of a typical laboratory blackbody (quartz-tungsten halogen light filament). The filters are custom-designed to possess high transmission in band and sharp cutoff to minimize their optical ghosting potential and to have large out-of-band blocking. The wide-band filters are intended for multiple purposes:

1. To acquire disk-integrated color observations of the asteroid's phase function and light curve and determine how these relate to previous Hubble Space Telescope and ground-based studies of Bennu as a point source
2. To provide high-spatial-resolution color band-ratio maps that differentiate and classify the asteroid's surface into color-units relatable to the more detailed classification performed by OVIRS
3. To use the color-unit classes to rate the various prospective sample sites as higher, or lower, science-value targets
4. To connect disk-integrated color observations of asteroids directly to their surface resolved properties, thereby aiding in the spectral interpretation of asteroid color data from ground-based observations

The B filter is responsible for gauging the degree of radiation damage that the asteroid has experienced in its lifetime. There is some indication that that long-term space weathering on organic asteroids may make them slightly bluish (Chapman 1996; Kaňuchová et al. 2012; Lantz et al. 2013; Fornasier et al. 2014; Moroz 2004). Bennu's color, spectral shape, and albedo are consistent with B-type asteroids, a class characterized by spectral enhancement of its blue reflectance (Clark et al. 2011). The V and X filters are intended to establish a continuum through the visible and near-infrared light reflected from Bennu's surface. The flux observed in the W filter compared to V–X provides a diagnostic for the presence of iron-bearing phyllosilicates (hydroxyl sheet silicate minerals) in Bennu's regolith and is well-aligned with an established broad, solid-state absorption feature near 710 nm (Vilas 1994; Clark et al. 2011).

MapCam is mounted on the –X, +Y quadrant of the OSIRIS-REx science deck, similarly to PolyCam, but 10 cm closer to the high-gain antenna (+X) and 25 cm further out toward the +Y edge of the deck (Table 3, Figure 11). MapCam is passively cold-biased, but unlike PolyCam and SamCam, the MapCam uses two radiators (Figure 16). These radiators are counterbalanced by three survival heaters controlled by the spacecraft flight software (CCD, lens assembly, and read-out electronics) and four operational heaters controlled by the OCAMS controller software (filter wheel motor, lens assembly, filter wheel housing, CCD). Of the two radiators, one is a dedicated plate, and one is a painted area on the upper portion of the riser housing. Predicted flight temperatures can be compared in Table 4.

## 2.3 SamCam

SamCam is a 24-mm F/5.5 radiation-hardened refractive system (Figure 18). It possesses a plate scale of 354 µrad/pixel and a field of view of 20.8°. Its primary imaging activity occurs when the spacecraft is very close to the asteroid. (Its nominal best-focus range is ~3–30 m), as shown in Figure 1.

This relatively low-resolution imager features:

1. A five-element refractive system based on a double-Gauss design (Figure 3)
2. A six-position filter wheel positioned externally to the optical cell (Figure 8). This filter wheel houses:
    a. Three identical Pan filters (500–800 nm bandpass)
    b. Two blocking filters, one of which doubles as a calibration tracing system
    c. One diopter lens that allows imaging at a distance of 2 m
3. The common OCAMS detector package

[Insert Figure 18]

Through one of its three Pan filters SamCam images the 15–20 minute sampling event at the highest cadence possible in order to capture the moment of sample acquisition, including the activation of the nitrogen canisters. Post-TAGSAM, the camera acquires images to verify the presence of sample external to the sampling head and determine if particles larger than 5–6 mm are present. Such particles could interfere with the insertion of the TAGSAM head into the sample return capsule (SRC).

SamCam is mounted closer to the SRC than either of its siblings, reflecting its primary task of observing the deployed TAGSAM head and sample site during sample acquisition. Unlike either the MapCam or PolyCam, which are collinear with the +Z axis, the SamCam is tilted 9.4° within the Y–Z plane toward the –Y direction in order to maintain a view centered on the TAGSAM head when the sample arm is fully extended.

SamCam is not thermally coupled to space as well as the other two cameras, allowing its thermal inertia to keep it cooler for longer when the spacecraft approaches the hot surface of Bennu. Otherwise, its thermal design is much like that of the other two cameras, using radiators and heaters to maintain a temperature range for the optics, electronics, and detector around 0°C. It makes use of two survival heaters (CCD and read-out electronics) and four operational heaters (lens assembly, filter wheel housing, CCD, and filter wheel motor). Predicted flight temperatures can be observed in Table 4.

## 2.4 Electronics and Software

OCAMS consists of several electronics subsystems controlled by software and firmware that:

1. Act to record an optical image of an external scene, digitize it, packetize it, and transfer it to the spacecraft's Data Telemetry and Command Interface (DTCI);
2. Operate a network of housekeeping sensors monitoring temperature, voltage, current, mechanical position, and software state to record their values at periodic intervals, package them, and submit them to the spacecraft both as separate housekeeping telemetry packets and as ancillary housekeeping information bundled with each recorded image; the latter is incorporated into the header of each OCAMS image, whose native format is FITS (Flexible Image Transport System);
3. Actuate three motors that operate three mechanisms, one for each camera; two of them are filter wheels; one is a focus mechanism;
4. Supply and regulate power to visible Light-Emitting Diodes (LEDs) as part of an in-flight calibration tracing system;
5. Supply and regulate power to infrared LEDs which illuminate fiducials to enable the deterministic control and positioning of the camera mechanisms.

Major electronic subsystems include the detectors, the Detector Assemblies (DA), and the CCM itself, which is controlled by its field-programmable gate array (FPGA). The detector assemblies and detectors are combined with the optics into the three sensor heads. A block diagram depicting system functionality and subsystem relationship is displayed in Figure 19. It can be seen that 3 × 2 boards occupy the CCM Chassis: two identical computer processing unit (CPU) boards, two Motor/Heater boards, and two low-voltage power supplies (LVPS).

[Insert Figure 19]

### 2.4.1 Camera Control Module

The CCM (Figure 10) is the electronic servicing system for the three OCAMS sensor heads. It uses commands submitted through the spacecraft virtual machine language (VML) engine to operate the three OCAMS imagers throughout the mission (the list of commands is shown in Table 4). The CCM can operate through either a primary or redundant side (Figure 19). In between commands, it regulates power, monitors the health of all OCAMS subsystems, maintains control of all operational (as opposed to spacecraft-controlled survival) heaters, and, every 2 seconds, collects and sends housekeeping data to the spacecraft to relay to the ground. Running on the CCM, the OCAMS flight software accepts commands from the spacecraft and controls the operation of the various subsystems: the detector assemblies, mechanisms, operational heaters, and calibration tracing lamps. It collects and transmits the housekeeping and imaging packets. Spacecraft commands initiate detector assembly power-on and entry into standby mode. Imaging can be collected through either of the detector's two taps or through both at once. Using the detector assembly's own digital processing front-end, the CCM can record detector read-out diagnostic information and image test patterns. It can read and write detector read-out registers and dump memory locations. And it can provide data that can be used to characterize both vertical and horizontal charge transfer efficiency through an extended-pixel edge response (EPER) technique.

The CCM communicates to each camera detector assembly (DA) through a synchronous, Low-Voltage Differential Signaling (LVDS), serial interface as defined in ANSI/TIA/EIA-644 (Figure 20). It provides each DA with ±24 V and +5 V regulated power rails. Each DA is functional within 10 seconds of turn-on. It is controlled with a menu of 4-byte commands.

[Insert Figure 20]

The CCM itself is constructed around an 8051 microcontroller IP processing core hosted on an Actel RTAX-2000 Field-Programmable Gate Array (Figure 20). This architecture was selected for its simplicity, speed, and heritage. It is implemented on two redundant sets of three boards: a Data Processing Unit (DPU), a Low-Voltage Power Supply (LVPS), and a motor/heater interface/driver. Both high-speed data links, between DA and CCM, and between CCM and spacecraft, are LVDS interfaces. They are referenced to 20- and 24-MHz clocks, respectively, for DA/CCM and CCM/spacecraft. Internally, the CCM FPGA is clocked at 36 MHz.

The parts used by both CCM and DA were chosen under a strict regime controlled by Goddard Space Flight Center (GSFC) EEE-INST-002 and PEM-INST-001 (Rev 2007). A parts control board (PCB) was established, and a Parts Identification List (PIL) and a Project-Approved Parts List (PAPL) were developed and maintained. Certain parts were rejected for insufficient radiation tolerance, either to proton radiation or low-dosage ion radiation, or insufficient thermal tolerance; others, including the detectors, were tested in order to demonstrate sufficient radiation tolerance to solar proton radiation.

As mentioned above, the CCM outputs some 97 housekeeping variables every 2 seconds as an independent data set Application Process ID (APID) 75 or attached to an imaging data set (APIDs 76–78). Temperatures, voltage and current states, lamp and mechanism states, software state, and key software variables are included.

In contrast to the OCAMS sensor heads, the CCM is located on the underside of the spacecraft instrument deck and is thermally coupled to it.

## 2.4.2 Detector Assembly

The DAs were provided by Space Dynamics Laboratory at Utah State University in Logan, UT. A block diagram describing their operation is shown in Figure 21. They consist of three boards each: one hosting the interface, LVDS drivers, and Actel RTAX-1000 FPGA; one containing the CDS (Correlated Double Sampling) Analog-to-Digital (A/D) converter signal processing chip, the clock drivers, regulators, and analog ground; and the third, a D-shaped board, mounted to intercept the optical axis, which contains the CCD. The position of the light-sensitive portion of the chip is precisely measured (using a Coordinate Measuring Machine) to titanium inserts bonded to the DA board. The D-board provides CCD bias voltages, video signal buffers, and the CCD interface.

[Insert Figure 21]

The analog front-end device implementing the CDS read-out of the CCD output, and the A/D conversion, is a Texas Instruments LM98640W-MLS dual-channel 14-bit signal processing chip. Its pairing with the FPGA which controls the analog signals expected by the CCD allows significant flexibility in the operation of the OCAMS detector through the manipulation of registers. Spacecraft commands can initiate DA power-on and imaging through either of two detector taps, record diagnostics, test patterns, read and write registers, dump memory locations, and measure both vertical and horizontal charge transfer efficiency through an extended-pixel edge response technique.

The OCAMS CCD (Figure 9) requires a choreographed power-on/power-off sequence in order to forestall burn-through damage to its substrate due to premature loss of substrate bias. Opto-couplers are implemented to monitor CCD bias voltage status while the detector assembly FPGA reacts to possible faults. A bias fault condition causes the FPGA to immediately remove power from the CCD.

The OCAMS sensors are used without windows to minimize the possibility of optical ghosting. They are equipped with an internal platinum resistance thermometer (PRT) temperature sensor. An external thermistor to the rear of the device allows a redundant CCD temperature monitor.

The DA FPGA controls the read-out electronics through five top-level blocks. The command executor handles all host commands, populates and monitors the control and status registers, and controls the flow of data between blocks. The host interface handles communication protocols. The CCD controller executes imaging commands and handles CCD clocking patterns. The ADC (Analog-to-Digital Converter) controller constructs the interface logic for the analog-to-digital converter; it synchronizes the FPGA control-status register values with the ADC via the SPI (Serial Peripheral Interface). Finally, the SDRAM (Synchronous Dynamic Random Access Memory) arbiter controls the flow of data between the FPGA and the 128 MB SDRAM.

### 2.4.3 Detector

The frame-transfer detector at the heart of OCAMS is a 1k×1k CCD array, with $6.5 \times 8.5$ µm pixels on an 8.5 µm pitch, provided by Teledyne-DALSA's Custom Division in Waterloo, Ontario. It was selected based on a series of ranked criteria informed by mission imaging requirements (Table 6). The detector exhibits a relatively high dynamic range, a quantum efficiency spectrum well-aligned to that of the solar spectrum, acceptable read noise, low dark current, and short shutter time. Its format minimizes downlink needs as well as the distortion. The chip also has a powerful anti-blooming capability, known ionizing radiation tolerance against gamma radiation, and recent flight heritage.

[Insert Table 6]

The detector itself consists of 1024 by 1024 active pixels surrounded by a:

1. 4-pixel transitional zone
2. 24-pixel-wide covered columns left and right
3. 6-pixel-deep covered rows top and bottom
4. 16-pixel-wide electronic lead-in columns left and right (bias columns) (Figure 22)

TD Waterloo was contracted to provide all necessary device qualification and validation, except for proton radiation, which was performed by the UA team in separate tests at UC Davis' Crocker Nuclear Labs.

[Insert Figure 22]

# 3 Instrument Performance

As a requirements-driven design and mission system, OCAMS' ground calibration was fully embedded within an extensive, but focused, verification effort (Merrill and Williams 2016). This program measured optical and radiometric performance, spectral responsivity, and stray light exclusion. The measurement of absolute responsivity was derived as an outgrowth of the verification of minimum detectable radiance for all three cameras. (Bennu's surface environment is believed to be one of the darkest in the solar system.) The dependence of optical and radiometric performance on temperature was derived from environmental testing. Stray light sensitivity followed upon the verification of stray light requirements. Geometric distortion calibration is relegated to the in-flight calibration for MapCam and SamCam. This calibration was performed for the PolyCam as a post-environmental ATLO (Assembly, Test, and Launch Operations) floor verification.

In summary, the OCAMS ground calibration measured:

1. Optical resolution performance in the form of modulation transfer function (MTF), which was trended throughout the assembly and testing of the cameras; related metrics, including ensquared energy and root-mean-square (RMS) spot size Full-Width at Half-Maximum (FWHM), were also derived.

2. Radiometric performance in the form of the minimum detectable radiance and maximum detectable magnitude; related metrics included vignetting and flat field uniformity.
3. Detector characterization, including read noise, photon transfer curve shape, and dark current generation rate.
4. Spectral responsivity.
5. Mechanism functionality and mechanical calibration.
6. Field of view/pointing.
7. Stray light characterization, including in-field and out-of-field stray light and the characterization of ghosts.
8. Distortion (for the PolyCam).

The advantages of embedding the calibration effort so completely within the verification effort outweighed the disadvantages. With eight boxes to calibrate (three cameras + electronics controller, EQM [Engineering Qualification Model], and FM [Flight Model] versions), and some 2000 requirements to verify, great emphasis was placed upon efficiency, straightforward metrics, and strict Configuration Management (CM) of alterations. The major disadvantage—the speed with which the calibration was conducted leaving little time to address inevitable flaws and deficiencies discovered during the calibration process—was remediated by a robust program of special tests, including at the spacecraft level, which upgraded verifications and calibrations in order to address deficiencies revealed after evaluating data produced by the standard calibration program. More than 130 special tests were conceived and performed in this effort. Special test requests were submitted through the CM process, rapidly reviewed, implemented, and incorporated into the standard calibration program.

After each instrument completed its standard calibration regime, a hardware analysis review was convened in which the results of the calibration were reviewed and a decision taken whether to proceed.

## 3.1 Optical Resolution Performance

Optical resolution performance for each of the systems was measured by imaging back-illuminated slightly slanted slit targets at nine points within each camera's field of view. From these high-contrast slanted-slit images, densely sampled line-spread profiles were derived in the direction perpendicular to the slit length (Golish et al. 2014). The amalgamated line-spread function (LSF) profiles are used to compute the modulation transfer function (MTF) of the system along the direction perpendicular to the slit. The system MTF is plotted as a function of spatial frequency and used as our optical resolution performance metric. The technique is flexible and can be used while the imagers are housed and tested in environmental chambers (system level) and while they were mounted on the spacecraft (observatory level) during ATLO. For the MapCam and PolyCam, laboratory test bench collimating systems were used to project the image of the tilted slits and point sources into each camera's field of view during ambient and environmental testing. For the SamCam, real images were observed at the camera's specified range. In all cases, projected or real range to target was altered in order to explore the dependence of optical resolution performance on range in order to verify its behavior accorded with predictions (Figures 23–26).

[Insert Figure 23]

[Insert Figure 24]

[Insert Figure 25]

[Insert Figure 26]

In general terms, to describe the resolution characterizing extended-object observations of Bennu, the rule of thumb was adopted that a small feature such as a pebble or distant boulder is considered fully resolved when its diameter subtends 5 pixels for a well-focused optical system. We define such a well-focused system as one with an optical resolution performance that is approximately Nyquist-limited, characterized by an MTF at Nyquist spatial frequencies of 20%. Writing the important resolution requirements in this way minimizes the possibility of aliasing while providing a significant amount of margin against degradation in both optical and radiometric performance because S/N (signal-to-noise) levels at Bennu for the three cameras are expected to routinely exceed 100.

This rule of thumb is formulated more precisely by requiring all 15 operational optical pathways (six for MapCam's five filters + focus plate, four for SamCam's three filters + diopter, and five for PolyCam's focal state at the five different ranges used to characterize its optical resolution performance during its fabrication and testing) to separate two spots described by Gaussian illumination profiles with a 3σ confidence (99.7%) and evaluating them on their ability to do so:

$$MTF(F) \geq \frac{3}{S/N} \quad (1)$$

where $F$ is the spatial frequency characteristic of the spot separation, usually Nyquist (59 lp/mm for our detectors) or half-Nyquist, $S/N$ is the signal-to-noise ratio characterizing the data, and $MTF$ is the modulation contrast at a given $F$. The 3 in the numerator refers to the 3σ confidence characterizing the separation. The above relationship states that if the S/N characterizing the observation is around 20, then an MTF at Nyquist of at least 15% is required to separate two points with a 99.7% confidence; if the S/N attains 100, then an MTF equal to only 3% is sufficient. If a smaller confidence on the detection is tolerable (1 or 2σ) then the MTF (contrast) is allowed to be correspondingly lower as scaled by the S/N characterizing the measurement. If an optical system records images with larger signal-to-noise due to its higher radiometric sensitivity, then a lower contrast system may be fielded. As ever, optical resolution performance and radiometric performance are traded one against the other. The end product, whether satellite or plume discovery, identification of cobbles from 3.5 km, or pebbles from 200 m, is always the final arbiter, after as many of the operational details governing the actual illumination geometry as can be learned in the early stages of mission planning are included. Each optical pathway, distinguished from the others by the use of a distinct filter or focus range, is associated with a driving case—usually an important imaging campaign with an anticipated and specific range, lighting condition, and ground-track velocity—that determines a single MTF criterion at a given spatial frequency. Some eight separate optical resolution performance requirements governing resolution, ensquared energy, depth of field, contrast, and LSF itself controlled the performance of MapCam. SamCam had six such requirements and PolyCam 13. These were the true instrument driving requirements, in the sense that they determined the optical and radiometric characteristics of the final designs.

The actual MTF measurements observed the slit target at nine locations within the field of view of each optical system (Figure 27). The slit targets are designed so their width projects to a sub-pixel (~1/10 pixel typical) feature when imaged onto that camera's focal plane. The slight angles, typically 5° to 7° to the nominal row-column alignment of each detector, allow a sub-pixel sampling of the camera's line-spread function (along two orientations, vertical and horizontal) to be acquired; more than 100 points could typically be acquired to trace out the line-spread function (Figure 28).

[Insert Figure 27]

[Insert Figure 28]

Optical resolution performance was measured for a range of simulated object distances; the simulated distance is set based on computer-controlled motion of the target at the focus of the collimator. The accuracy of the simulated range was then independently verified using a custom-designed pentaprism divergence-angle measurement system. The pentaprism is placed within the collimator beam and translated perpendicularly across the beam in order to measure the angles of the incoming rays by their degree of displacement from a reference position corresponding to a perfectly collimated ray ensemble (from ∞) (Figure 29). In the infinite reference case, the displacement measured by the pentaprism is essentially zero (limited by collimator aberrations). At all finite ranges, a nonzero displacement is measured and related to the range.

[Insert Figure 29]

The results for each camera are shown in the form of line-spread function full-width-at-half-max vs. range-to-target (Tables 7a-c) and also in the form of system MTFs for each camera during ambient conditions before and after vibration testing and during thermal-vacuum testing (Figures 30–34). Results for horizontal slits (the worst case) are also shown. The horizontal slits sample the widest dimension of each pixel's 8.5×6.5 µm active area. System performance in the other dimension is typically 25% better at Nyquist or ½ Nyquist.

[Insert Table 7(a–c)]

[Insert Figure 30]

[Insert Figure 31]

[Insert Figure 32]

[Insert Figure 33]

[Insert Figure 34(a–e)]

### 3.2 Radiometric Performance

Bennu orbits the Sun between 0.89 and 1.36 AU. Its mean diameter and expected density of 500 m and 1.3 g cm$^{-3}$, respectively, result in typical gravitational accelerations of 2–7 µ$g$ and orbital velocities of 5–15 cm sec$^{-1}$ (Chesley et al. 2014; Hergenrother et al. 2016). The range to target falls between several hundred to several thousand meters from the body's center, implying angular velocities of surface features of 0.01 to 0.75 mrad sec$^{-1}$. At even the most rapid relative velocity, and at most ranges to target, an exposure time of 20 msec (1/50 sec) barely produces a motion blur equaling a pixel. However, on several campaigns the demands of mission operations are better served by acquiring images while continually slewing the OSIRIS-REx spacecraft pointing at a rate of up to 2 mrad sec$^{-1}$ and at times accepting blurs on the order of a pixel or more (Mink et al. 2014).

The near-Earth asteroid Bennu is fully as dark as the nucleus of comets such as 1P/Halley, 81P/Wild, 19P/Borrelly, or 103P/Hartley and of asteroids such as 253 Mathilde (Hergenrother et al. 2013). In addition, Bennu possesses a disk-averaged phase function whose measured reflectance plummets steeply at high phase angle (Hergenrother et al. 2013, Figure 5). Various disk-integrated photometric models (Figure 35), derived based on different conventional functional-form assumptions, display a disparity between noontime observations at the equator at perihelion and near-terminator observations at aphelion that can be 3 orders of magnitude (Takir et al. 2015), 2 orders of magnitude from the reflectance itself and 1 order of magnitude due to the cosine effect itself. This large dynamic range is accommodated by combining a wide dynamic range detector with exposure times ranging from millisecond to tens of seconds. In addition to these considerations, the signal-to-noise ratio characterizing the resulting observations strongly affects the overall optical resolution performance, as shown in Eq. (1).

[Insert Figure 35(a–c)]

Many of the definitive campaigns that gather the information needed to enable successful sampling require the spacecraft to orbit near the solar-terminator (Berry et al. 2013, 2015; Beshore et al. 2015; Lauretta et al. 2015; Williams et al. 2017). In this attitude, the weak gravitational force balances the solar radiation pressure on the OSIRIS-REx spacecraft. Illumination at the terminator is reduced by a factor of more than 10 from that of the darkest parts of the sunlit hemisphere and by a factor of 1000 compared to the sub-solar point (Figure 35(a–c)). Incidence and emission angles at this vantage point span 45°–60° for the most observable locations on the asteroid, resulting in phase angles that routinely exceed 100° and radiances that are a fraction of the afternoon radiances of locations on Bennu's surface. Anticipating this illumination geometry, radiance requirements for each camera were among the most highly scrutinized. These 16 separate requirements on the MapCam, SamCam, and PolyCam Minimum Detectable Radiance (MDR) drove unusually fast designs for the MapCam and PolyCam, and a relatively fast design for the SamCam (Table 1). We say "unusually" because, after all, the spacecraft spends a good fraction of its lifetime at a solar range near 1 AU in very bright sunlight and should not require such fast cameras for any "normal" body with even twice the reflectance and surfaced with a regolith constructed from relatively smooth cobbles and pebbles (e.g., Itokawa) (Fujiwara et al. 2006; Saito et al. 2006; Yano et al. 2006; Miyamoto et al. 2007; Marshall and Rizk 2015). The combination of Bennu's very low albedo and steep phase function demands fast imagers: not only is its regolith of very low reflectance, but at the microscale, it is rough.

The radiometric performance of the OCAMS imagers was verified and calibrated relative to the internal surface of a 20-inch integrating sphere illuminated by a 3400 K QTH (quartz-tungsten-halogen) laboratory blackbody source (Figure 36) for the MapCam and SamCam and by a self-illuminated Alnitak Flat-Man flat panel (Figure 29) for PolyCam. The source's calibration is traced to a NIST source using a proxy reference: a calibrated standard detector (OL DH-300C) provided by Gooch & Housego. This detector is mounted perpendicular to the sphere's 8-inch exit port, centered, and placed at a measured distance, (Figure 36). The sphere's surface is simultaneously observed by a second calibrated standard detector reference (OL 730-5A) permanently mounted to the sphere, with responsivity identical to the first to within 1%, in order to monitor temporal changes in light level. Both detectors were freshly calibrated by the manufacturer before use.

[Insert Figure 36]

The combination of the calibrated detectors, one positioned in front of the integrating sphere (proxy detector) with its well understood solid angle view of the source, and the other detector at the sphere's equator with a direct view of the interior, allows the precise determination of the solid angle of the reference detector's observation geometry when mounted to the sphere:

$$\Omega_{proxy} = \pi \sin^2 \theta = \pi \frac{r^2}{r^2 + d^2} \quad (2)$$

where $\Omega_{proxy}$ is the solid angle of the open integrating sphere port as seen from the proxy detector, $\theta$ is the half-angle of the solid angle, $r$ is the radius of the integrating sphere port, and $d$ is the distance from the detector to the integrating sphere port (Figure 37). Other quantities above are presented in Table 8.

[Insert Figure 37]

[Insert Table 8]

The results are quoted both relative to the laboratory source and also relative to a solar blackbody (5780 K). The derived responsivities are presented in Table 9. If the OCAMS detectors, by analogy to photographic film, are assigned an ISO rating scaled to a working signal level and resulting signal-to-noise ratio, then one derives the curve displayed in Figure 38. At a 40:1 signal-to-noise ratio (SNR), the ISO rating is around 830, and at a 10:1 SNR, the ISO rating is 4700.

[Insert Table 9]

[Insert Figure 39]

It is worth noting that typical signal levels for laboratory light (e.g., the illumination level inside of the Lockheed Martin high-bay where the OCAMS imagers were integrated onto the spacecraft) are 6000 DN (data number) for a 100 ms SamCam exposure. The high-bay was measured to have an illuminance of 160 lux, corresponding to 0.51 $Wm^{-2}$ $sr^{-1}$, assuming a typical luminous efficacy of 100 lm/W. Bennu's surface is somewhat brighter under the Sun's full illumination at 1 AU, with an estimated illuminance level around 410 lux and a radiance of 1.4 W $m^{-2}$ $sr^{-1}$ at 65° incidence and 0° emission requiring a 40 ms SamCam exposure to generate a peak signal level of 6000 DN.

### 3.3 Spectral Performance

The relative spectral responsivity for all seven distinct optical paths was measured to verify seven spectral-related requirements. The configuration used is shown in Figure 39. A single-pass monochromator (Acton SL-150), backlit with a QTH source, filled an integrating sphere with light over a 12-nm-wide band. It was observed by the unit-under-test (UUT) and a standard detector (Gooch and Housego OL 730-5A) calibrated for spectral responsivity. The scans of the MapCam color filters were performed at both 1-nm and 10-nm spacings. The scans of the Pan filters for all cameras were performed with a10-nm spacing.

In an effort to ensure that MapCam is capable of detecting the presence of the 710 nm absorption feature indicative of phyllosilicates, MapCam's ability to detect relative radiometric differences equal to 2% or less was demonstrated in a simulated observation in our laboratory.

[Insert Figure 39]

Order-blocking filters are used to prevent second-order diffraction on the monochromator grating from introducing light at one-half the wavelength required. For wavelengths from 350 to 700 nm, a UV-blocking filter is used. For wavelengths above 700 nm, an order-blocking filter is used which blocks wavelengths lower than about 665 nm. The overlap region between the two datasets is used to tie the short- and long-wave datasets together.

The results are shown in Figures 40–42.

[Insert Figure 40]

[Insert Figure 41]

[Insert Figure 42]

### 3.4 Stray Light Performance

During the design phase, in order to simulate and better exclude stray light from each system's focal plane, stray light models were created for each camera in FRED, Photon Engineering's ray-tracing and optical design utility. Both out-of-field and in-field stray light for the most demanding illumination geometries during various mission campaigns were predicted and analyzed. When not already known, surface reflectances for all critical camera surfaces were measured, modeled, and incorporated into the relevant model. The results guided and verified each camera's baffle design.

The resulting baffle structure for all three cameras began from the outside with external sunshades cut at a narrow angle to the optical axis—the resulting structures are termed sugar-scoop baffles—sufficient to exclude the direct solar beam during the campaign when its closest angle to the Sun was expected. This geometry occurs during the Detailed Survey phase (Sect. 5.4), when Bennu is surveyed for dust and gas plumes in forward-scattered solar light and the high sides of the three sugar-scoops are oriented toward the Sun. This operation was facilitated by mounting the three OCAMS imagers so that all of their sugar-scoop baffle high-sides are oriented toward the +X side of the spacecraft. The +X side is also where the spacecraft's high-gain antenna is mounted and used as a de facto sunshade throughout the mission, aided by several perpendicular baffles erected alongside it on the science deck.

During ground verification and calibration, the success of the design was tested by several stray light testbeds: a bright point source that measured out-of-field rejection along two perpendicular axes and a uniform bright field target surrounding designated dark regions that measured in-field rejection. The results for this testing are shown in Figures 43–47. Ghosting was also tested by imaging a very bright source at different positions in and out of the field of view of the cameras.

[Insert Figure 43]

[Insert Figure 44]

[Insert Figure 45]

[Insert Figure 46]

[Insert Figure 47]

Glints from the sample head into the SamCam images of the TAGSAM event—especially specular reflections—have been extensively studied. The original metallic surface finish of the sampling arm and head was reworked and shielded in order to remove or reduce the effects of single-bounce interfaces threatening to swamp the desired signal of the dark asteroid surface with the series of SamCam images acquired during the sampling maneuver.

## 3.5 Detector Performance

The CCD detectors residing at the focal plane of each of the OCAMS imagers were selected because they provided a relatively high dynamic range with low dark current, acceptable format, read noise, shutter speed, pixel pitch, anti-blooming, and a heritage design with radiation-tolerant features. Their fabrication process was relatively mature, and they were qualified at delivery.

### 3.5.1 Noise Results

Maintaining the electronic detector readout noise (read noise) level to the required level of 50 electrons at room temperature (less than 12 DN) was a primary concern of the detector readout and operation within its active electronic environment. Common-mode chokes and point-to-point grounding were employed to suppress power supply noise and grounding loops through the design's implementation to a level of 8.2–8.6 DN at room temperature, rising to 9–10 DN at colder detector temperatures. Each camera possesses an individually recognizable noise pattern, which was maintained essentially unchanged throughout all electromagnetic susceptibility testing.

*3.5.2 Linearity*

At low signal levels, a nonlinear response is observed in the processed images. An investigation is ongoing to characterize this effect and understand its origin.

*3.5.2 Radiation*

The OCAMS detectors had to successfully pass a radiation screening. This screening occurred on three separate sessions at Crocker Nuclear Labs in Davis, CA, where several Grade A and B engineering versions of the Teledyne DALSA detector were subjected to proton irradiation. These engineering and flight-lot devices were irradiated with 64-MeV protons to levels comprising two to six times the dose expected during the full mission at the position of the OCAMS focal planes, corresponding to a total ionization dose (TID) equivalent of 0.6, 1.2, 1.4, and 2.0 krad.

Irradiation of the focal planes produced hardly any change in the photon transfer curves, a negligible decrease in both vertical and horizontal charge transfer efficiency, and an increase in bulk dark current by a factor of one to two.

The irradiated detectors do display a significant increase of hot pixels. Hot pixels show a dark current generation rate greatly elevated relative to the rest of the detector plane. However, they seem readily removable, using dark frames taken at similar temperatures, reasonably close in time (e.g., minutes to hours) to the target frames.

More significant is the radiation-induced onset of flicker noise in the dark frames. This random telegraph signal noise (RTS), has a temporal impermanence that could potentially make it problematic. Its distinctive signature, that of an individual pixel whose signal is elevated abruptly over that of its neighbors, has the appearance of a cosmic ray strike. Therefore it is removable using cosmic-ray-like removal algorithms based on statistical comparisons with neighboring pixels.

The flickering pixels, the hot pixels, and the bulk dark current all decline with decreasing exposure times and decreasing temperatures. Since the majority of the imaging campaigns occur at temperatures and exposure times well below those used in the pre- and post-irradiation testing, one of the main conclusions of the radiation testing was that the OCAMS detectors have significant margin toward radiation exposure when used as the focal plane for the three OCAMS cameras.

## 3.6 Focus Mechanism Performance

In practice and as mentioned in Sect. 2.1.1, the operational range of the PolyCam focus mechanism is divided into 80 valid positions within the PolyCam operational range (i.e., the range in which the PolyCam is actually in focus at some real distance to target). Forty settings outside of this range position the refocusing lens—lens 1—outside any current valid focus range. They represent design margins on the PolyCam optical system that allow it to recover gracefully from a massive shift of focus, should one be experienced in flight, due to either thermal or mechanical disruptions.

The PolyCam focus table is based on a model of the optical system constructed in Zemax, the optical ray-tracing tool. The tool is used to generate a third-order functional relationship between L1 offset $m$ from the infinity position $m_\infty$ (slightly dependent on temperature) and best focus object distance (Range $R$). The coefficients $a$, $b$, and $c$ are determined and $m_\infty$ adjusted to fit focus test data. The fit weights errors inversely with the depth of field. Since the depth of field at large ranges approaches ½ to 1 times the range itself, while the depth of field at low ranges (near 200 m) is less than 5% of the range itself, the low-range data gets weighted much more significantly and the curve tied much more tightly to these points. This analysis predicts the functional form of the relationship between linear focus mechanism travel and range of optimum focus to be adequately described by an inverse cubic.

$$R = \frac{0.001}{-8.67 \times 10^{-7} \left[\frac{m - m_\infty}{s}\right] + 1.75 \times 10^{-8} \left[\frac{m - m_\infty}{s}\right]^2 - 2.64 \times 10^{-14} \left[\frac{m - m_\infty}{s}\right]^3} \quad (3)$$

where

$R$: Range (dependent variable) [in meters]

$m$: Motor Step Position (independent variable) [in motor steps]

$m_\infty = 18336$: Motor Step Position at Infinity (Fitted Parameter, dependent upon temperature)

$s = 2539.84$: L1 steps per travel [in motor steps per revolution]

$a = -8.67 \times 10^{-7}$: linear coefficient in inverse cubic fit

$b = 1.75 \times 10^{-8}$: second-order coefficient in inverse cubic fit

$c = -2.64 \times 10^{-14}$: third-order coefficient in inverse cubic fit

The correspondence to actual range is determined by fitting motor step position to laboratory virtual ranges measured using the pentaprism-based rangefinder that travels across the laboratory collimator to directly assess the ray angles (Figure 29). The result generated the focus table cited in Table 2.

### 3.7 Thermal Performance

#### 3.7.1 Focus vs. Temperature (PolyCam)

PolyCam's measured focus dependence on temperature, as well as its shift due to mechanical vibration, is shown in Figure 15. It is characterized by the value of $m_\infty$, the fitted motor-step position at infinity. From cold to hot temperatures, lens 1, the first lens of the field-correcting doublet, must be shifted toward the secondary and higher motor-step positions by roughly 1000 motor steps (about two shutter rotations) in order to achieve best focus at the hotter temperatures. If the ambient PolyCam focus-mechanism possesses an $m_\infty$ of near 17371, then the "Cold" (–20 C) and "Hot" (+30 C) curves are characterized by $m_\infty$ values of 17060 and 18050, respectively. The "Hot" and "Cold" curves in Figure 15 are based on behavior observed within a thermal-vacuum chamber during environmental testing, while the "Post-Vibe" curve is based on ambient testing near room temperature after random vibration testing. Neither of these is exactly flight-like when it comes to the temperature dependence of the focus mechanism. A more definitive model of this dependence awaits actual temperature-dependent flight data, planned for 6 months after launch.

## 4  In-Flight Calibration

The OCAMS in-flight calibrations characterize and track the three imagers' optical and radiometric performance throughout the OSIRIS-REx mission to the asteroid Bennu. Table 10 summarizes important calibrations planned at the time of this manuscript, including image loading, timing, and targets. Instrument self-calibrations measure changes in detector responsivity, dark current and bias level, charge transfer efficiency, and calibration lamp positioning. Notable star calibrations gauge absolute and spectral responsivity, geometric boresight and distortion, focus changes, and stray light. To improve efficiency, calibrations may be split into individual test campaigns, or several calibrations gathered into a single campaign (e.g., the instrument health check). This process may continue up until the actual epoch of the calibration.

[Insert Table 10]

The cruise phase extends from launch in September 2016 until the staged approach to the asteroid begins in August 2018. During this period, the OCAMS team characterizes the distortion map for all three imagers and their filters, acquires an absolute and relative radiometric calibration independent from the pre-flight ground calibration, and updates their detector responsivities by monitoring internal calibration tracing lamp illumination profiles. Along the way, the health of the several OCAMS modules is regularly monitored, ensuring that existing operating procedures and flight rules maintain their validity.

The harsh launch and space environment presents a number of challenges to the operation of a camera, including ionizing radiation–induced damage to the focal-plane solid-state detectors, to the electronics, and to the mechanism LEDs, contamination, vibration and other launch-associated performance alterations, extreme temperatures, and rapid temperature changes. Of these, one the most impactful—Random Telegraph Signal pixel creation in the focal plane—can be readily calibrated by recording the light pattern of a dark source. Other effects include radiation-induced changes in the CCD's charge transfer efficiency (CTE), the dark current, the calibration of the housekeeping temperatures, voltages, and index-lamp responsivities (index lamps provide light fiducials that aid in controlling the operation of all three mechanisms).

## 4.1    Self-Calibration

The primary means of monitoring OCAMS' general performance is the Instrument Health Check. At a minimum, it is run post-launch and every 6 months starting in March 2017. It is based on a sequence used during ground testing that operated all copper pathways in the cameras, without excessively exercising limited-life mechanisms. It tests heater, motor, lamp, and detector functionality, basic optical resolution performance, dark current, and detector responsivity relative to the on-board lamps. All heaters are activated and deactivated; all mechanisms are operated. Anomalies detected by the Instrument Health Check are triaged, and changes in detector or optical responsivity, dark current, and bias level are tracked.

After launch, the primary goals are to record launch-induced changes from the ground-based calibration and functional tests. Dark frames and images illuminated by the in-flight calibration lamps are logged. EPER (Extended-Pixel Edge Response) frames are logged, allowing the post-launch estimation of the charge transfer efficiency. Post-launch imaging benefits from minimal radiation-induced changes to the three detectors. Effects due to the launch itself (optics or detector-related physical shifts) can be isolated.

The OCAMS health check does not require an external target, but is enhanced if external targets are present in the field of any of the three cameras.

## 4.2 Distortion Calibration

The distortion calibration uses observation of open star clusters repeated at multiple locations within a camera's field of view to establish the observational basis of a camera distortion model. The observed location differences of star positions relative to each other, especially at the center and corners, are logged by precise column/row coordinate centroids. They are modeled both as relative and as absolute differences from each other and from a reference provided by the catalogued star positions to derive the error in observed location vs. actual feature absolute position against location within the camera field of view, informing a camera distortion model. Similarly to focus and optical resolution performance, distortion calibrations are planned at both hot and cold optics and detector temperatures, anticipating the temperature range to be experienced during proximity operations. Target temperatures are informed by predicted optics and detector temperature extremes during important proximity-operations observational campaigns, such as Orbital A and B, Detailed Survey, and Reconnaissance. For the MapCam and SamCam, unified lens stack and detector temperature are monitored and controlled. For the PolyCam the primary and secondary mirrors, the unified refractive element/focus mechanism assembly, and the detector are monitored and controlled. If it is difficult to simultaneously achieve all temperature extremes for these subsystems, the goal is to get within 3–4°C of the targets.

For the PolyCam, which possesses the largest physical assembly with the most complicated thermal profile, observations are recorded at focus positions from infinity to near 800-m range (the lowest range for which star point-spread envelopes are centroidable). Two full rotations of the 56-tooth lens gear (Figures 5 and 6) are recorded, at approximately 120° intervals, a spacing corresponding to 672 motor steps. These lens gear fiducials are realized at focus mechanism motor step positions 17371 (32A), 16650 (30C), 16110 (29C), 15390 (28B), 14669 (27A), and 13950 (25C), corresponding to the nominal ranges 39 km–∞, 3.3–3.7 km, 2.1–2.2 km, 1.3–1.5 km, 1.03–1.07 km, and 0.80–0.82 km. A baseline of 13 positions within the field of view (with a potential descope to 5 positions) are recorded, depending upon the degree of fidelity desired and the time and effort available to prepare the command sequencing for the calibration. The target list for the initial calibration is shown in Table 11.

[Insert Table 11]

For the MapCam, observations are recorded through all six filters (Pan, Pan-30, B, V, W, X) and at five positions within the FOV; potential targets include Hyades and Coma clusters; a total of about 360 images are recorded. For the SamCam, observations are recorded through one of the three Pan filters (1, 2, or 5).

## 4.3    Absolute Responsivity Calibration

In addition to the two clusters listed in Table 11, two solar twins were included as spectral reference targets and observed at multiple positions within the MapCam field of view. For SamCam and PolyCam, these targets are placed at the center of each FOV when observed. Five distinctive light paths are recorded for MapCam and one each for SamCam and PolyCam. Similarly to the open cluster observations, the calibration is repeated for hot and cold extremes.

## 4.4    Earth Gravity Assist Observations and Calibrations

The Earth Gravity Assist observations provide an opportunity for OCAMS to observe terrestrial and lunar surfaces after closest approach. During the encounter, the spacecraft approaches the Earth on its dark side, ducks under the South Pole and emerges on the day side over the Pacific Ocean. Imaging begins several hours after closest approach, at a range around 100,000 km. These observations continue until the spacecraft reaches a range of 200,000 km, and then they are repeated at a range of 2.8 M km. The Moon is observed at a range of around 1.1 M km. Highest priority is to derive absolute and color responsivities from the Earth/Moon imaging. Once acquired, the images can be compared with the measured signals of (1) simultaneous independent observations by Earth-orbiting spacecraft in the case of the Earth and (2) calibrated lunar radiances, in order to derive an absolute calibration. Such a calibration can be compared to those acquired from stellar standard candles.

At the planned observational ranges, rarely does a target fill the field of any of the OCAMS imagers. Useful flat fields are very unlikely, even for the PolyCam. For the lunar disk, which does not fill the field of view of any of the cameras, the campaign attempts to image it in as many locations within the cameras' field of view as feasible.

For PolyCam, observations are recorded at long and short exposure times, near infinity (motor step position 17371, 32A), at phase angles between 15° and 24° and around 33° for the Earth, and around 41° for the Moon. An Earth image is recorded at location P50 within the PolyCam FOV (Figure 27); for the Moon, center-of-bright-disk images are recorded at locations P10–P90. For the other cameras, observations at long and short exposure times are recorded through five filters (Pan, B, V, W, X), centered at P50, for MapCam and through one of the Pan filters (1, 2, or 5), for SamCam. For SamCam, images are centered at locations R7, R9, R17, and R19 within the FOV.

# 5 Concept of Operations

The OCAMS imager designs are optimized to serve the goals of the OSIRIS-REx mission. In much the same way, the design of the OSIRIS-REx concept of operations for OCAMS is strongly affected by the imperative to gather critical asteroidal information with sufficient fidelity (Mink et al. 2014) to enable (1) the low-risk, and successful, acquisition and return of the sample from Bennu's surface, (2) the selection of a scientifically interesting sample site, and (3) to document both the asteroid's surface properties and the nature of the sample site itself. Spacecraft trajectories are designed to provide appropriate spatial resolution; mission phases and campaigns are tailored to ensure comprehensive coverage; the spacecraft's orientation and pointing permit the acquisition of high-signal image data across desired photometric geometries. Products derived from OCAMS images decrease the risk to sample acquisition by resolving Bennu's physical parameters (Antreasian et al. 2016). PolyCam and MapCam are also used to directly perform optical navigation from ranges greater than 7.6 km in order to verify Bennu's ephemeris and confirm the spacecraft's relative position. Although primarily the responsibility of TAGCAMS, OCAMS can also provide images for optical navigation at ranges less than 7.6 km, if necessary. For example, on the rehearsal approaches to the sample site, after the Matchpoint maneuver, MapCam and SamCam record images of the surface in order to confirm the reduction of the spacecraft's lateral velocity with respect to the surface to less than 2 cm sec$^{-1}$, the maximum permissible rate for a successful sampling attempt (Berry et al. 2015).

The ability of MapCam to spectrally image terrain, which is also observed by OVIRS and OTES but at lower spatial resolutions, allows the team to identify scientifically valuable regions from which to collect the sample. The characteristics determined by OVIRS are extended to higher spatial resolution by MapCam after cross-calibrating on regions observed simultaneously by both instruments. The thermal inertias determined by OTES are similarly extended by the same observations. Bennu, a B-type asteroid and possible CM-chondrite spectral analog, is believed to be primitive and volatile-rich (Clark et al. 2011), and potentially derived from the inner regions of the main belt (Campins et al. 2010; Bottke et al. 2015). The physical state of its surface, unresolvable by ground-based observations, can be constrained and used to determine whether key spectral slopes are due to compositional or grain-size effects (Müller et al. 2012; Emery et al. 2014; Binzel et al. 2015).

The Design Reference Mission (DRM) (Mink et al. 2014) contains ten operational phases, nine of which are illustrated in Figures 48 and 49. To support the primary objective of returning a pristine sample of asteroid Bennu to Earth, each mission phase increases the integrity of knowledge of Bennu (or of the calibration and performance of the imagers). OCAMS acquires data in nearly all mission phases to inform the key decision points faced by the OSIRIS-REx team. As the mission progresses, the fidelity of the asteroid shape and distribution of the surface features on Bennu gradually improves with corresponding increases in image spatial resolution. In particular, knowledge of the surface slopes, regolith grain size, and composition of candidate sample sites is refined until a sampling site is chosen with confidence.

[Insert Figure 48]

[Insert Figure 49]

The early phases (Approach, Preliminary Survey) focus on acquiring Bennu as a point source, gathering long-range optical navigation, and shape-model imaging with the PolyCam, as well as performing astrometry and disk-integrated photometry and conducting a natural satellite survey with the MapCam. Following the Approach and Preliminary Survey phases, an early shape model is generated that, combined with the radio science (McMahon et al. 2017), informs the gravitational field and establishes whether or not the environment near Bennu is debris free and safe for orbital insertion. Successfully entering orbit immediately following the Preliminary Survey represents a significant mission milestone and retires a great deal of risk.

After leaving orbit, the spacecraft is guided through multiple campaigns in Detailed Survey. This phase focuses on improving the resolution of Bennu's shape model, mapping the distribution of TAGSAM head-sized hazards across the asteroid, and identifying candidate sample sites that can be examined in more detail during subsequent phases. Bennu's geological and color units are globally mapped using OCAMS data from this campaign. The data acquired from Detailed Survey also enable the team to identify up to a dozen candidate sample sites.

The completion of Detailed Survey initiates Orbital B, where the 12 sites are imaged at higher resolution and more subtle illumination geometries to support a down-selection to two candidate sites. Successive mission phases (Reconnaissance and TAG Rehearsal) acquire the highest-resolution data at the two prospective sample sites. This information is used to plan the TAG (touch and go) maneuver and decrease the risk associated with carrying out TAG. Finally, the Sampling Phase collects the sample from the prime candidate site and records the sampling event.

## 5.1 Approach Phase

During the Approach phase, the OCAMS imagers verify the location of the asteroid and confirm the trajectory of the spacecraft, identify any near-asteroid dust clouds from a range of 1 M km, search for natural satellites from 1000-km and 200-km ranges, and begin to assemble images for a preliminary shape model (Lauretta et al. 2017). On approach to Bennu, MapCam also measures both the disk-integrated phase function and asteroid light curve, observations that can be compared to similar ground-based and Hubble Space Telescope studies (Hergenrother et al. 2013). The approach trajectory is shown in schematic form in Lauretta et al. (2017, Figure 26). While the expected position of Bennu has an uncertainty of a few km (Milani et al. 2009; Chesley et al. 2014), a campaign to acquire the asteroid from a range of some 2 million km begins in August 2018. During the campaign Bennu is imaged against the star field background to precisely determine its position, and a regular series of PolyCam frames are acquired to allow optical navigation to verify the asteroid's trajectory (18 images every 3 days for a month and a half, then 18 every day for a full month).

Both PolyCam and MapCam are capable of performing the campaign to acquire the asteroid: while the PolyCam is expected to acquire Bennu in late August, the MapCam should acquire it by mid-September, well in advance of the trajectory correction maneuver AAM1 (Lauretta et al. 2017, Figure 26) planned for the beginning of October 2018. Bennu becomes an extended object for all three cameras in October: near October 12 for PolyCam at a range around 37,000 km; near the 15$^{th}$ for MapCam at a range around 7400 km; and not until the 26$^{th}$ for SamCam at a range of 1400 km.

In mid-September a campaign to survey the dust environment proximal to Bennu is conducted by both the PolyCam and the MapCam, with 64 images each planned. These observations capture Bennu as a point source (range nearly 1,000,000 km), imaging the asteroid's photopic profile and comparing it to comet-like profiles where the signature of a coma is evident.

Further observation programs during this phase include:
1. A full suite of MapCam images (1152) to measure Bennu's light curve in four colors
2. 400 MapCam images to measure Bennu's phase function in four colors
3. 120 MapCam images devoted to two satellite-search campaigns, one from a 1000-km and one from a 200-km range
4. 1142 PolyCam images devoted to shape-model determination acquired in the last week and a half before spacecraft arrival in proximity to Asteroid Bennu, as the range decreases from 140 to 25 km

## 5.2 Preliminary Survey

Several campaigns across the mission are devoted to the development of accurate digital terrain models of the asteroid using MapCam and PolyCam. During the Preliminary Survey phase the asteroid shape and gravity models are significantly enhanced. Three hyperbolic trajectories cross over and under Bennu's north and south poles and move along the equator at a close-approach range of 7 km (Lauretta et al. 2017, Figure 27). The team uses the spacecraft's radio track and improved shape to refine the Bennu's gravitational field, allowing for entry into a safe orbit (Orbital A). A large number of PolyCam images are acquired in three transits over the north and south poles as well as an equatorial flyby. MapCam and PolyCam deliver a 75-cm-resolution shape model (Digital Terrain Model, or DTM) and associated maplets/landmarks by the end of Preliminary Survey to allow for transition to landmark-based OpNav to occur during Orbital A (Berry et al. 2013, 2015; McMahon et al. 2014, 2017; Beshore et al. 2015; Antreasian et al. 2016; Mario and Debrunner 2016). At least 100 landmarks will be identified by the end of this period.

## 5.3 Orbital A

During Orbital A, the spacecraft takes up a 1- to 1.5-km-radius 50-hour terminator orbit around Bennu, the first time a manmade object assumes an orbit around a body so small (Lauretta et al. 2017, Figure 28). Currently no OCAMS images are planned during Orbital A; this phase provides a 4-week period for the Flight Dynamics team to transition from starfield-based to landmark-based optical navigation (Williams et al. 2017).

## 5.4 Detailed Survey

Detailed Survey is one of the most important scientific imaging surveys during the OSIRIS-REx mission. During Detailed Survey, >80% of Bennu's surface is mapped by separate Pan and color imaging campaigns across a series of hyperbolic flybys (Lauretta et al. 2017, Figures 29 and 30). The range to the surface and illumination geometries of images acquired during this phase allow surface hazards to be readily identified and catalogued and also provide the first digital base maps of Bennu, while simultaneously improving the resolution of the asteroid shape model from 75 to 35 cm. Surface-resolved phase and disk functions are derived from images acquired during Detailed Survey, updating the photometric model for Bennu. A high-resolution search for dust and/or gas plumes is performed to identify any potential source regions on the asteroid. Finally, regions of interest (ROIs), which are areas of regolith free from hazards, are identified and Bennu's surface divided into "go" and "no-go" areas. The ROIs identified during this period are subjected to further study at higher resolution.

During the Baseball Diamond campaign, a comprehensive set of PolyCam images is acquired from a range of 3.5 km, where the instantaneous field of view (IFOV) is 5 cm/pixel, with the goal of identifying all hazards >21 cm on the surface of Bennu (Lauretta et al. 2017, Figure 29). These images are mosaicked into digital base maps and also improve Bennu's shape model to an accuracy of better than 35 vertical cm per facet.

The second part of Detailed Survey includes seven equatorial stations located at distinct local asteroid times ranging from 12:30 p.m. to 3 a.m. Observations are acquired at each station during south-to-north hyperbolic flybys at a range of 5 km from the surface of Bennu (Lauretta et al. 2017, Figure 30). As the spacecraft passes through the equatorial plane of Bennu, the field of view of MapCam sweeps from pole to pole while the asteroid rotates over a full rotation period, thus providing global coverage. At four of these equatorial stations, each corresponding to a different phase angle, MapCam acquires images in all four color filters. These data provide sufficient information to derive a surface-resolved phase function of Bennu per wavelength and update the disk-integrated photometric model of the asteroid acquired from approach and ground-based observations (Takir et al. 2015). Bennu's phase function is likely to be interesting for two reasons that are strongly correlated: composition and micro-texture. As observed in Hergenrother et al. (2013), as asteroid albedos decrease, the magnitudes of the slopes of their phase functions increase, especially for carbonaceous asteroids (Li et al. 2006). Throughout the mission, OCAMS observations at different scales clarify this correlation and potentially elucidate the cohesive properties of Bennu's carbonaceous regolith.

Detailed Survey MapCam data, especially those acquired at low phase angle, reveal the spectral properties of Bennu at high spatial resolution. Using these images, band-ratio composites can be generated to determine the strength of visible–near IR slopes as well as the presence of the 700-nm absorption feature indicative of the Fe-bearing phyllosilicates in carbonaceous chondrites (e.g., Johnson and Fanale 1973; King and Clark 1989). Global band-ratio maps of Bennu are assembled from these data in an effort to illustrate the distribution of color-units on Bennu. Color-units are a powerful diagnostic for understanding if spectral variations are associated with differences in surface composition (e.g., Delamere et al. 2010; Le Corre et al. 2013), grain size and particle size-frequency distribution (e.g., Jaumann et al. 2016), or space weathering (Chapman 1996). Correlating these observations with spectrometer measurements allows for enhanced identification of scientifically compelling ROIs—perhaps indicating areas with a maximum likelihood of volatile content.

A MapCam X-filter reflectance map acquired from these images also serves as a proxy for the performance of the GNC LIDAR (Light Detection and Ranging), used to guide the spacecraft toward Bennu's surface during the TAG rehearsal and TAG maneuvers. Specifically, these data (acquired at 860 nm) can be extrapolated to a 1064-nm reflectance map using OVIRS' spectral results as a key. This map is used to ensure that the surface of Bennu is within the range of reflectance values in which the LIDAR is designed to operate.

Bennu is believed to be a transitional object between comets and asteroids and may have active dust and/or gas plumes. Accordingly, MapCam (backed up by PolyCam) searches for plumes during a campaign of Pan observations of the asteroid limb at high phase angles (~135°) in forward-scattered light. If and when plumes are observed, modeling of the plumes is conducted to identify their surface origin(s). Finally, geological maps based on the MapCam data can be used to study four important feature types: craters, linear features, boulders, and regolith in order to classify Bennu's surface, as well as revealing its topography sufficiently to understand its geological history (Walsh and Richardson, 2006; Delbo and Michel 2011; Walsh et al. 2012, 2013; Connolly 2015).

### 5.5 Orbital B

After Detailed Survey, the spacecraft transfers to a 1.0-km-radius orbit with an average range to surface of 0.75 km, and begins the Orbital B phase (Lauretta et al. 2017, Figure 31). During various campaigns within this phase, PolyCam acquires a series of images at 1 cm per pixel to characterize up to 12 candidate sample sites by identifying features with a maximum size of 5 cm. The high-phase-angle illumination geometries characterizing these observations can be expected to enhance detectability of pebbles and small cobbles by the relatively long shadows they cast. These detailed images are used in tandem with context imaging from earlier phases to further catalogue the grain size-frequency distribution on Bennu and ensure that surface material at the candidate sites can be ingested by TAGSAM. Prior to Orbital B, a second DTM to be delivered is expected to be accurate to 35 cm.

The spacecraft trajectory during Orbital B provides a "Safe Home" state from which all sampling sorties can depart and return. The 12 candidate sites identified in Orbital B are reduced to two sites: a primary and secondary. Armed with a high-resolution asteroid shape model, spin state, gravity models, well-calibrated propulsion and attitude control systems, and the experience gained from performing precise, close-proximity maneuvers from this orbit, the OSIRIS-REx science team embarks on campaigns to inspect specific potential sample sites at closer range in the Reconnaissance phase.

## 5.6   Reconnaissance

During the Reconnaissance phase, the short-listed primary and backup sample sites are imaged at increasingly higher resolution to verify their viability. Sampleable regolith consists of particles and grains 2 cm or smaller in diameter: this is the diameter than can safely fit through the throat of the TAGSAM head. PolyCam (backed up by MapCam) verifies the presence and maps the extent of such grains for prospective sample sites.

The Reconnaissance phase consists of four 225-m-altitude flyovers over the illuminated side of Bennu, out of the 1-km terminator orbit plane, to collect sampleability data for two candidate sites, followed by two 525-m flyovers of both sites to characterize their science value with MapCam color imaging (Lauretta et al. 2017, Figure 32). The six low-range sorties are spaced by two-week intervals in the 1-km Safe Home orbit. Between sorties, the science data is processed, analyzed, and interpreted in order to evaluate the selection of the prime site. The 525-m sorties are performed to collect spectral data with favorable illumination (40°–70° solar phase angles) to aid in determining the site's composition and identifying whether phyllosilicates are present.

For the four 5-hour flyovers at 225-m range, solar phase angle varies between 30° and 50°. This viewing geometry provides both sufficient light and shadows to identify 2-cm-sized pebbles. During the transit to each site the spacecraft retains a thermally favorable attitude, with the High Gain Antenna (HGA) pointed at the sun. Prior to reaching the site, the spacecraft reorients to a nadir-pointing attitude and slews back and forth perpendicular to the flight direction, collecting images with PolyCam and altimetry data with OLA. The trajectory targets ranges to the surface between 200 m and 287 m, the altitudes at which PolyCam provides sub-cm spatial resolution if each target pebble is assumed to be at least 5 pixels wide. Following each flyover, the spacecraft returns to a sun-point attitude until it performs a maneuver to return to the 1-km-radius orbit. While in the 1-km orbit, spacecraft operations follows a similar pattern of 16 hours of nadir and nadir-relative pointing for optical navigation image collection, and 8 hours of downlink and tracking. OVIRS solar calibrations also occur from the 1-km orbit. Following two weeks of data evaluation and planning, the 225-m flyover maneuver and observing sequence is repeated until both sites have been surveyed.

During the 525-m flyovers, MapCam acquires Pan and color filter images at each of the candidate sample sites. Images are concentrated around the site, with a slower image cadence on approach and return. Only Pan images are collected during the approach to the sample site and on return to safe home orbit to give context to the color data and to minimize the data volume per sortie. Images acquired during this phase are used to construct band-ratio composites of the final two sites to aid in an assessment of their scientific value.

## 5.7   TAG Rehearsal

OSIRIS-REx's trajectory to the surface is a three-segment pathway in space and time synchronized with the rotation of Bennu (Lauretta et al. 2017, Figure 33). This trajectory is designed to deliver the spacecraft to the selected sample site and allow it to contact the surface for sampling. The spacecraft travels autonomously along a predefined corridor to a final leg. Two predefined maneuvers punctuate the journey—at the waypoints known as Checkpoint and Matchpoint. Passing through these successfully requires the spacecraft to achieve a particular spacecraft state (location plus velocity) at a specific time.

Preparing for this scenario consists of rehearsing these two critical maneuvers in a mission phase called TAG Rehearsal. The corridor to the primary sample site is traversed, and the spacecraft practices flying through each of the two control points. The ability to deliver the spacecraft to the targeted sample site with an accurate touchdown velocity and minimum attitude and rate errors is then evaluated, which includes acquiring MapCam and SamCam images and comparing them to the expected image set.

## 5.8   Sampling

During the Sampling campaign, the primary sample site is approached in a deliberate maneuver that safely harvests at least 60 grams of Bennu's regolith from its surface (Lauretta et al. 2017, Figure 34). The MapCam and SamCam imaging pre- and post-Checkpoint and pre- and post-Matchpoint are important images. Nearly all post-Matchpoint images provide unique views of the sample site. These data not only document the pre-sampled state of the sample area, but also capture frames from the touch-and-go sampling maneuver, which helps to assess its success. SamCam frames, acquired at a range of 3–5 m and a rate of three every 5 seconds, optically verify the release and effects of the flow of gas that activate the sampled regolith. It is planned to acquire some 200 MapCam images during the entire TAG sequence, and about 150 SamCam images.

## 5.9 Post-Sampling Verification

Immediately after sampling, the spacecraft retreats to a range from Bennu of 20 km. Here, the spacecraft performs a rotation to measure its moment of inertia, which is compared to the pre-sample value in order to verify that the required mass was collected. SamCam then images the TAGSAM head to document the presence of the sample and thus to determine and verify the success of both the bulk and surface-sampling efforts. If the sample event is determined to have been successful, the TAGSAM head is eventually placed in the Sample Return Capsule (SRC), and at the appropriate maneuver window, OSIRIS-REx begins its journey back to Earth.

Three series of images are collected with the SamCam following the collection event(s). These images are used for three different verifications. In each instance the SamCam images the TAGSAM head using the 2.1-m diopter plate. The TAGSAM arm is repositioned in order to achieve this distance. The TAGSAM head is reoriented so that images are collected at a minimum of three different positions. Each series is taken at a different sun angle and includes the images shown in Figure 50.

[Insert Figure 50]

Several aspects of the post-sample TAGSAM head are of interest: (1) the images provide optical documentation of the total area of the TAGSAM surface sample collection discs that is evidently covered with sample; this area is assessed and used to determine whether sufficient surface sample mass has been acquired; (2) the TAGSAM head is recorded optically backlit by solar light in order to observe the silhouette of sample contained within the chamber seen through the head's screens; (3) a series of images along the head's bottom surface is used to determine if there are any particles hanging off the TAGSAM head large enough to cause interference with entry into the sample return capsule (SRC); the largest diameter of the adhering particles must be less than 6.0 mm to not interfere with SRC storage.

# 6 Conclusion

Of more than 100 exploratory deep-space missions sent beyond low-Earth orbits by various space agencies, very few have attempted to collect a sample from a planetary body and return it to Earth on a single visit. Missions have typically relied upon an object's prior exploration, survey, mapping, and even landed investigations (Siddiqi 2002) to gather actionable details about that body's environment and physical state. For example, the U.S.S.R. Luna program launched at least 24 separate spacecraft to the Earth's Moon, whose primary tasks included intentionally crash-landing, orbiting, and landing on the surface and documenting its conditions before one returned a sample from its surface. Space missions often failed, requiring multiple attempts to increase the probability of a single success.

In contrast, OSIRIS-REx will perform sample collection and sample return on humanity's first physical encounter with Bennu. The three OCAMS imagers, and the well-chosen complement of instruments mounted on OSIRIS-REx science deck, greatly contribute toward mitigating the risk of this encounter. Combining low- (SamCam), medium- (MapCam), and high-resolution (PolyCam) optical systems with a radiometric sensitivity tuned to the low albedo of the target, OCAMS provides long-range Bennu acquisition, sub-centimeter surface sample site reconnaissance, global mapping and shape determination, optical navigation during the approach phase and as a back-up during proximity operations, sample-site characterization, sample-acquisition documentation, and post-sample verification of sampling success. In addition, various critical direction and velocity changes can be independently documented, up to and including the TAG (touch-and-go) maneuver that acquires the sample itself, providing both a check and a backup to the mission's navigation cameras. At every stage, the cameras identify possible hazards: satellites, potential dust plumes, extreme slopes, large boulders, and smaller cobbles.

The imagers—especially MapCam and PolyCam—provide redundant capabilities for each other's key tasks and provide context for one another by imaging the asteroid's surface from alternate ranges. The medium-angle MapCam can, if necessary, provide both long-range Bennu acquisition and sub-cm sample site reconnaissance of the surface of Bennu. The PolyCam can enable global mapping and shape determination and sample-site reconnaissance, and can support optical navigation during proximity operations. The wide-angle SamCam can perform sample-site characterization and optical navigation.

The OCAMS concept of operations was an important aspect of designing the Design Reference Mission. The DRM organized the mission's stages into various phases characterized by the spacecraft's mean distance from Bennu, the character of its orbit, whether hyperbolic or elliptical, and the intensity and direction of solar illumination. These characteristics allowed for the specification of camera requirements which were traded off against altering the DRM in its key aspects in order to minimize both technical and development risk across mission, spacecraft, and instrument implementation.

In addition to their role as a mission system, the OCAMS imagers increase the overall scientific return of the mission by documenting the surface of a microgravity object at unusually high resolution over a high proportion of its surface. These observations provide insight into the microgravity physics of Bennu, a body that is representative of the primitive planetesimals that acted as the building blocks of the solar system. Images acquired by OCAMS reveal the physical processes at work on Bennu's surface, its surrounding environment, and subsequently the asteroid's geological history. They provide several clues to detail the nature of mass wasting and material ponding processes, particle size-frequency distributions at sub-centimeter scales, and cohesion of surface regolith as it interacts with the TAGSAM during sampling. These clues are vital for understanding the strength and structure of small bodies and geological processes that exist within microgravity environments.

Pairing high-resolution OCAMS images with other OSIRIS-REx instrument results provides complimentary datasets that enhance one another. Correlating MapCam color images with OVIRS spectra at bands similar to the MapCam filters extends OVIRS' spectral identifications to higher spatial resolutions. PolyCam images acquired during Reconnaissance detail sub-centimeter grain sizes, providing some "ground truth" to OTES measurements of thermal inertia at the surface of Bennu. Comparing image-derived stereophotoclinometric shape and terrain models with the OLA terrain data also provides a unique cross-validation of image-based and LIDAR observation techniques. Finally, comparing images of Bennu as a point source to surface-resolved images acquired during proximity operations provides a rare opportunity for evaluating disk-integrated observations of asteroids, especially ground-based ones. The requirements and mission plan defined for OSIRIS-REx have been driven by the need to enable and reduce the risk of an asteroid sample-return mission; however, it is clear that these plans have resulted in a robust set of instrumental assets to maximize the science return of the OSIRIS-REx mission.

## Acknowledgments

We thank the many people throughout the OSIRIS-REx project, at the University of Arizona, at Goddard Space Flight Center, and at Lockheed Martin Space Systems for the tremendous assistance and support they have provided to the development of this camera system. This material is based upon work supported by NASA under Contracts NNM10AA11C, NNG12FD66C, and NNG13FC02C issued through the New Frontiers Program. Copy editing and indexing provided by Mamassian Editorial Services.

P.H. Smith, M.G. Tomasko, D. Britt, D.G. Crowe, R. Reid, H.U. Keller, N. Thomas, F. Gliem, P. Rueffer, R. Sullivan, R. Greeley, J.M. Knudsen, M.B. Madsen, H.P. Gunnlaugsson, S.F. Hviid, W. Goetz, L.A. Soderblom, L. Gaddis, R. Kirk, The imager for Mars Pathfinder experiment. J. Geophys. Res. 102, 4003–4025 (1997)

C. Stevens, B. Williams, A. Adams, C. Goodloe, Cleared for Launch – Lessons Learned from the OSIRIS-REx System Requirements Verification Program. Proceedings of the IEEE Aerospace Conference, Big Sky, MT. March (2017).

D. Takir, B.E. Clark, C. Drouet d'Aubigny, C.W. Hergenrother, J.-Y. Li, D.S. Lauretta, R.P. Binzel, Photometric models of disk-integrated observations of the OSIRIS-REx target Asteroid (101955) Bennu. Icarus 252, 393–399 (2015)

E.F. Tedesco, D.J. Tholen, B. Zellner, The Eight-Color Asteroid Survey: Standard stars. Astronomical J. 87, 1585–1592 (1982)

M.G. Tomasko, D. Buchhauser, M. Bushroe, L.R. Doose, A. Eibl, C. Fellows, M.J. Pringle, B. Rizk, C. See, P.H. Smith, K. Tsetsenekos, The Descent Imager/Spectral Radiometer (DISR) instrument aboard the Huygens entry probe of Titan. Space Sci. Rev. 104, 469–551 (2002)

K.J. Walsh, M. Delbó, W.F. Bottke, D. Vokrouhlicky, D.S. Lauretta, Introducing the Eulalia and new Polana asteroid families: Re-assessing primitive asteroid families in the inner Main Belt. Icarus 225(1), 283–297 (2013)

K.J. Walsh, D.C. Richardson, Binary near-Earth asteroid formation: Rubble pile model of tidal disruptions. Icarus 180(1), 201–216 (2006)

K.J. Walsh, D.C. Richardson, P. Michel, Spin-up of rubble-pile asteroids: Disruption, satellite formation, and equilibrium shapes. Icarus 220(2), 514–529 (2012)

B. Williams, P. Antreasian, E. Carranza, C. Jackman, J. Leonard, D. Nelson, B. Page, D. Stanbridge, D. Wibben, K. Williams, M. Moreau, K. Berry, K. Getzandanner, A Liounis, A. Mashiku, D. Highsmith, B. Sutter, D.S. Lauretta, OSIRIS-REx flight dynamics and navigation design. Space Sci. Rev. (2017) – this volume

F. Vilas, A cheaper, faster, better way to detect water of hydration on Solar System bodies. Icarus 111(2), 456–467 (1994)

H. Yano, T. Kubota, H. Miyamoto, T. Okada, D. Scheeres, Y. Takagi, K. Yoshida, M. Abe, et al., Touchdown of the Hayabusa spacecraft at the Muses Sea on Itokawa. Science 312, 1350–1353 (2006)
56

**Figure Captions**

**Fig. 1** OCAMS operational range resolution. Regions of OCAMS imager operation at different mission phases are plotted by resolution and range. Colored lines show the operating in-focus range of each camera: SamCam (blue); MapCam (green); PolyCam (purple). Vertical lines intersect with the solid colored lines and indicate a camera observation is possible at a specific range. A horizontal line intersecting at the same point gives the 3-pixel resolution at that range for that camera. Note: the SamCam focal range includes ∞, but is not shown

**Fig. 2** MapCam optical design schematic. MapCam optical design displayed along with traces of key field rays. It is a 125-mm focal length, F/3.3 five-element rad-hard telephoto design

**Fig. 3** SamCam lens design schematic. SamCam optical design displayed along with traces of key field rays. It is a 24-mm F/5.5 six-element rad-hard double-Gauss design

**Fig. 4** PolyCam optical design schematic. PolyCam optical design displayed along with traces of key field rays. It is a 20-cm-wide aperture, F/3.15 Ritchey-Chretien telescope with a 629-mm focal length at infinity. It possesses a focusing mechanism which works by actuating the first lens of the field-correcting doublet

**Fig. 5** PolyCam focus mechanism exploded view. The PolyCam focus mechanism contains a threaded hub, holding a lens, that leverages itself against grooved rollers as it is advanced back and forth along an optical axis by a 56-tooth gear. This gear is driven by a 28-tooth idler gear which is itself driven by a smaller (20-tooth) geared shaft rotated by a stepper motor. The large gear, in its turn, rotates a shutter gear (15-tooth)

**Fig. 6** PolyCam focus mechanism disassembled. The PolyCam focus mechanism is shown partially disassembled, allowing a view of the mounting arrangement for the various components. Lens 1 on the right-hand side, one of a field-correcting doublet, appears blue in the image, mounted to the baffled threaded hub (not seen). Lens 2 is seen on the left, next to the 15-tooth gear that enables the shutter to be driven

**Fig. 7** MapCam filter definition. MapCam filters are identified by index position, label, and motor step position. The filter wheel complement is represented by two blocking positions, one Panchromatic (Pan) position, four color filters, and a refocusing plate which allows in-focus operation at a range of 30 m

**Fig. 8** SamCam filter definition. SamCam filters are identified by index position, label, and motor step position. The filter wheel is mounted externally to the optical system. Three identical Pan filters allow imaging during three sampling attempts even if an earlier attempt causes contamination severe enough to blanket the glass plates with regolith dust

**Fig. 9** OCAMS detector (Teledyne DALSA). The Teledyne DALSA detector, used at the focal plane of all three OCAMS imagers, is shown in this microscope image. A frame transfer device, its active area is the brighter square in the image, while the masked storage region is located just below. Tiny bottle-shaped output gate structures can be seen left and right near the bottom. Typically provided with a cover glass, the detectors were used in OCAMS without the glass cover in order to reduce scattered light

**Fig. 10** OCAMS Camera Control Module (CCM). The OCAMS electronic control module is shown in this image, taken during its final assembly. It is comprised of a dual-redundant three-board set, mounted horizontally, primary triplet above and redundant triplet below. Within each group of three, the data processing unit is mounted on top, the motor/heater board in the middle, and the power board on the bottom. Connections are provided (1) to each side of the spacecraft Command and Data Handling (C&DH) from both primary and redundant, allowing four separate parallel links, if needed; (2) to all three cameras' data interfaces; (3) to all three cameras' motors and heaters; (4) to spacecraft power and all three cameras' power connections. Spacecraft connectors are larger; camera connectors are smaller

**Fig. 11** OCAMS mounting configuration. The OSIRIS-REx Camera Suite shown as mounted to the spacecraft's science deck. The spacecraft's coordinate system is defined with the +Z axis pointing normal to the science deck, upward through the center of the Sample Return Capsule (SRC) (item f in the image) and the +X axis pointing outward through the high-gain antenna (item g). OCAMS (item d) is located on the –X, +Y quadrant of the science deck

**Fig. 12** OSIRIS-REx Polyfunctional Camera (PolyCam). PolyCam is shown—minus its thermal dressing—in a perspective view that emphasizes the sharply angled sugar scoop baffle. The back face of the secondary mirror is visible within the telescope's dark interior

**Fig. 13** Simulated PolyCam image detail. This simulated PolyCam image section, acquired of a target with an albedo similar to the asteroid Bennu and illuminated in a flight-like manner, represents a zoomed-in view of a dark asteroid surface at a 0.33 cm/pixel plate scale. The pixel boundaries are evident. Features of 4–5 pixel size are identifiable as pebbles and measurable, indicating a capability to verify that such pebble sizes—smaller than the TAGSAM throat—are present within the regolith

**Fig. 14** Focus mechanism functional gear relationship. Each circle is meant to represent a focus mechanism gear with the number of teeth per gear indicated at the center of each circle. The number of motor steps corresponding to a single rotation of the gear is shown below each gear. The largest gear is located normal to the optical axis

**Fig. 15** Focus mechanism range (m) dependence upon motor step position. The relationship between focal range and focus mechanism absolute motor step position is shown here. The dependence is a nonlinear one, where the focus changes most dramatically with motor position at near ranges. The PolyCam can be said to be in good focus at a particular range if the motor is positioned within a horizontal range corresponding roughly to half a division, or around 500–600 motor steps

**Fig. 16** OCAMS Mapping Camera (MapCam). The MapCam's cylindrical housing contains its lens cell and filter wheel. It mounts atop a base similar to that of the other two imagers, which contains the detector assembly and supports a sugar-scoop baffle and two white radiating areas

**Fig. 17** OCAMS MapCam filter transmissions. OCAMS MapCam filter transmission curves (B = dark blue, V = green, W = red, X = brown, Pan = black) compared to measured solar relative spectral irradiance (gold), general detector quantum efficiency (magenta), and theoretical laboratory relative spectral irradiance (light blue)

**Fig. 18** OCAMS Sample Acquisition Camera (SamCam). The SamCam cylindrical housing contains its lens and filter wheel and is mounted on a base containing the detector assembly similar to that of the other two imagers. The radiator is the vertical plate; the filter wheel motor protrudes upward from the center of the housing; and the entire assembly is mounted at angle of 9.4° from vertical in order to have an optimum view of the TAGSAM arm while it is elongated into its sampling position

**Fig. 19** OCAMS system block diagram. Six boards occupy the CCM chassis: primary and redundant CPU boards, Motor/Heater boards, and low-voltage power supply (LVPS) boards. They control operation of the detector assembly, mechanisms, lamps, and heaters of the three sensor heads and receive input from dozens of sensors that monitor temperature, voltage, current, mechanical position, and software state. All information is digitally stored and packaged and then sent to the spacecraft's command and telemetry interface

**Fig. 20** OCAMS CCM block diagram. The OCAMS CCM is controlled by a data processing unit board housing an 8051 microcontroller hosted by an Actel RTAX-2000 field-programmable gate array and implemented as an intellectual property (IP) processing core. The architecture is shown as implemented through the three redundant CCM boards: the DPU, a motor/heater interface/driver (MHD), and a low-voltage power supply (LVPS). Internally, the FPGA is clocked at 36 MHz

**Fig. 21** OCAMS detector assembly (DA) block diagram. The detector assemblies enable control of OCAMS detector operation through manipulation of register parameters to directly influence detector operation. Spacecraft commands can initiate DA power-on and imaging through either of two detector taps, record diagnostics, test patterns, read and write registers, dump memory locations, and measure both vertical and horizontal charge transfer efficiency through an extended-pixel edge response technique

**Fig. 22** Format of the OCAMS CCD defined by region. The active region is surround by a 4-pixel-wide transitional zone which may receive light, 24-pixel-wide covered columns left and right, and 6-pixel-deep covered rows top and bottom. Sixteen-pixel-wide electronic lead-in columns left and right are also present. The image is finally packaged and released as a fits file as shown

**Fig. 23** MapCam MTF slant line target. Backlit slant line target projected through collimator and imaged by MapCam in order to measure modulation transfer function (MTF) and verify optical resolution performance

**Fig. 24** PolyCam MTF slant target. Backlit slant line target and point source projected through collimator and imaged by PolyCam in order to measure modulation transfer function (MTF) and verify optical resolution performance

**Fig. 25** SamCam MTF slant target. Backlit slant line target and point source imaged as real image by SamCam in order to measure modulation transfer function (MTF) and verify optical resolution performance

**Fig. 26** SamCam MTF slant target on ATLO floor. Slant targets were used at ATLO after spacecraft environmental testing to assess changes in optical resolution performance

**Fig. 27** OCAMS field of view relative location positions. OCAMS relative optical and radiometric performance can be referenced to these relative field-of-view positions

**Fig. 28** Measured line-spread function. The ensemble of points transverse to the core of the slant line used in the MTF measurement are shown in green here, interpolated by a blue line. They are plotted against the transverse distance from the slant line peak. The MTF is derived from the line-spread function by a Fourier Transform

**Fig. 29** Pentaprism range verification system. The pentaprism range verification system is circled in this image; it is mounted in front of the Alnitak self-illuminated flat panel which is used for PolyCam flat-fielding. The system observes reflected light from the collimator beam and records the deviation from a null result (which corresponds to infinite range). Finite deviations correspond to finite ranges

**Fig. 30** MapCam System MTF for Pan filter vs. Spatial Frequency: ambient conditions. MapCam System MTF for the Pan filter at different ranges to target is shown at ambient temperature and pressure. Horizontal slit result is shown (worst case) before Thermal Vacuum TVAC and vibrational testing. Results are shown at 125-m, 500-m, 950-m, 3500-m, and infinite ranges

**Fig. 31** MapCam System MTF for Pan filter: cold vacuum. MapCam System MTF is shown for the Pan filter vs. range to target at flight-like cold temperature and vacuum pressure for horizontal slits (worst case) after TVAC and vibrational testing. Performance improves somewhat at longer ranges and decreases somewhat at shorter ranges due to the colder optical temperatures. In addition, after vibration and thermal testing, optimum camera optical resolution performance shifted slightly toward longer ranges for all temperatures. Results are shown at 125-m, 500-m, 950-m, 3500-m, and infinite ranges

**Fig. 32** SamCam System MTF for Pan filter: ambient conditions. SamCam System MTF is shown for the Pan filter vs. range to target at ambient temperature and pressure using horizontal slits (worst case). Results are shown at 2.8- and 5-m ranges

**Fig. 33** SamCam System MTF for Pan filter: cold vacuum. SamCam System MTF for the Pan filter vs. range to target is shown at flight-like cold-temperature vacuum conditions using horizontal slits (worst case). Results are shown at 2.8- and 5-m ranges

**Fig. 34(a–e)** PolyCam System MTF at 200-, 290-, 440-, 900-m, and infinite ranges: ambient vs. cold vacuum. PolyCam System MTF performance at 200-m **(a)**, 290-m **(b)**, 440-m **(c)**, 900-m **(d)**, and infinite **(e)** ranges using horizontal slit comparing pre-vibe, pre-TVAC performance at ambient temperature and pressure on the left with post-vibe, cold-temperature performance in a vacuum on the right. In general, the horizontally oriented MTF (as opposed to the vertically oriented MTF) represents the worst-case resolution performance for the detector

**Fig. 35(a–c)** Bennu phase functions. Based on work reported in Takir et al. (2015), these curves describe Bennu's phase function with increasing phase angle at different incidence angles when modeled with functional form using **(a)** modified Lommel-Seeliger, **(b)** modified Minneart, and **(c)** modified ROLO model dependencies at different incidence angles. When disk-integrated its predicted point-source phase function matches that from Hergenrother et al. (2014). (Driss Takir, personal communication, 2016.)

**Fig. 36** OCAMS radiance calibration testbed. The OCAMS radiance calibration integrating sphere; units under test mount opposite the main port and view the interior. Simultaneously, a calibrated standard detector views the interior from an equatorial port

**Fig. 37** Radiance calibration illumination geometry. A schematic of the radiance calibration testbed as seen from above is shown. Proxy calibrated standard detector mounted to view the integrating sphere's main port, while a calibrated standard detector monitors the radiance within the sphere. Light is injected into the sphere from a small sphere illuminated by a quartz-tungsten-halogen (QTH) lamp

**Fig. 38** OCAMS ISO rating. Signal-to-noise ratio plotted against OCAMS detector equivalent ISO rating. At a 40:1 signal-to-noise ratio, the ISO rating is around 830, and at a 10:1 SNR, the rating is 4700

**Fig. 39** Spectral responsivity illumination geometry. A schematic of the spectral responsivity laboratory testbed as seen from above is shown. A single-pass monochromator injects light, and a calibrated standard detector monitors the radiance of the sphere's interior

**Fig. 40** MapCam relative spectral responsivity. MapCam relative spectral responsivity for all six optical paths, using data acquired at coarse (10 nm) and fine (1 nm) resolutions shown as dark- and light-shaded blue (B), green (V), red (W), and gold (X) for the color filters. Black indicates the Pan and Pan-30 plates which plot on top of each other; they were measured with a 10-nm resolution

**Fig. 41** Spectral responsivity of the PolyCam Pan filter. Spectral responsivity of the PolyCam Pan filter is shown

**Fig. 42** Spectral responsivity of the SamCam Pan filter. Spectral responsivity of the SamCam Pan filter is shown

**Fig. 43** Degree of PolyCam scattered light (X-axis). The magnitude of stray light scattered onto the PolyCam focal plane is shown as a relative quantity in this diagram, the Point Source Normalized Irradiance Transmittance (PSNIT). The PSNIT is the ratio of the detector irradiance to the irradiance incident on the camera. It magnitude is the result of a theoretical model of scattered light involving the tracing of billions of rays from an off-axis point source. The lateral position along the PolyCam X-axis of this point source is plotted along the graph's horizontal axis; the relative signal from this point source which is scattered onto the focal plane is shown along the vertical axis

**Fig. 44** Degree of PolyCam scattered light (Y-axis). The magnitude of stray light scattered onto the PolyCam focal plane is shown as a relative quantity in this diagram, the PSNIT. The lateral position along the PolyCam Y-axis of this point source is plotted along the graph's horizontal axis; the relative signal from this point source which is scattered onto the focal plane is shown along the vertical axis

**Fig. 45** Degree of MapCam scattered light (X-axis). The magnitude of stray light scattered onto the MapCam focal plane is shown as the PSNIT. The lateral position along the MapCam X-axis of this point source is plotted along the graph's horizontal axis; the relative signal from this point source which is scattered onto the focal plane is shown along the vertical axis

**Fig. 46** Degree of MapCam scattered light (Y-axis). The magnitude of stray light scattered onto the MapCam focal plane is shown as the PSNIT. The lateral position along the MapCam Y-axis of this point which is scattered onto the focal plane source is plotted along the graph's horizontal axis; the relative signal from this point source is shown along the vertical axis

**Fig. 47** Degree of SamCam scattered light (X-axis). The magnitude of stray light scattered onto the SamCam focal plane is shown as the PSNIT. The lateral position along the SamCam X-axis of this point source is plotted along the graph's horizontal axis; the relative signal from this point source is shown along the vertical axis

**Fig. 48** OSIRIS-REx Bennu proximity operations. A staged series of mission phases gather the knowledge to maintain a low-risk posture and ensure the success of a sample acquisition. Sample images from each of the phases are shown around the edge of the figure

**Fig. 49** OSIRIS-REx cruise operations. OSIRIS-REx mission operations during the first 732 days after launch. The outbound cruise, punctuated by the Earth-Gravity Assist, includes camera operations that are devoted to refining the calibrated radiometric, optical, and geometric performance

**Fig. 50** Post-sample verification imaging. SamCam images are acquired of the TAGSAM head during the post-sampling phase from three configurations through the diopter filter (from Mink et al. 2014)

# Tables

**Table 1** OCAMS optical and radiometric properties. The OCAMS imagers span a factor of 25 in resolution and 1650 in point-source sensitivity. They are relatively fast by planetary standards, appropriate to the dark body they are to be imaging. The PolyCam's nominal aperture diameter is 200 mm, but its effective diameter smaller than that due to obstruction by the secondary and the mounting vanes ("spiders")

**Table 2** Sample section of PolyCam focus table. This excerpt from the PolyCam focus table highlights the motor step positions at the 200-m and infinite ranges. For simplicity of operation, each of the PolyCam's 40 accessible shutter rotations is divided into three valid focus positions, 120° of shutter rotation apart

**Table 3** OCAMS locations in spacecraft coordinate frame. Entrance pupil locations and instrument origins (in.) for all three OCAMS Cameras within the spacecraft coordinate frame

**Table 4** OCAMS FSW/CCM commands. The OCAMS command lexicon is summarized in this table. Commands can be issued to operate all mechanisms, heaters, detectors, and calibration lamps, directly address memory, reset the system, access a palette of action sequences, and write to the Field-Programmable Gate Array (FPGA)

**Table 5** Predicted OCAMS flight temperatures. Flight temperature extremes currently predicted by the OSIRIS-REx thermal model and reported in the thermal memo, in °C by camera subsystem

**Table 6** Detector properties vs. requirements. Requirements for the OCAMS detectors written during the design phase are compared to their realized performance. The detectors exhibit relatively high dynamic range, a quantum efficiency well-aligned to solar, acceptable read noise, low dark current, a short shutter time, and a useful anti-blooming capability. They had a measured ionizing radiation tolerance and were the product of an ongoing fabrication process. They were flight-qualified before delivery

**Table 7(a–c)** OCAMS measured line-spread function vs. range to target. Line-spread function full-width-at-half-max (FWHM) for all three OCAMS imagers are expressed in pixels at different target ranges for the widest (worst-performing) one-dimensional axis

**Table 8** Geometric factor parameter values. Values of parameters used in the determination of geometric factors in the OCAMS integrating sphere-aided radiometric calibration are presented here

**Table 9** OCAMS calibrated radiometric performance. OCAMS absolute responsivities to laboratory and solar radiation reported from the radiometric ground calibration for each of the distinct optical modes

**Table 10** Planned in-flight imaging calibrations. OCAMS calibrations planned in-flight monitor detector responsivity, optical resolution performance, dark current and bias levels, and charge transfer efficiency, and calibrate absolute and relative color responsivity, as well as geometric distortion and boresight. In addition, calibrations are planned hot and cold in order to assess optical and radiometric dependence on temperature

**Table 11** OCAMS combined absolute and distortion calibration targets. Two solar analogues (18 Sco and HD 138573) and two well-known open clusters were down-selected from many candidates in order to satisfy the competing demands of imager radiometry and geometry, spacecraft pointing, operations, and downlink, and ground data system sequence construction

Table 1

| OPTICAL | Units | MapCam | SamCam | PolyCam @ 200 m | PolyCam @ ∞ |
|---|---|---|---|---|---|
| Focal Length | mm | 125 | 24 | 610.2 | 628.9 |
| F/# |  | 3.3 | 5.6 | 3.5 | 3.5 |
| Aperture Diameter | mm | 38 | 4.3 | 175 | 175 |
| IFOV | mrad | 0.068 | 0.354 | 0.0139 | 0.0135 |
| IFOV | arcsec | 14.0 | 73.1 | 2.9 | 2.8 |
| FOV | mrad | 69.6 | 363 | 14.3 | 13.8 |
| Range near which Bennu becomes an extended source | km | 7353 | 1412 | N/A | 36994 |

Table 2

| FOCUS MECHANISM RANGE TABLE | Position ID A B C | Operating Area (Optical Path Open) These Ranges (Motor Steps) are Directly Accessible; All Other Ranges Indirectly Accessible at these Motor Step Positions due to Depth of Field | | | | | | | |
| --- | --- | --- | --- | --- | --- | --- | --- | --- | --- |
| | Shutter | Shutter Area (Optical Path Blocked!) These Ranges (Motor Steps) Are Not Directly Accessible and must Default to Nearest Accessible Range (Motor Step) | | | | | | | |
| REV # (Operating) | Position ID | Near Range (m) | Far Range (m) | Motor Position (Step #) | Motor Position (hex) | Depth of Field (m) | Focal Length (mm) | Focal Length (mm) | IFOV (mrad) | IFOV (mrad) |
| 7 | Shutter | 192.0 | 194.8 | 3781 | EC5 | 12.8 | 609.49 | 609.75 | 0.0139 | 0.0139 |
| | A | 194.8 | 196.3 | 3871 | F1F | 13.1 | 609.75 | 609.90 | 0.0139 | 0.0139 |
| | B | 196.3 | 199.3 | 4051 | FD3 | 13.4 | 609.90 | 610.16 | 0.0139 | 0.0139 |
| | C | 199.3 | 200.6 | 4231 | 1087 | 13.8 | 610.16 | 610.27 | 0.0139 | 0.0139 |
| | | • | | • | | • | | | |
| | | • | | • | | • | | | |
| | | • | | • | | • | | | |
| 31 | Shutter | 3657.1 | 4732.5 | 16739 | 4163 | 3803.8 | 627.68 | 627.94 | 0.0135 | 0.0135 |
| | A | 4732.5 | 5797.5 | 16830 | 41BE | 5779.7 | 627.94 | 628.11 | 0.0135 | 0.0135 |
| | B | 5797.5 | 9258.6 | 17010 | 4272 | 7902.3 | 628.11 | 628.39 | 0.0135 | 0.0135 |
| | C | 9258.6 | 11420.0 | 17190 | 4326 | 15381.2 | 628.39 | 628.48 | 0.0135 | 0.0135 |
| 32 | Shutter | 11420.0 | 39036.2 | 17280 | 4380 | 20275.8 | 628.48 | 628.77 | 0.0135 | 0.0135 |
| | A | 39036.2 | ∞ | 17371 | 43DB | 85854.0 | 628.77 | 628.90 | 0.0135 | 0.0135 |

Table 3

| | Cameras | | | SamCam | MapCam | PolyCam | OTES |
|---|---|---|---|---|---|---|---|
| Coordinates (in.) | Instr. Origins (S/C Frame) | | X | 1.25 | −12.64 | −28.81 | |
| | | | Y | 27.00 | 42.25 | 31.92 | |
| | | | Z | 0.00 | 0.00 | 0.00 | |
| | Pupil Loc. (Instr. Frame) | | X | −1.25 | −6.1 | 5.81 | |
| | | | Y | −6.53 | 0.75 | 1.08 | |
| | | | Z | 3.82 | 7.27 | 5.13 | |
| | Pupil Loc. (S/C Frame) | | X | 0.00 | −18.74 | −23.003 | −38.08 |
| | | | Y | 20.47 | 43.00 | 33.00 | 25.12 |
| | | | Z | 3.82 | 7.27 | 5.13 | |

Table 4

| OCAMS | COMMANDS | Op Code |
|---|---|---|
| Motor Commands | Motor Enable | 10 |
| | Motor Move Relative | 11 |
| | Motor Abort | 13 |
| | Motor Home | 14 |
| | Motor Safe | 15 |
| | Motor Move Absolute | 16 |
| | Motor Move Index | 17 |
| | Motor Index LED Enable | 18 |
| Heater Commands | Heater All Off | 30 |
| | Heater Duty Cycle | 32 |
| Dynamic Command Processing | Dynamic Camera Setup | 40 |
| | Dynamic Camera Setup and Image | 41 |
| | Dynamic Camera Image | 42 |
| | Dynamic Motor Move Absolute | 45 |
| | Dynamic Motor Move Index | 46 |
| Camera Commands | Calibration Tracing Lamp Enable | 50 |
| | Camera Power | 51 |
| | Camera Debug | 52 |
| | Camera Image | 53 |
| | Camera Image | 54 |
| | Camera Setup and Image | 55 |
| | Camera All Off | 56 |
| | Camera Select | 57 |
| | Camera Setup | 58 |
| Memory Operations Commands | Memory Load | 60 |
| | Memory Dump | 61 |
| | Memory Copy | 62 |
| | Memory Checksum | 63 |
| | SRAM Test | 64 |
| | Memory Load Table | 66 |
| | Memory Dump Table | 67 |
| | EEPROM_ENABLE | 68 |
| System Commands | System Reset FPGA | 72 |
| | System Execute Safe Config | 73 |
| | Set Parameter | 76 |
| | Set Analog | 77 |
| | System Unsafe | 78 |
| | System Execute Action Sequence | 80 |
| Debug Commands | Motor Index Scan | 20 |
| | Read FPGA Register | 90 |
| | Write FPGA Register | 91 |
| | EDAC Info | 102 |

Table 5

|  (°C) | Cruise | | Orbital A & B | | Det. Surv. | | Recon | | Overall | |
|---|---|---|---|---|---|---|---|---|---|---|
|  | Min | Max | Min | Max | Min | Max | Min | Max | Min | Max |
| PolyCam Detector | −24 | −23 | −13 | 23 | −25 | 13 | −25 | −22 | −25 | 13 |
| PolyCam M1 | −28 | −27 | −22 | 6 | −28 | −28 | −28 | −24 | −28 | 6 |
| PolyCam M2 | −33 | −32 | −33 | −32 | −33 | −32 | −33 | −26 | −33 | −26 |
| PolyCam L1/L2/FM | −17 | −12 | −22 | 6 | −19 | −18 | −24 | −19 | −22 | 6 |
| PolyCam Chassis | −31 | −28 | −17 | 19 | −31 | 9 | −28 | −24 | −31 | 19 |
| MapCam Detector | −24 | −23 | −22 | 11 | −25 | 18 | −24 | −21 | −25 | 18 |
| MapCam L1/L2 | −25 | −23 | −25 | 0 | −25 | −20 | −24 | −20 | −25 | 0 |
| MapCam Chassis | −31 | −25 | −8 | 28 | −31 | 32 | −24 | −18 | −31 | 32 |
| SamCam Detector | −25 | −24 | −25 | −16 | −25 | 17 | −25 | −23 | −25 | 17 |
| SamCam Lens Cell | −35 | −30 | −34 | −26 | −30 | −29 | −30 | −25 | −35 | −25 |
| SamCam Chassis | −36 | −30 | −35 | −15 | −36 | 18 | −31 | −28 | −36 | 18 |

Table 6

| Parameter | Req | Units | Teledyne DALSA Trius |
|---|---|---|---|
| **Pixel Size** | 5 - 18 | µm | 8.5 |
| **Shutter Time** | ≤ 1-2 | msec | 1.044 |
| **Active Cooling** | No | | Not required |
| **Antiblooming** | ≥ 50 | × | Yes; > 100x |
| **Drk Cur +15C (Before Radiation)** | < 2500 | e/s/px | 33 |
| **Drk Cur -15C (Before Radiation)** | < 130 | e/s/px | 1.7 |
| **Dyn Range** | ≥ 1000 | | 2400 |
| **Gain Modes** | Single-Gain | | Single-Gain |
| **Read Noise** | 0-50 | e | 30 |
| **Full Well** | ≥ 50000 | e | 60000 |
| **Read-out Rate** | ≥ 2 | fr/sec or msec | 8.8 (dual mode) = 113.4 msec; 4.4 (single mod) = 226.8 msec |
| **QE** | ≥ 0.3 | peak | 0.4 |
| **Peak Wvl** | 650 | nm | 660 |
| **Thermal Range (Oper)** | -25 - +40 | C | -25 --> +40 |
| **Thermal Range (Survival)** | -55 - +70 | C | -55 --> +70 |
| **CCD Format** | 1k × 1k | px | 1024 × 1024 |
| **Shielded Columns** | Yes | | 16, 16, 6, 6, 4, 4, 4, 4 |
| **Shutter Architecture** | Interline or frame | | Frame Transfer |
| **Comp of Sup Elec** | | | 2-level vs. 3-level Clocking; Read-out Current Spikes |
| **Bits of Full Digit** | 12 - 14 | bit | 14 |

Table 7

**(a)**

| MapCam | 31.1 m | 32.9 m | 34.7 m | | |
|---|---|---|---|---|---|
| Pan 30 | 1.30 | 1.38 | 1.55 | | |
| | 125 m | 500 m | 950 m | 3500 m | ∞ |
| Pan | 1.60 | 1.40 | 1.50 | 1.59 | 1.63 |
| | | 500 m | 950 m | 3500 m | ∞ |
| | B Filter | 1.27 | 1.30 | 1.33 | 1.35 |
| | V Filter | 1.18 | 1.17 | 1.19 | 1.20 |
| | W Filter | 1.24 | 1.22 | 1.24 | 1.25 |
| | X Filter | 1.38 | 1.40 | 1.42 | 1.44 |

**(b)**

| SamCam | 2 m | 2.1 m | 2.2 m | |
|---|---|---|---|---|
| Pan Diopter | 1.32 | 1.27 | 1.23 | |
| | 2.8 m | 5 m | 33 m | ∞ |
| Pan 1 | 1.75 | 1.22 | 1.41 | 1.87 |
| Pan 4 | 1.76 | 1.23 | 1.40 | |
| Pan 5 | 1.84 | 1.24 | 1.36 | |

**(c)**

| PolyCam | 200 m | 290 m | 440 m | 900 m | ∞ |
|---|---|---|---|---|---|
| Pan | 1.35 | 1.29 | 1.38 | 1.34 | 1.36 |

Table 8

| Quantity | Value |
|---|---|
| $r$ (m) | 0.1016 (±0.0001) |
| $d$ (m) | 0.5421 (±0.0001) |
| Signal of proxy detector without filter holder (A) | $1.882 \times 10^{-5}$ ($\pm 0.001 \times 10^{-5}$) |
| Signal of monitoring detector (A) | $5.167 \times 10^{-5}$ ($\pm 0.001 \times 10^{-5}$) |
| Signal of proxy detector with filter holder (A) | $6.380 \times 10^{-7}$ ($\pm 0.05 \times 10^{-7}$) |
| Signal of monitoring detector (A) | $5.147 \times 10^{-5}$ ($\pm 0.001 \times 10^{-5}$) |
| $\Omega_{proxy,f}$ (sr) (proxy detector with filter holder) | 0.03291 |
| $\Omega_{mon}$ (sr) | 2.654 |
| $\Omega_{proxy}$ (sr) | 0.1066 |

Table 9

| Optical Path | Responsivity (DN sec$^{-1}$)/(W m$^{-2}$ sr$^{-1}$) between 250 and 1100 nm) | |
|---|---|---|
| | Laboratory (3000–3400 K) | Solar (5780 K) |
| PolyCam Pan | 580900 | 292100 |
| SamCam Pan & Pan-2 | 160300 | 145300 |
| MapCam Pan | 470400 | 426400 |
| MapCam Pan-30 | 464300 | 420900 |
| MapCam B | 15570 | 50420 |
| MapCam V | 40470 | 63020 |
| MapCam W | 130300 | 88610 |
| MapCam X | 128300 | 59270 |

Table 10

| Calibration | Purpose | How test performed? | # Im. | Campaign |
|---|---|---|---|---|
| Health Check | Instrument Functionality & and Performance Triage | Internal Target, Internal Blocking Plate | 82 | 14-day, 6-, 12-, 18-month, pre-Approach, Pre-Detailed Survey, 30-month, Pre-Recon, Mid-Recon, Post-Recon, Post-TAG |
| Cal Tracing Lamp | Flat Field Tracking; Motor repeatability | Internal Target | 28 | 14-day, 6-, 12-, 18-month |
| Dark Model | Tracks Hot and Random Telegraph Signal Pixels Dark and Bias Levels | Internal Blocking Plate | 126 | 14-day, 6-, 12-, 18-month |
| Solar Twin | Absolute/Spectral Responsivity | Solar Twin Targets | 1000 | 6-, 18-, 30-month |
| Open Cluster | Optical Distortion/Boresight Calibration; Camera PSF | Open Clusters Targets | 125 | 6-, 18-, 30-month |
| Glint Characterization | Stray Light and Sample Head Reference Reflectance | Extended Sample Arm and Sample Canister | 50 | Pre-TAG |
| CTE Characterization | Charge Transfer Efficiency Tracking | Internal Target | 48 | 14-day, 6-, 12-, 18-month |
| Earth/Moon | Absolute and Spectral Responsivity | Earth/Moon | 400 | EGA |

Table 11

| Target Name/Designation | RA | Dec |
|---|---|---|
| 18 Sco, HD146233 | $16^h\,15^m\,37.3^s$ | -08° 22′ 06″ |
| HD 138573, HIP 76114 | $15^h\,32^m\,43.7^s$ | +10° 58′ 05.9″ |
| M6, NGC 6405, Butterfly Cluster | $17^h\,40^m\,0.0^s$ | -32° 13′ 0.0″ |
| M7, NGC 6475, Ptolemy Cluster | $17^h53^m51.1^s$ | -34°47′34″ |

Figure 1

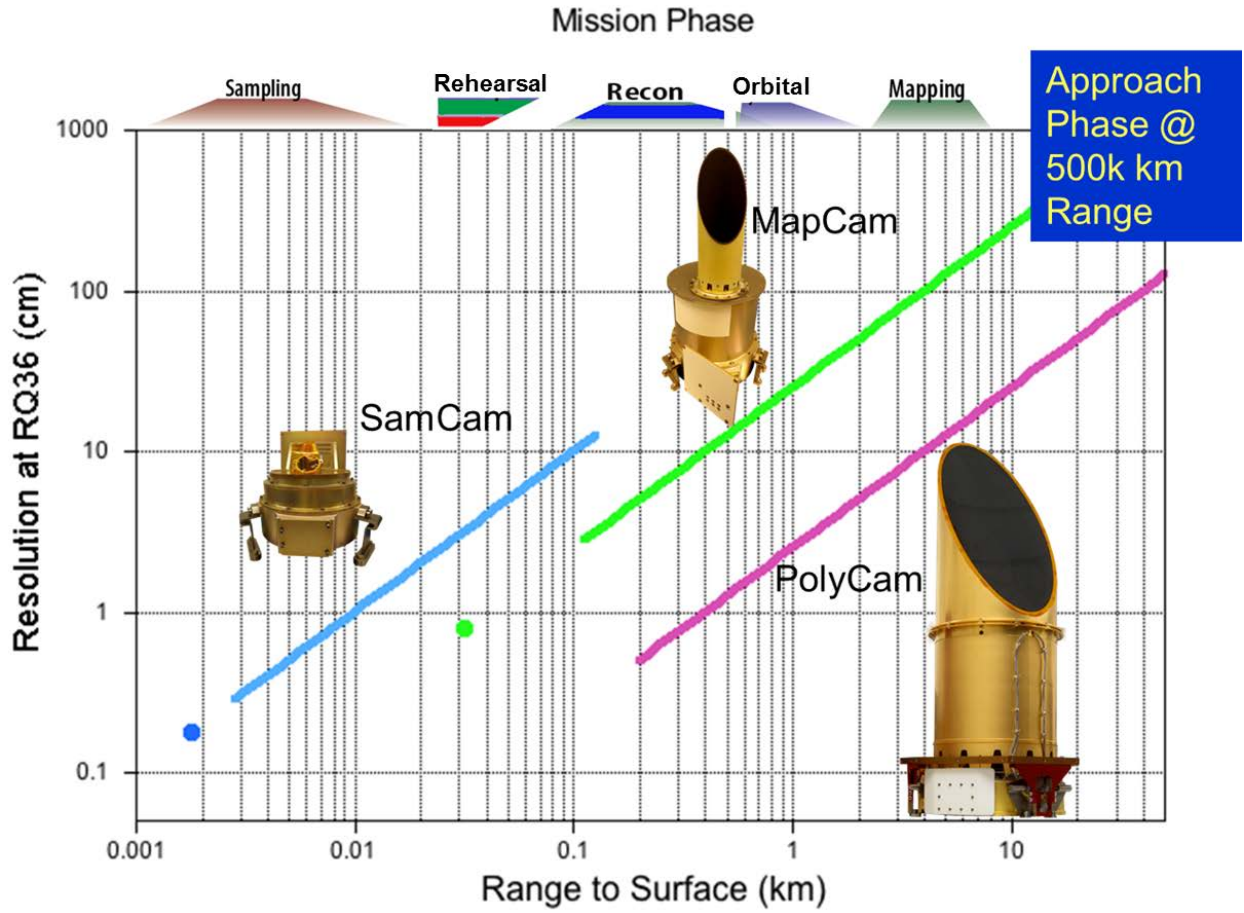

Figure 2

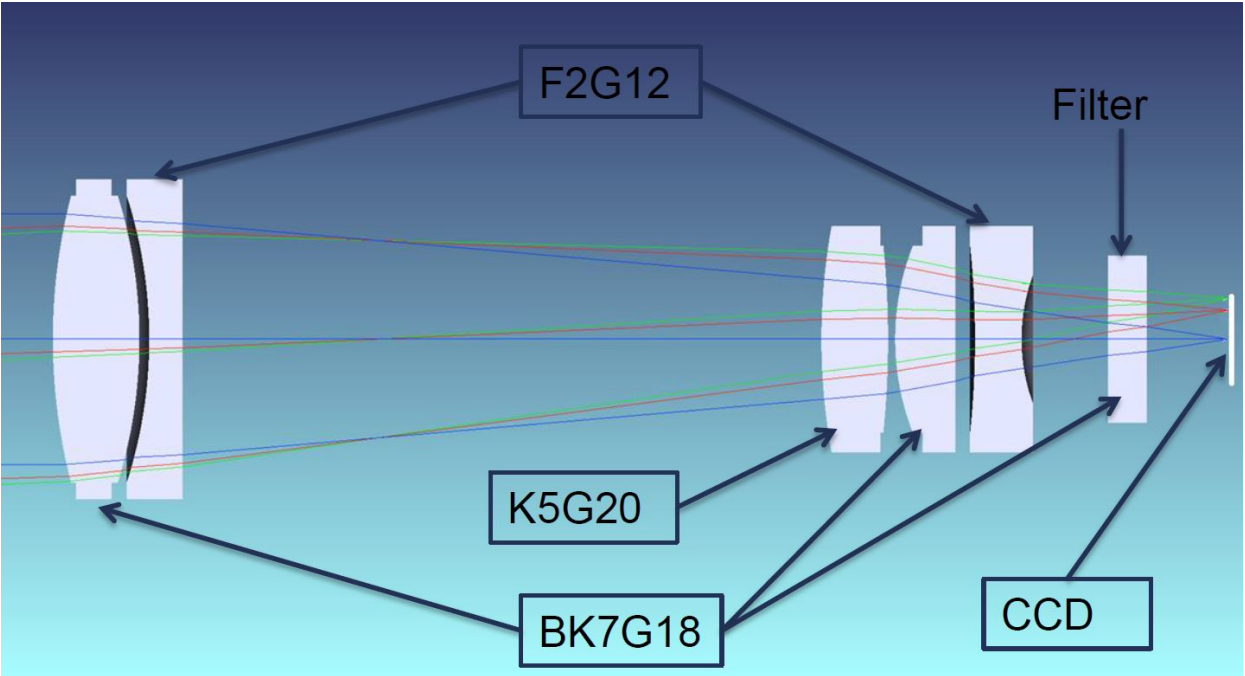

Figure 3

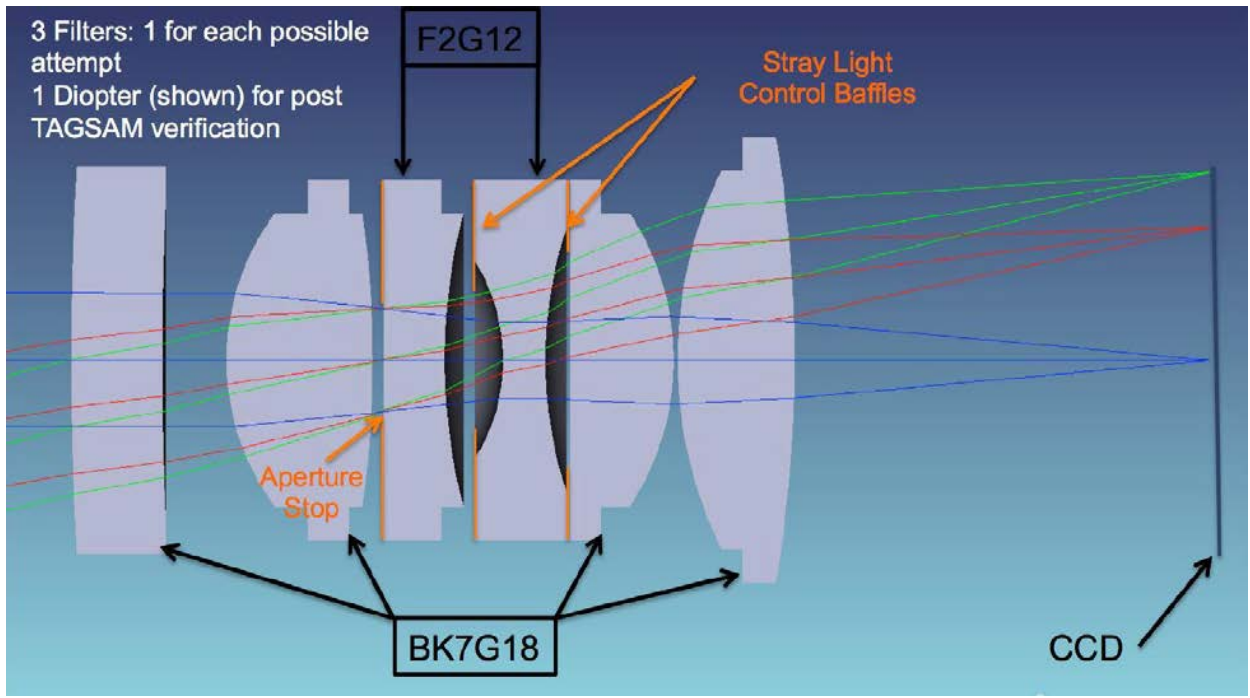

Figure 4

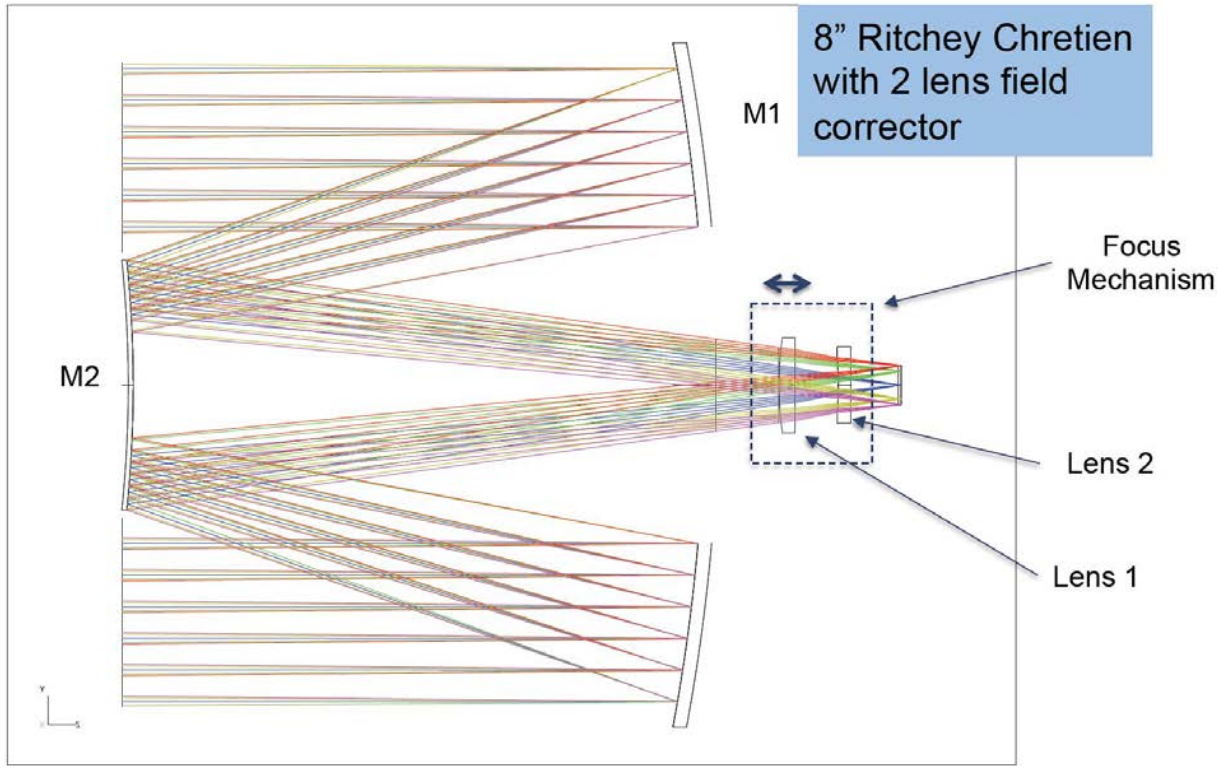

Figure 5

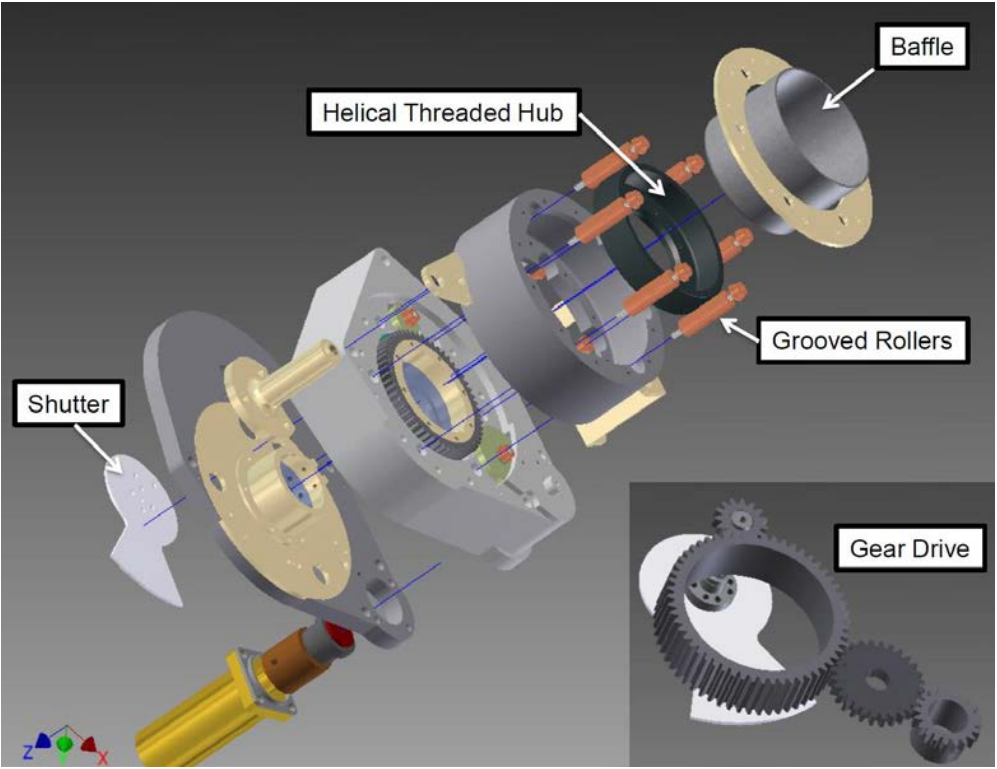

Figure 6

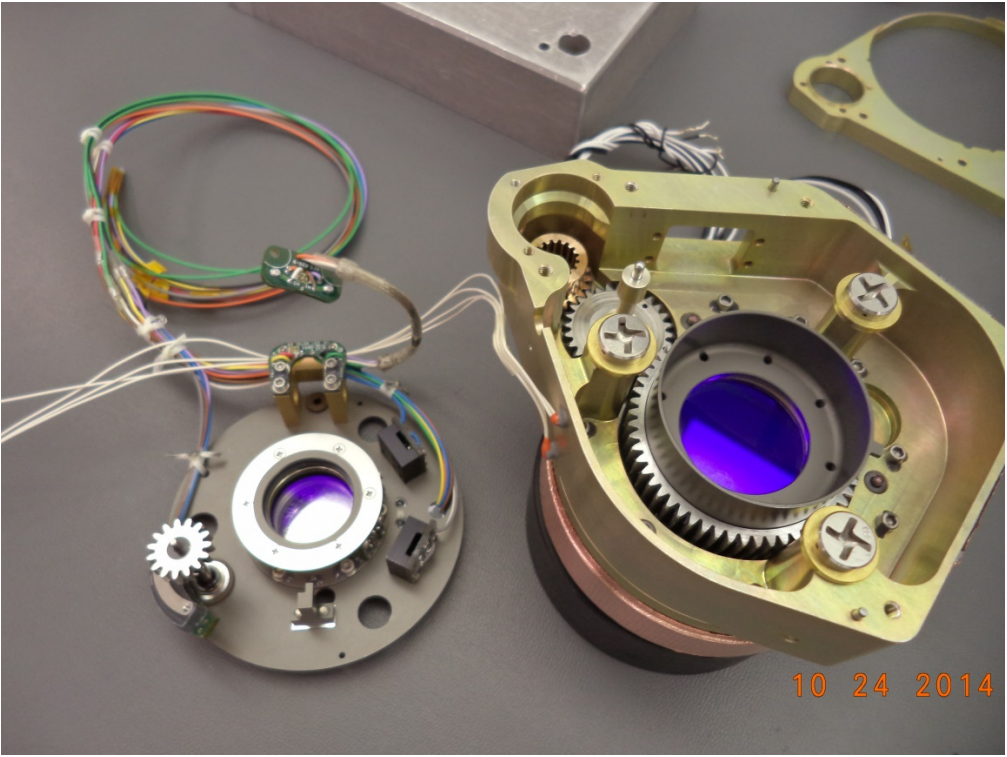

Figure 7

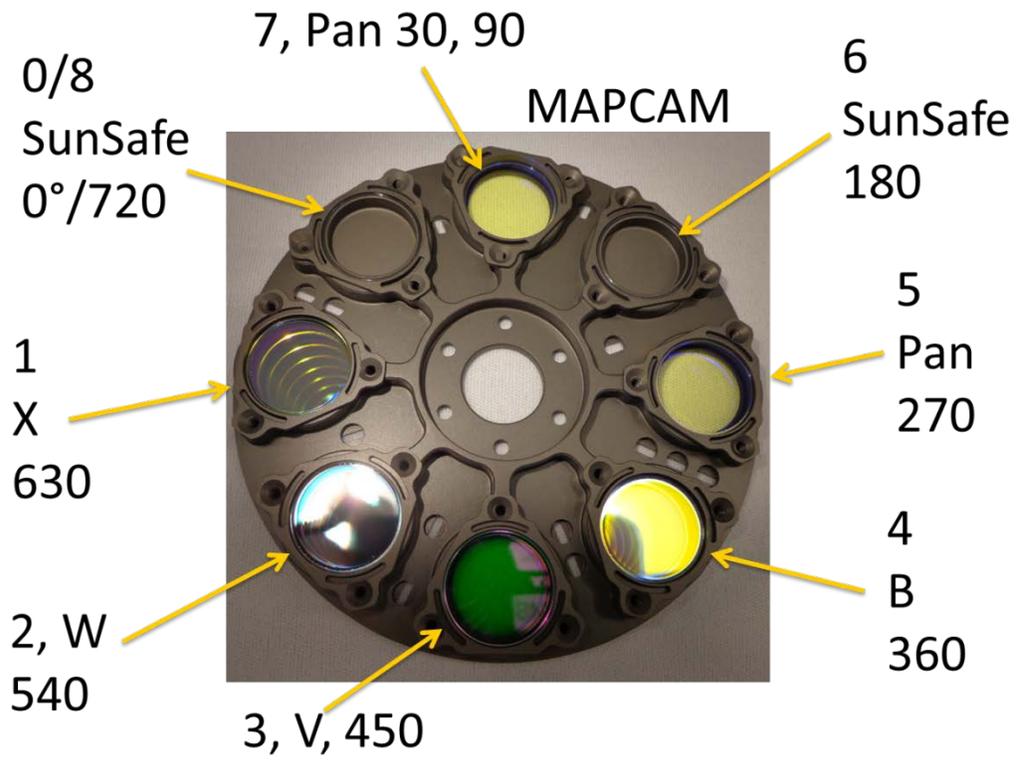

Figure 8

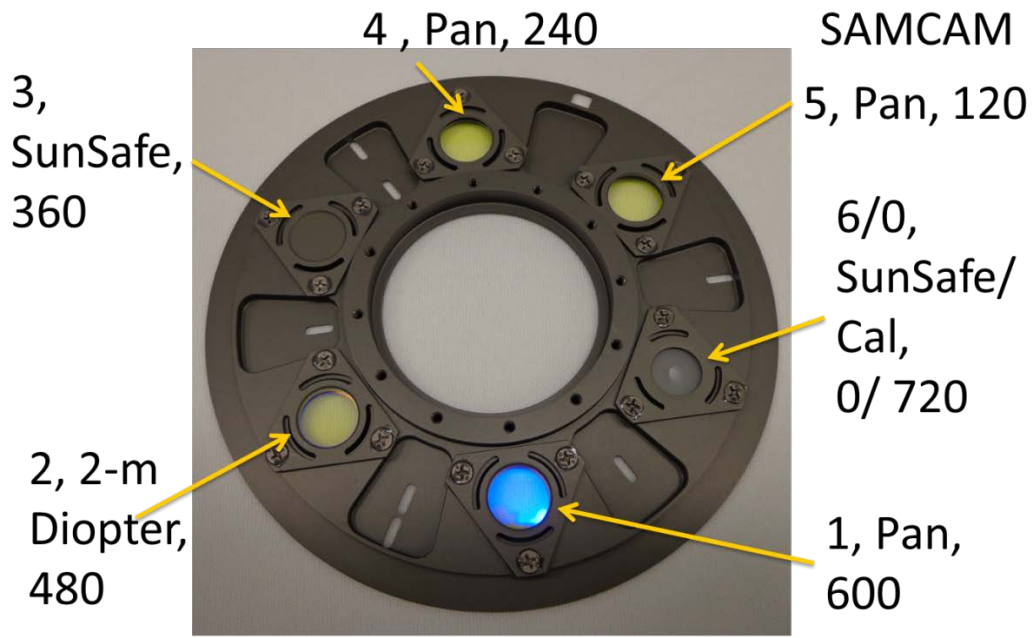

Figure 9

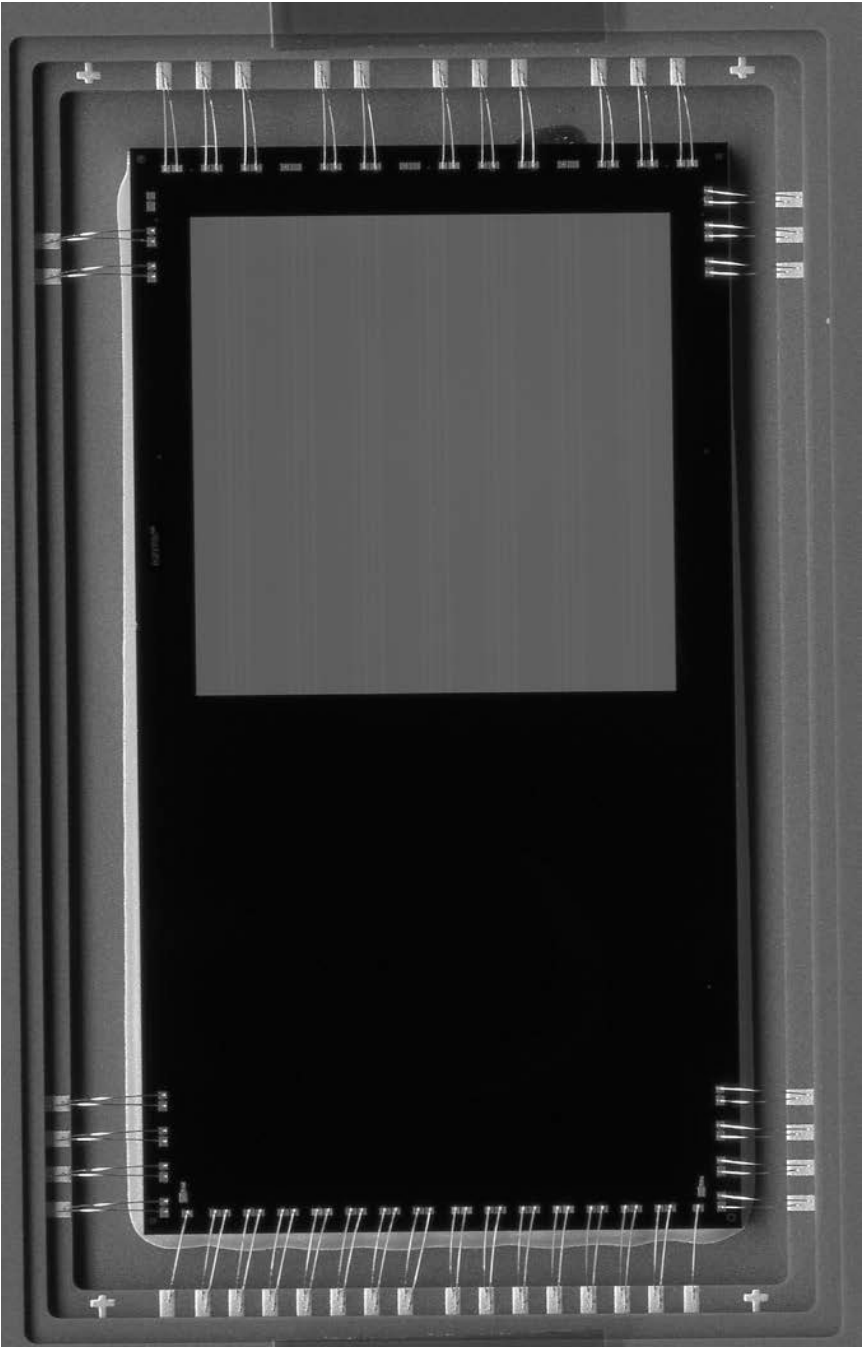

Figure 10

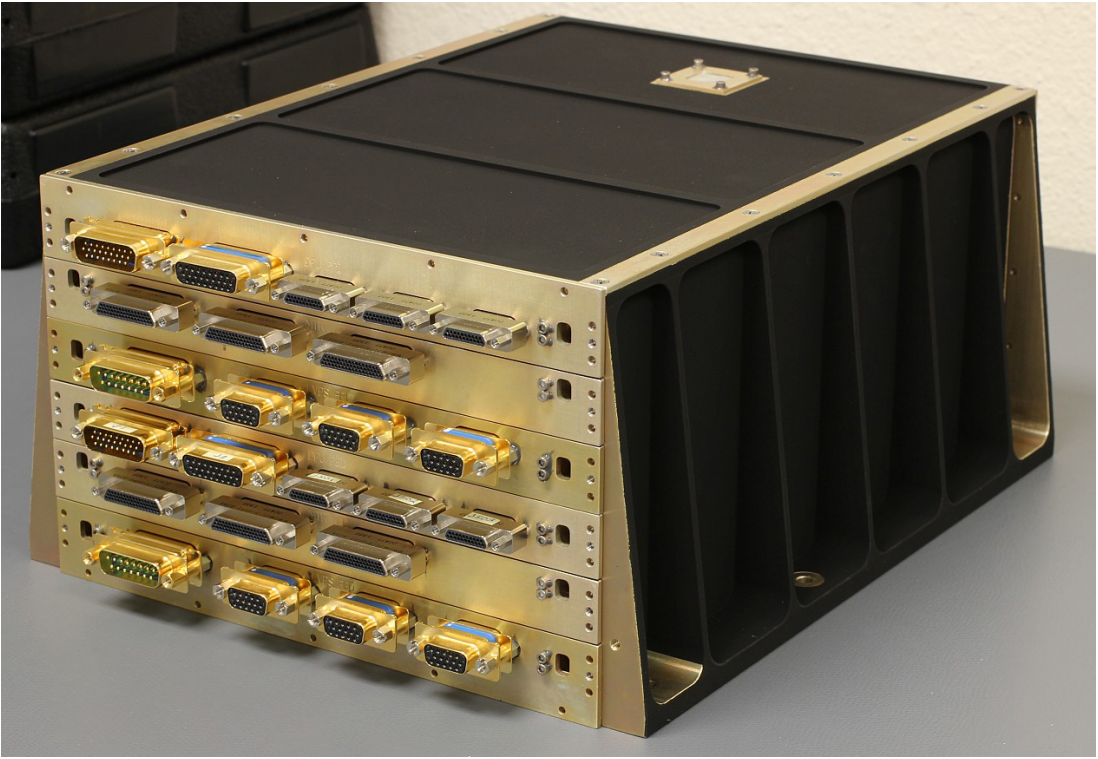

Figure 11

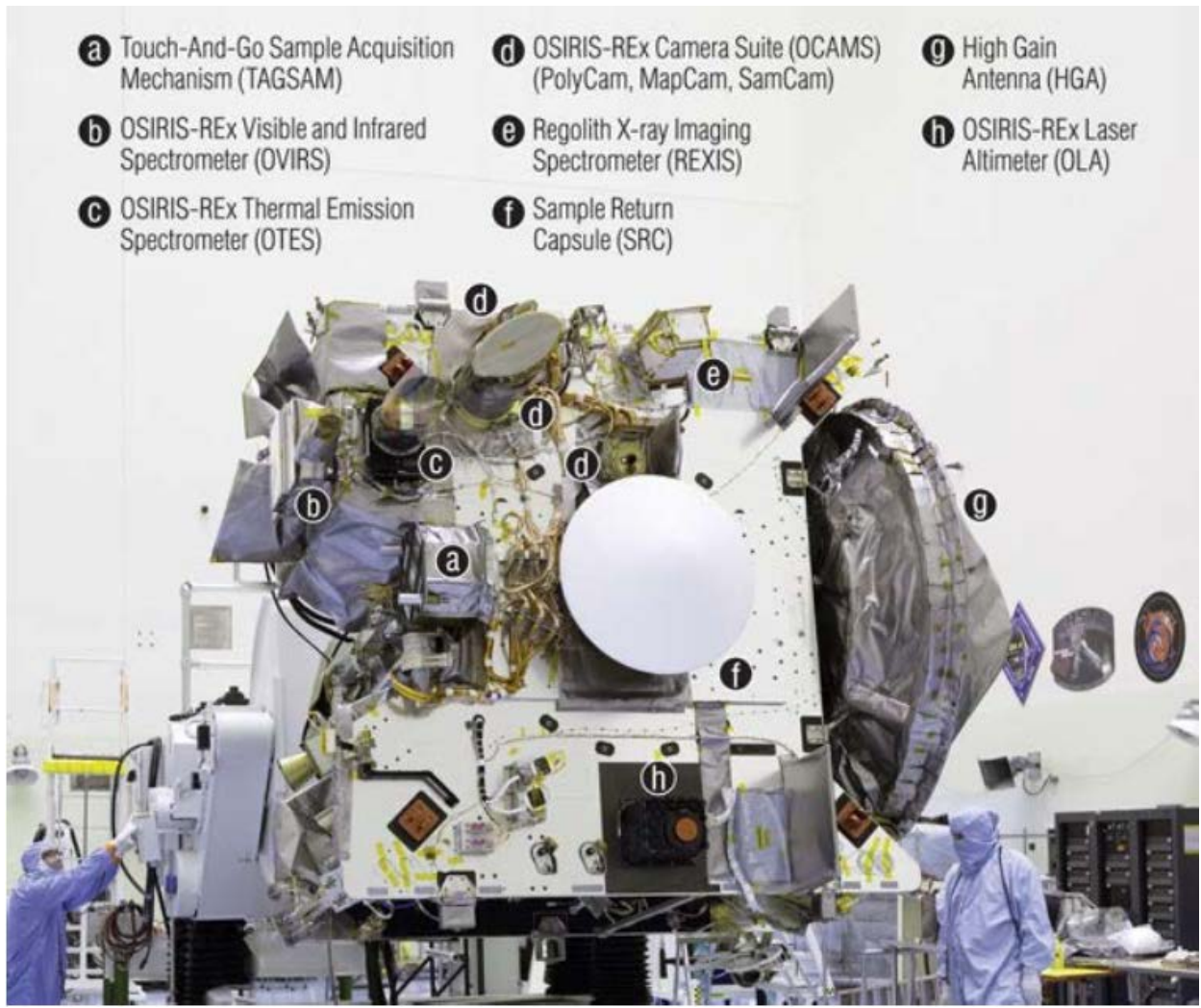

Figure 12

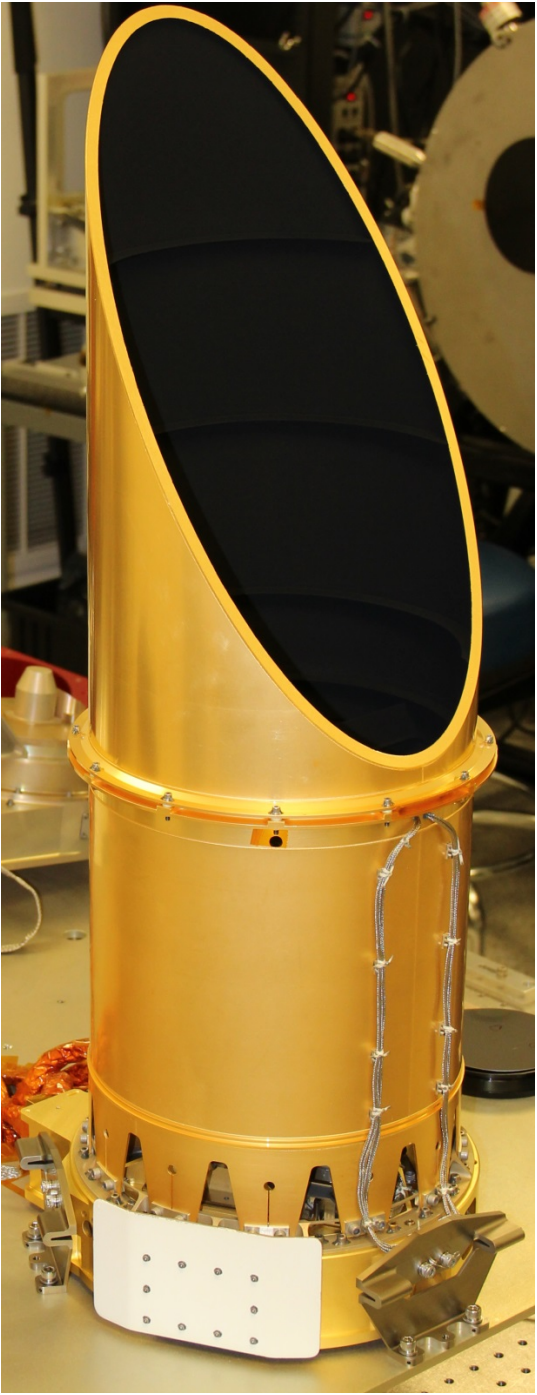

Figure 13

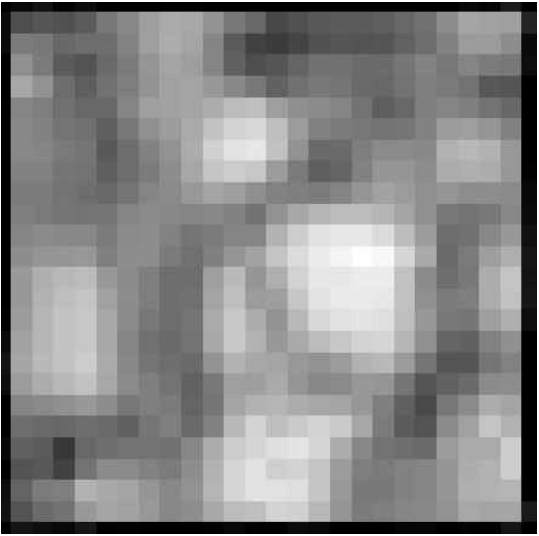

Figure 14

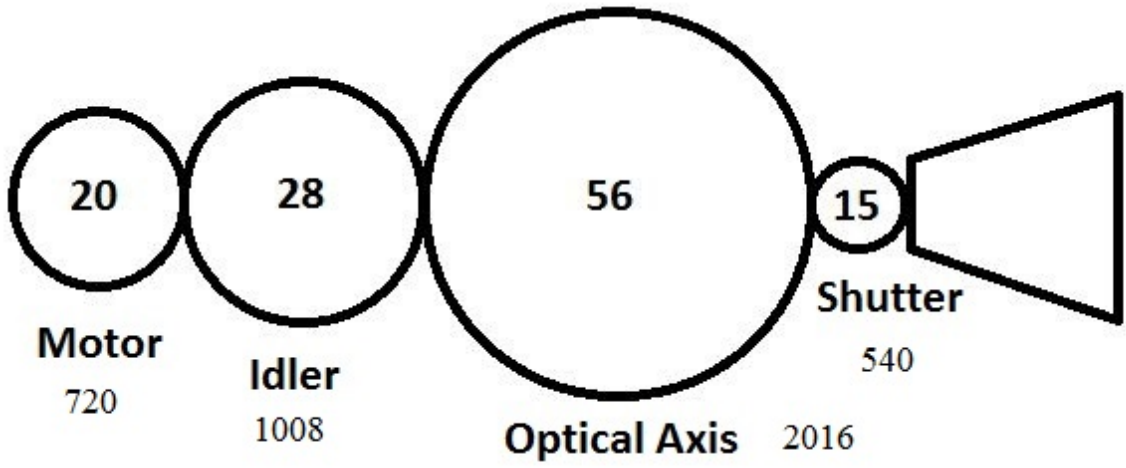

Figure 15

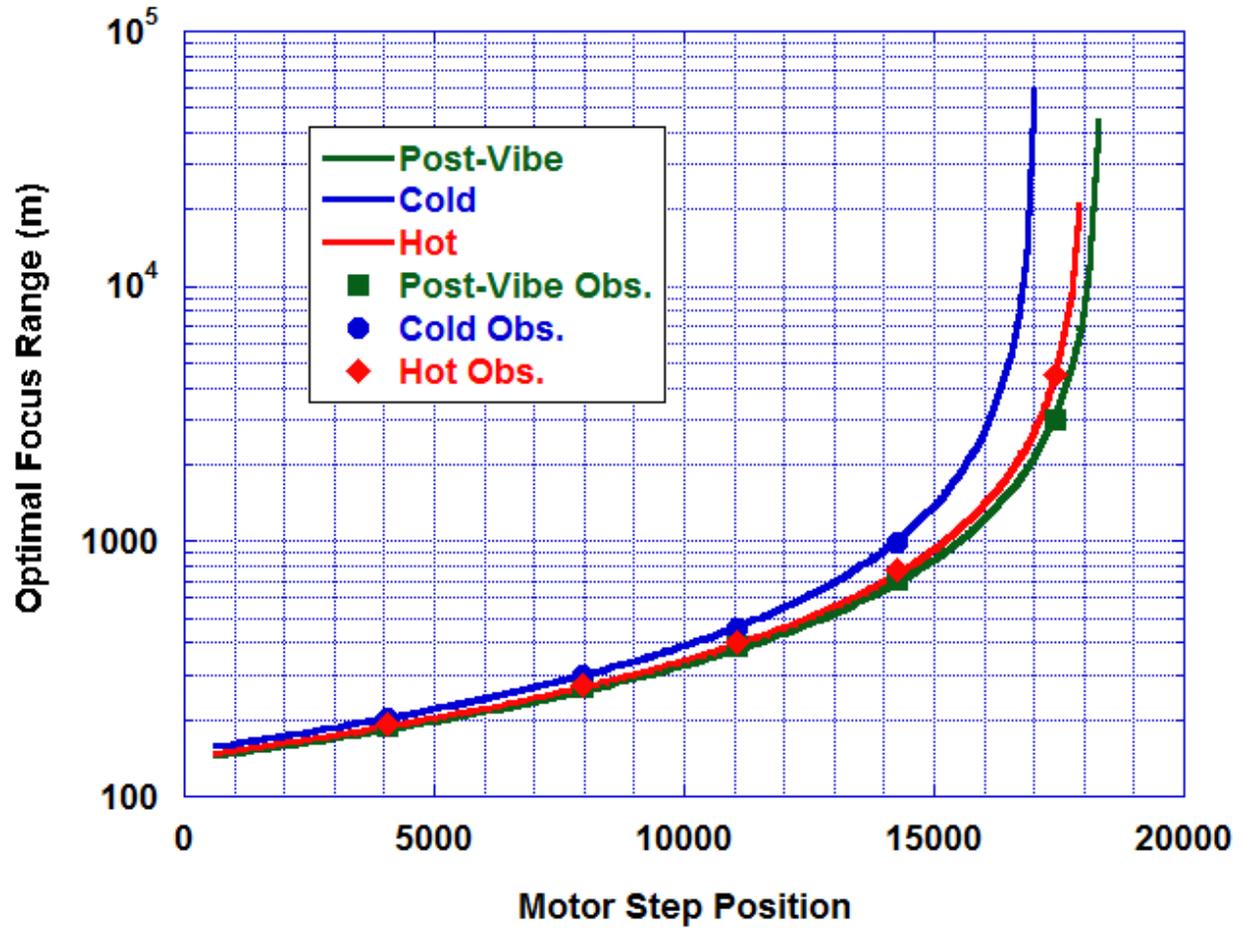

Figure 16

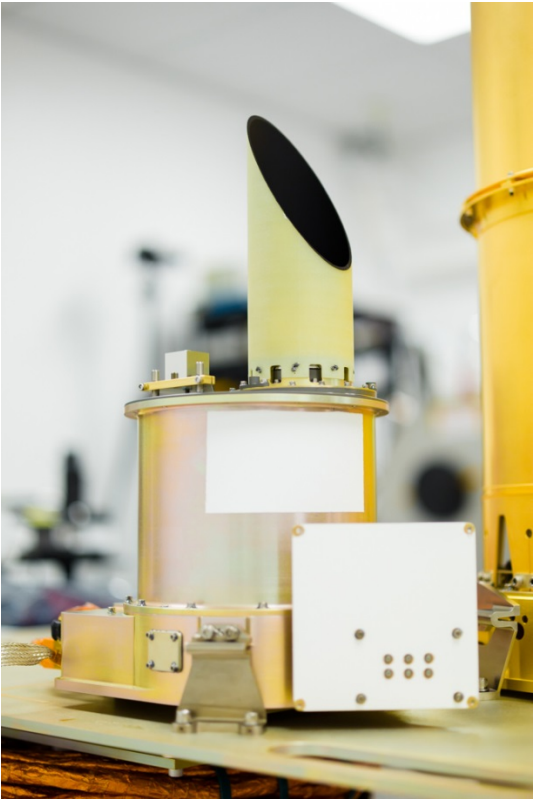

Figure 17

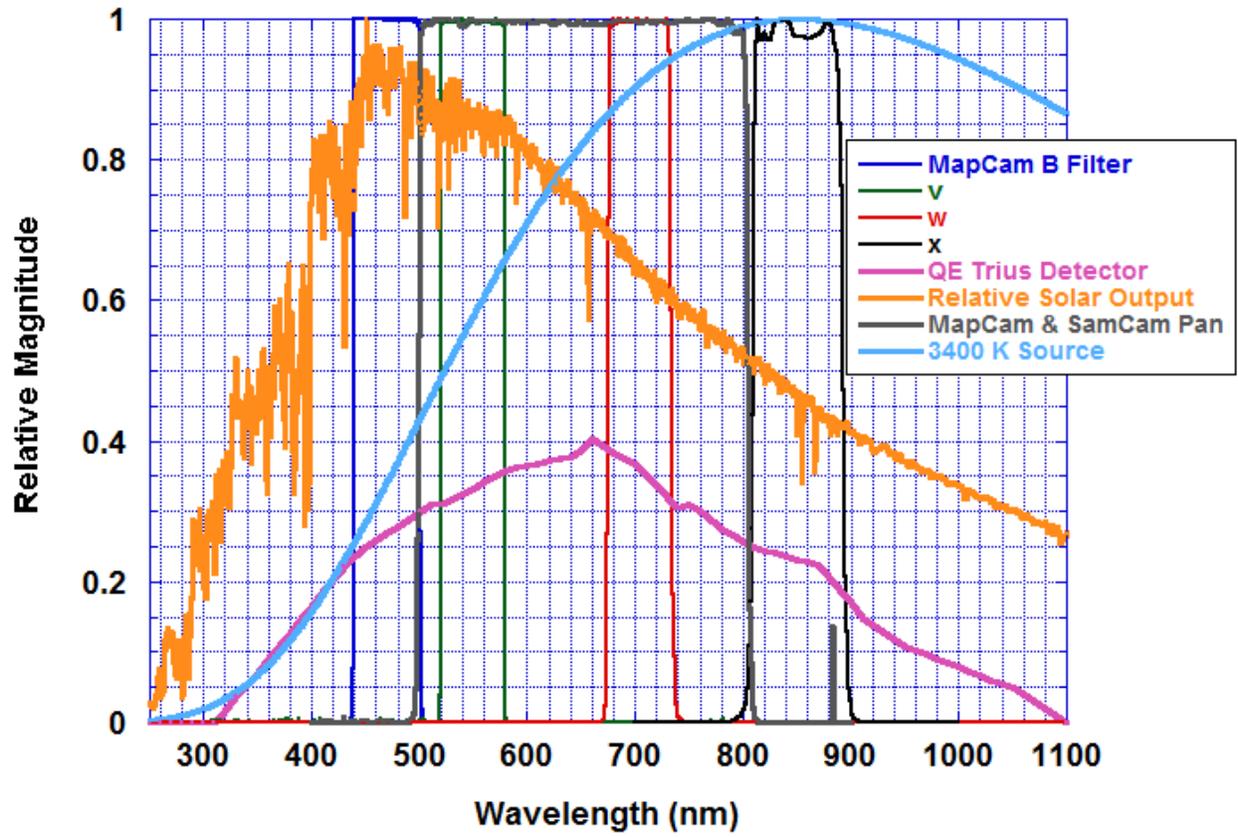

Figure 18

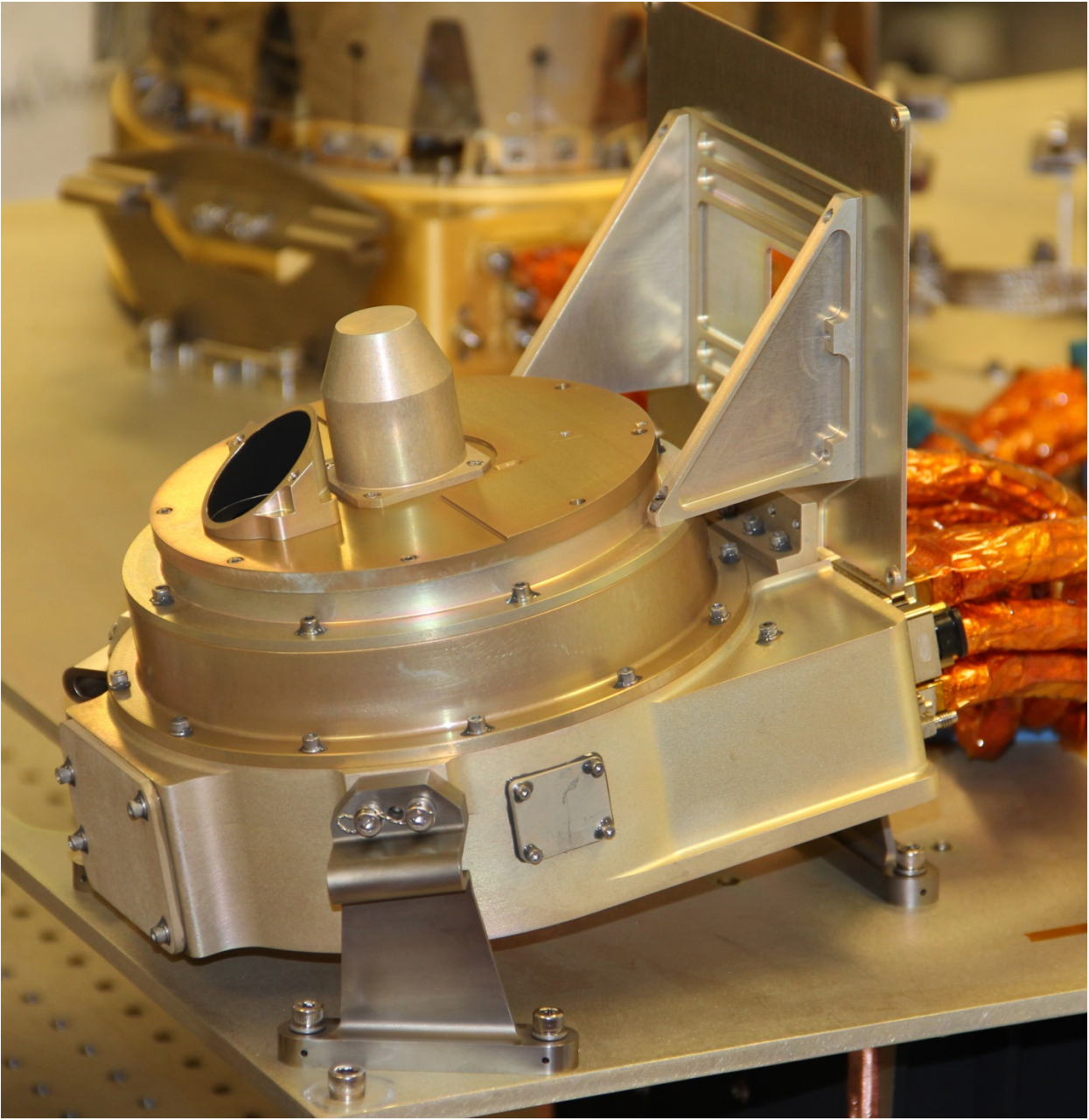

Figure 19

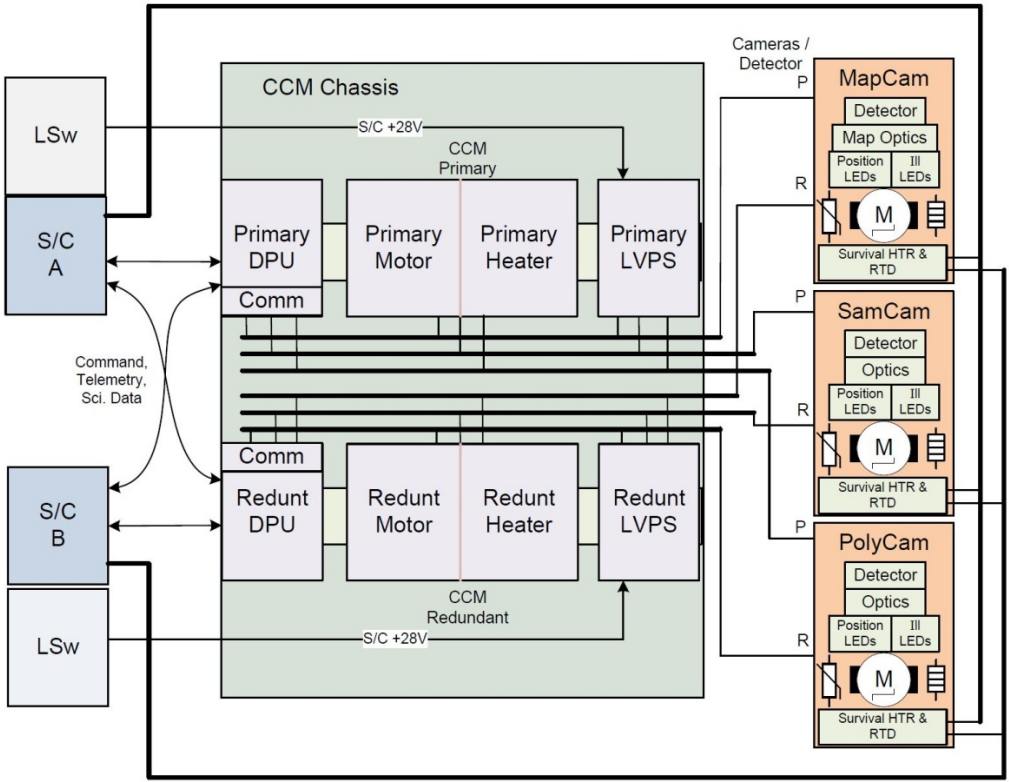

Figure 20

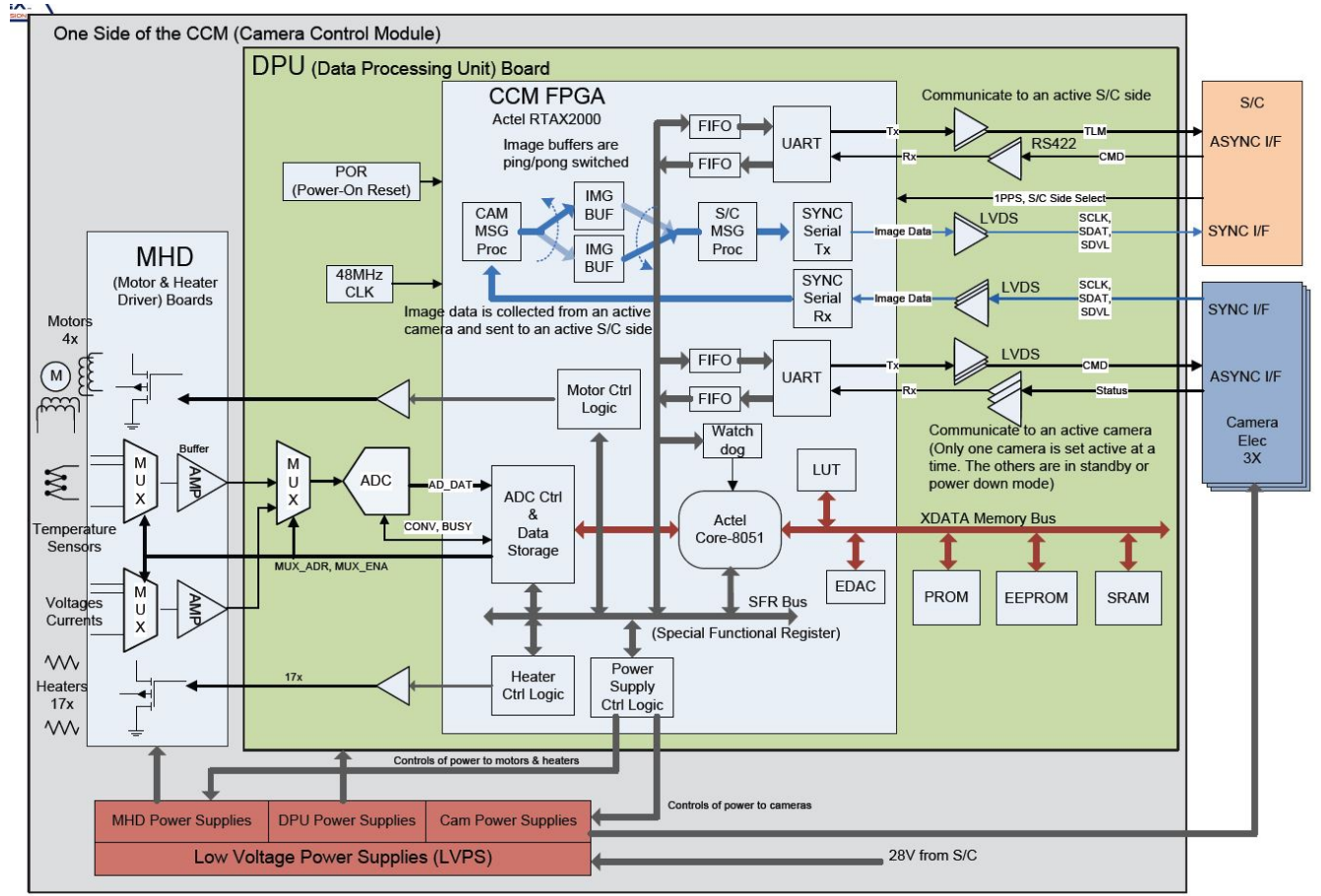

Figure 21

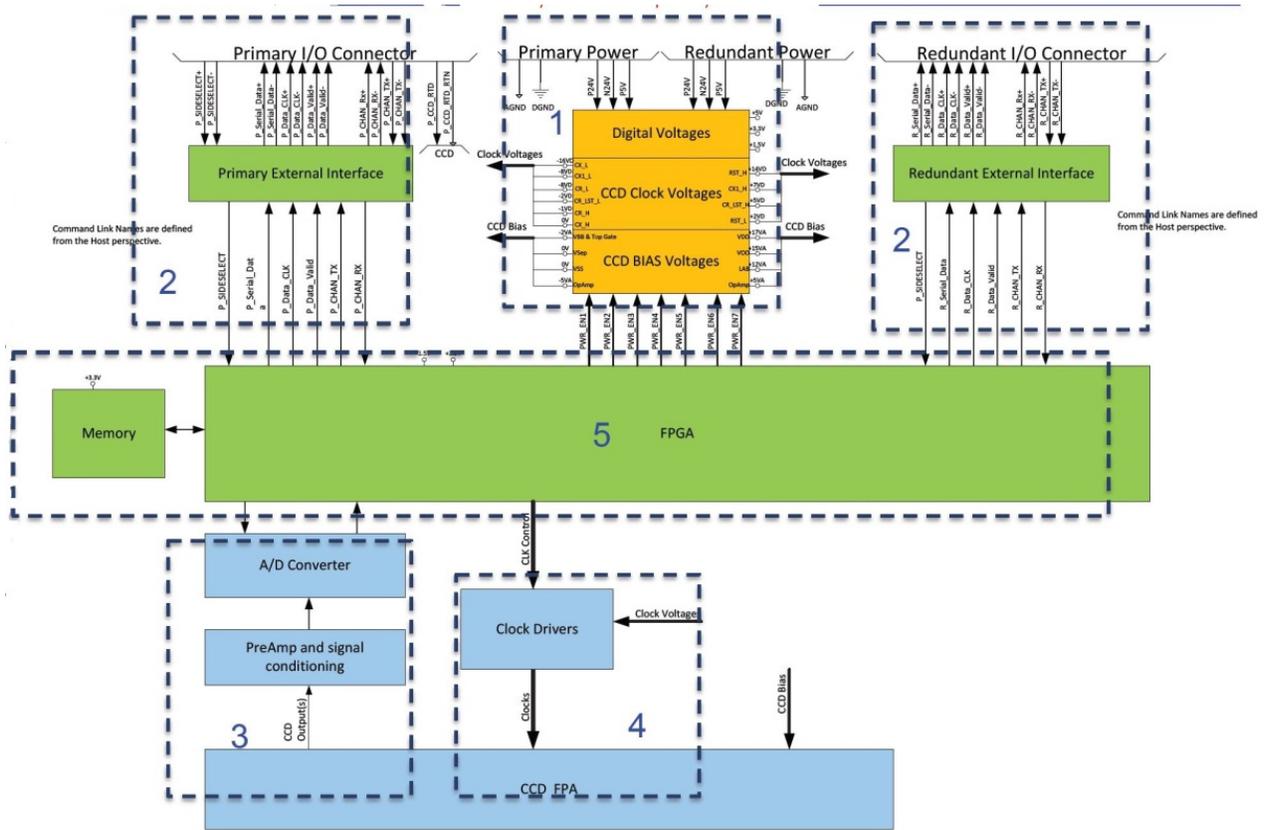

Figure 22

Full CCD image is 1112x1044
Active region – 1024x1024 region exposed to light
Covered columns – 24x1044 masked columns on left and right, used to measure dark current
Covered rows – 1032x6 masked rows on top and bottom, used to measure charge smear
Transition regions – Between active & covered, not used by default (but do receive light and could be used)
Overscan columns – 16x1044 virtual columns (empty-reads of electronics), used to measure bias
Isolation columns – 16x1044 virtual columns, read between physical and overscan to act as a buffer, not used

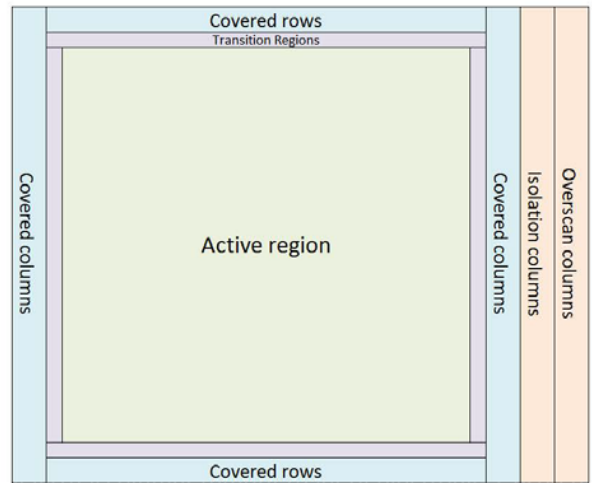

Figure 23

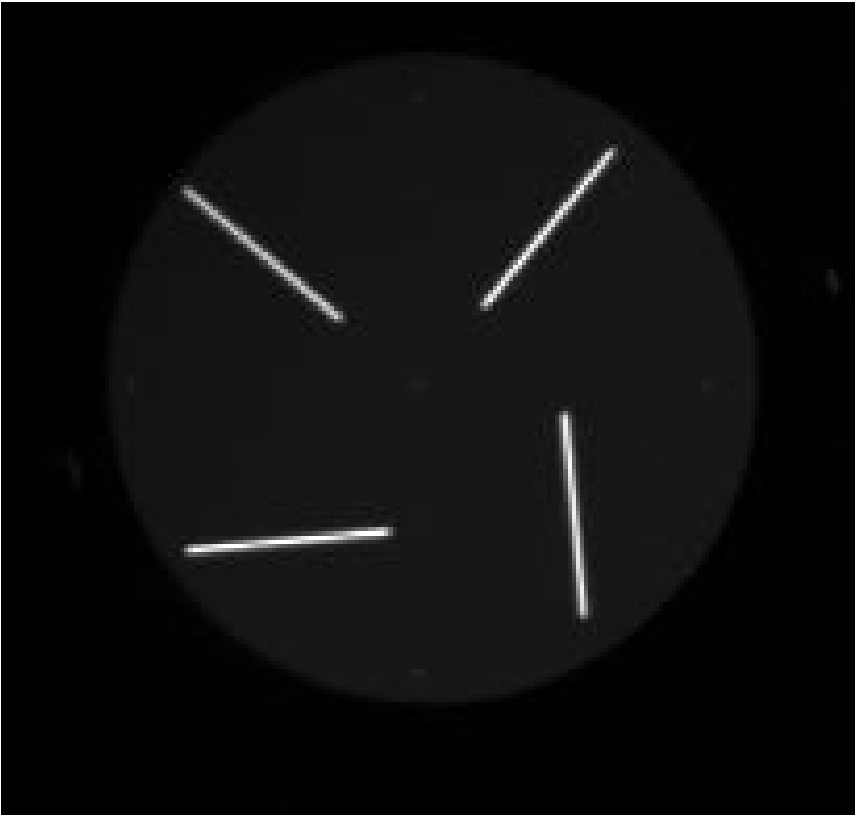

Figure 24

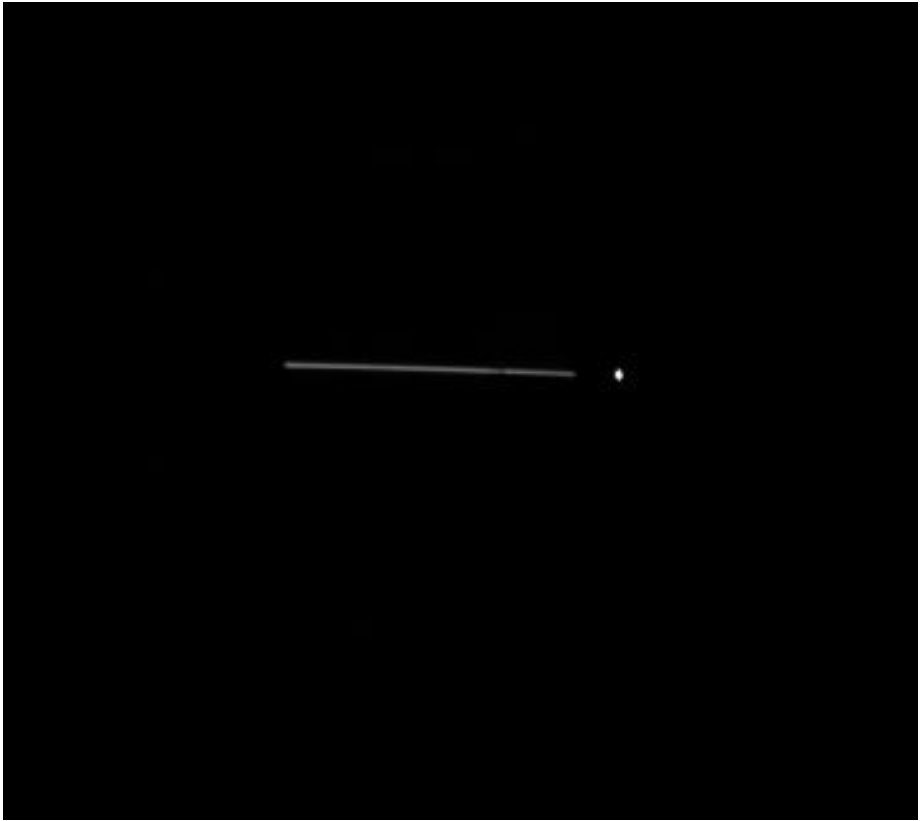

Figure 25

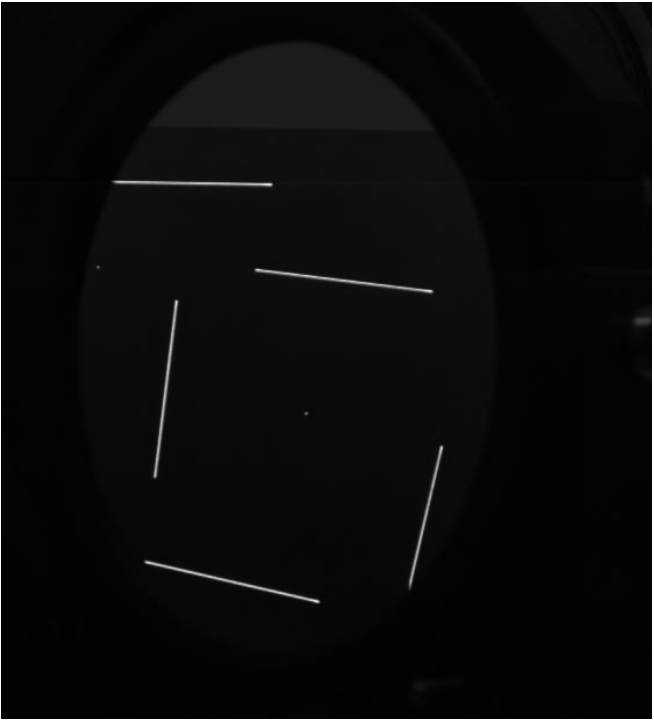

Figure 26

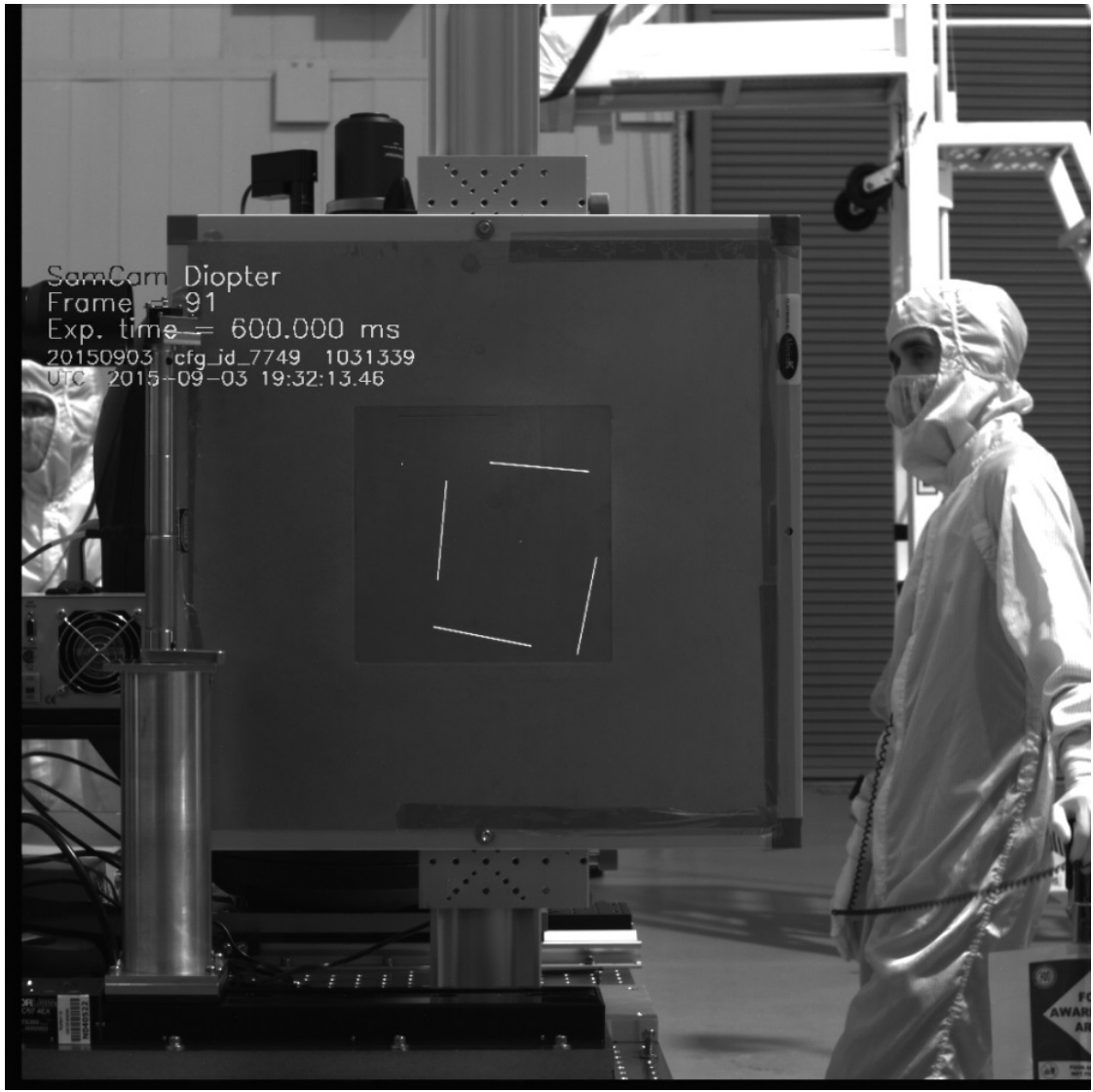

Figure 27

| Relative Location of Distortion Calibration Images within Camera Fields of View: Labels | | | | | |
|---|---|---|---|---|---|
| P10 |  | P20 |  | P30 |
|  | R7 |  | R9 |  |
| P40 |  | P50 |  | P60 |
|  | R17 |  | R19 |  |
| P70 |  | P80 |  | P90 |

Figure 28

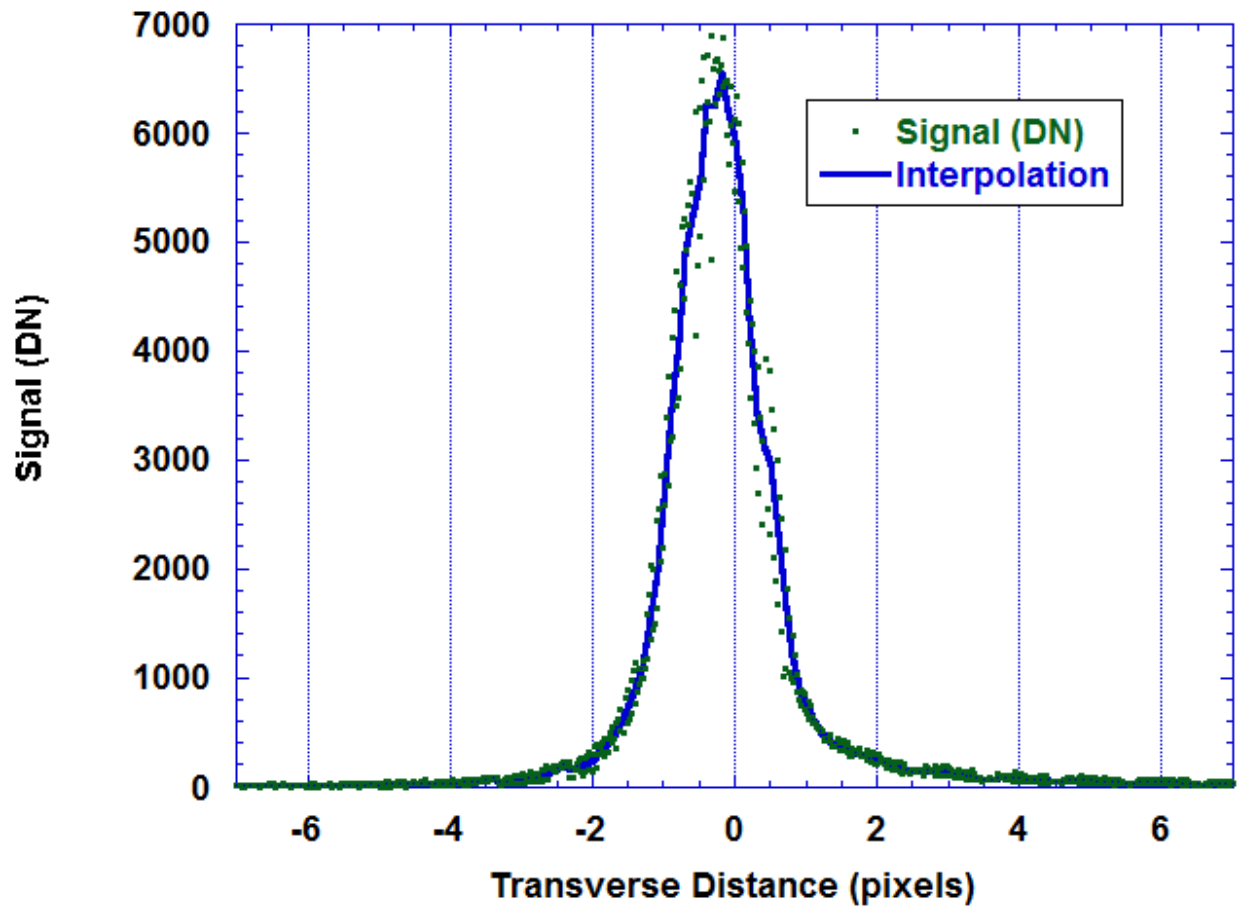

Figure 29

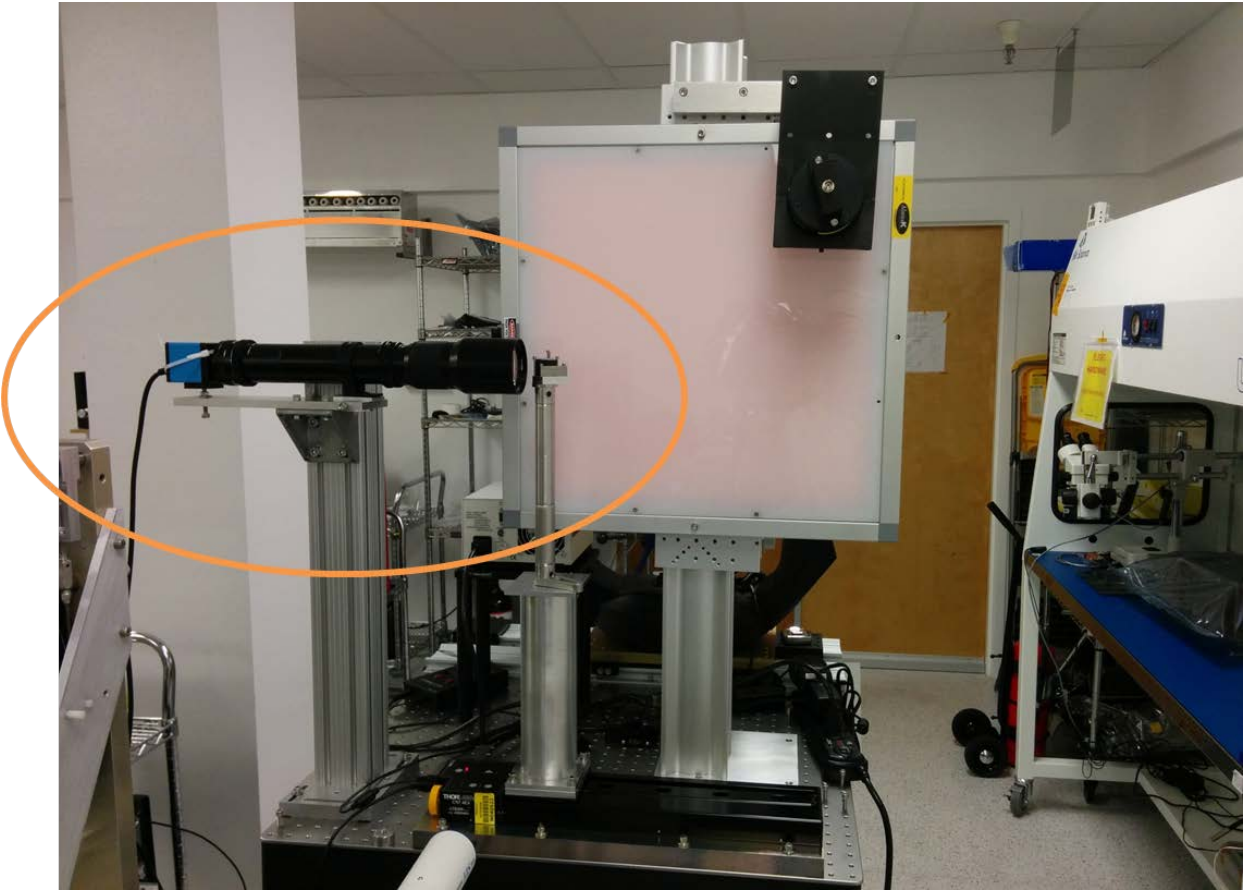

Figure 30

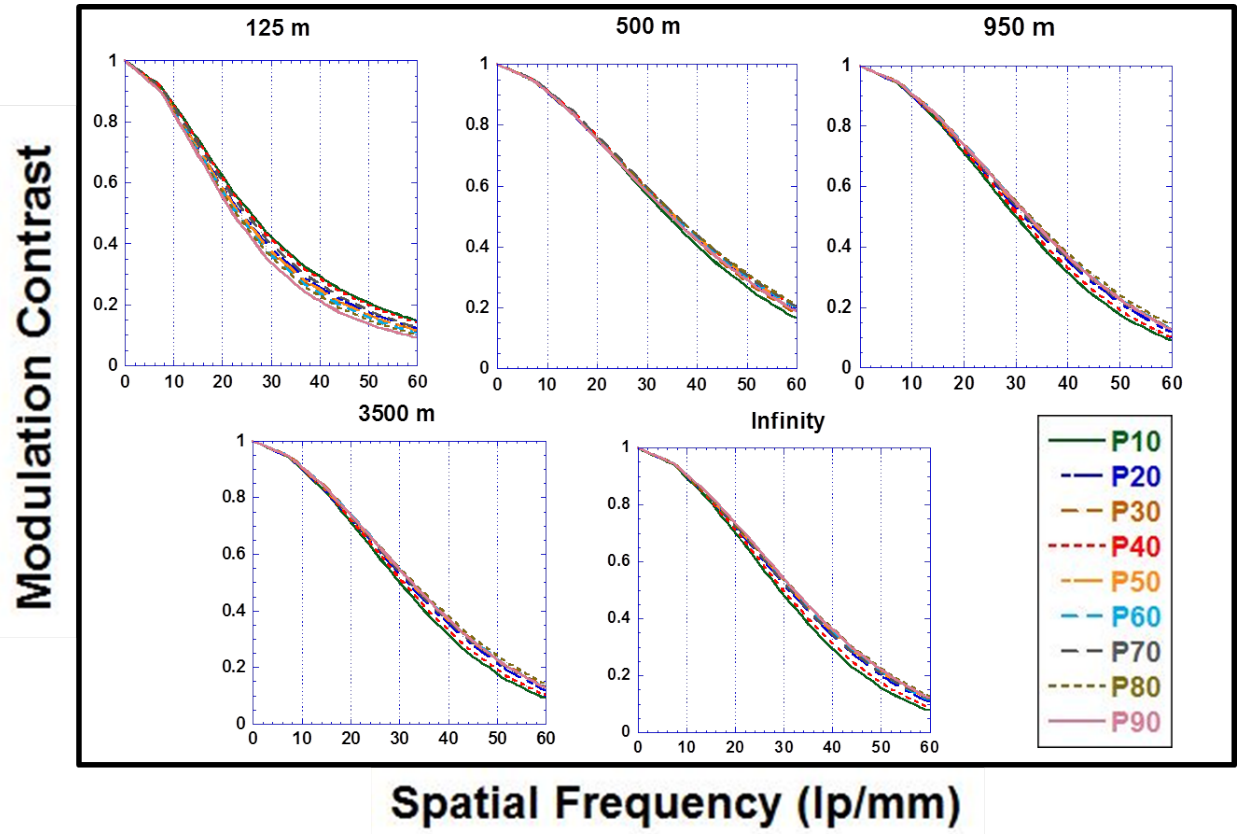

Figure 31

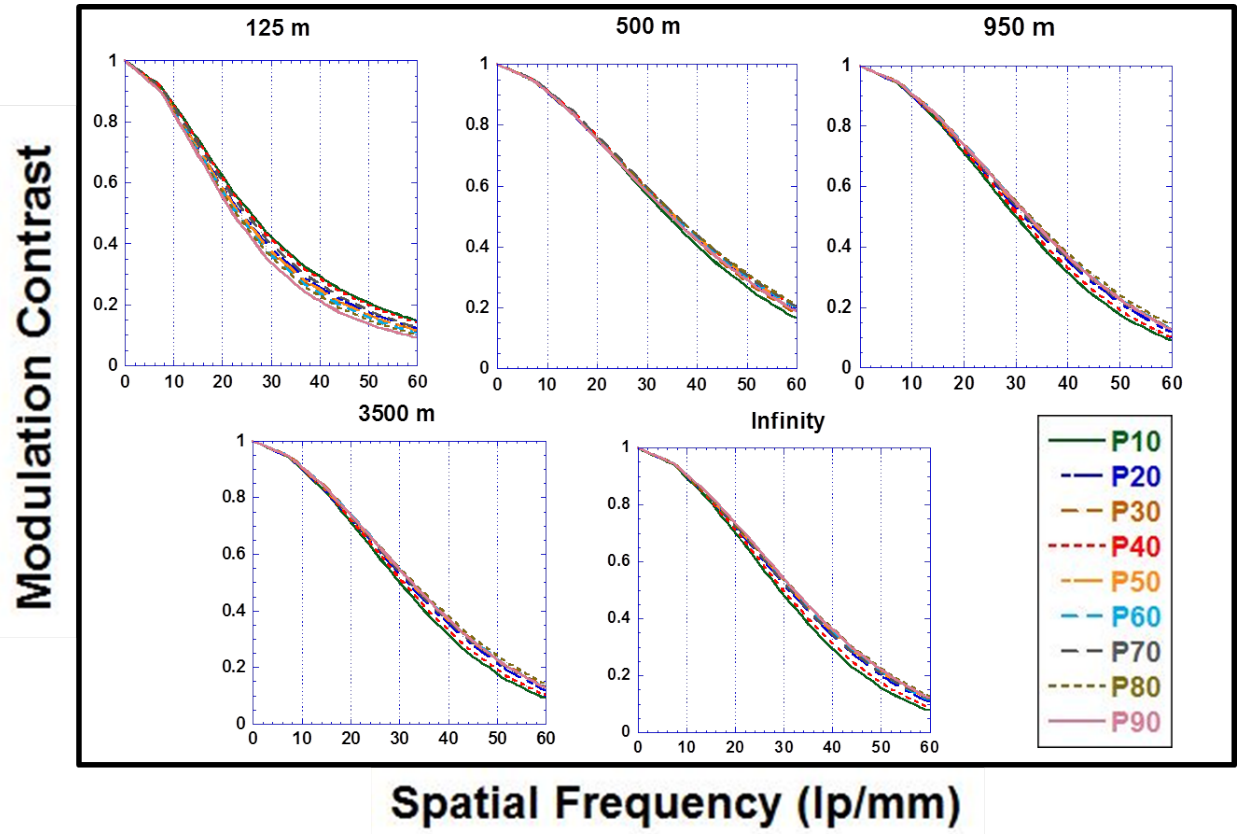

Figure 32

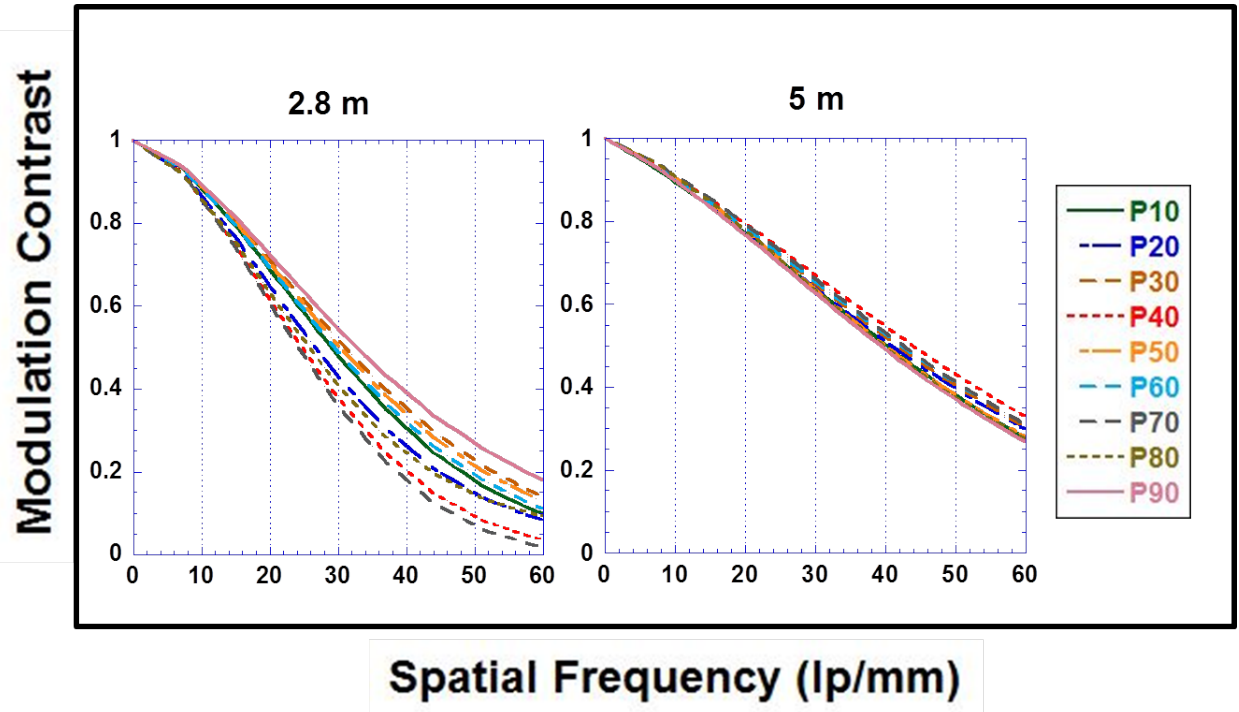

Figure 33

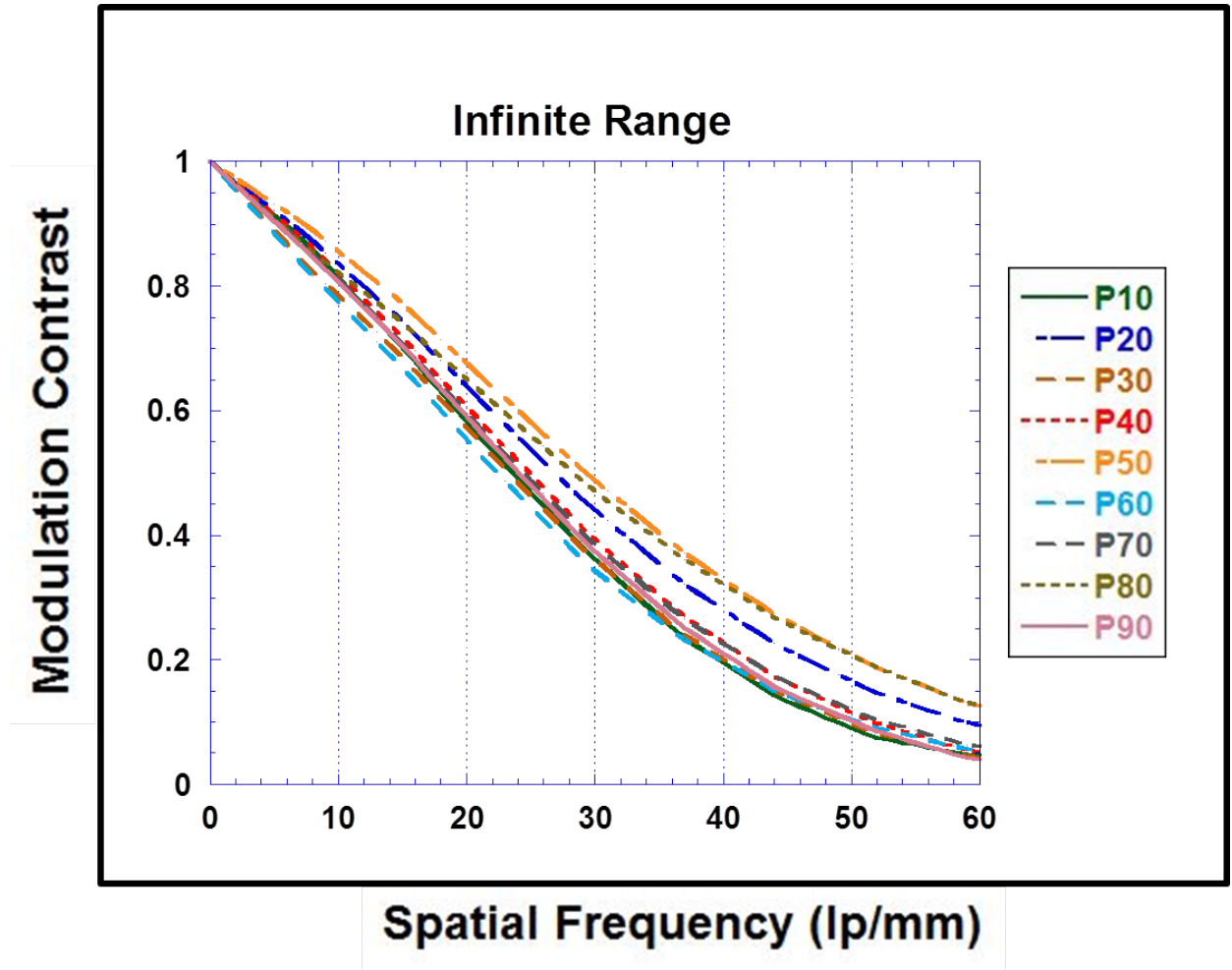

Figure 34

a)

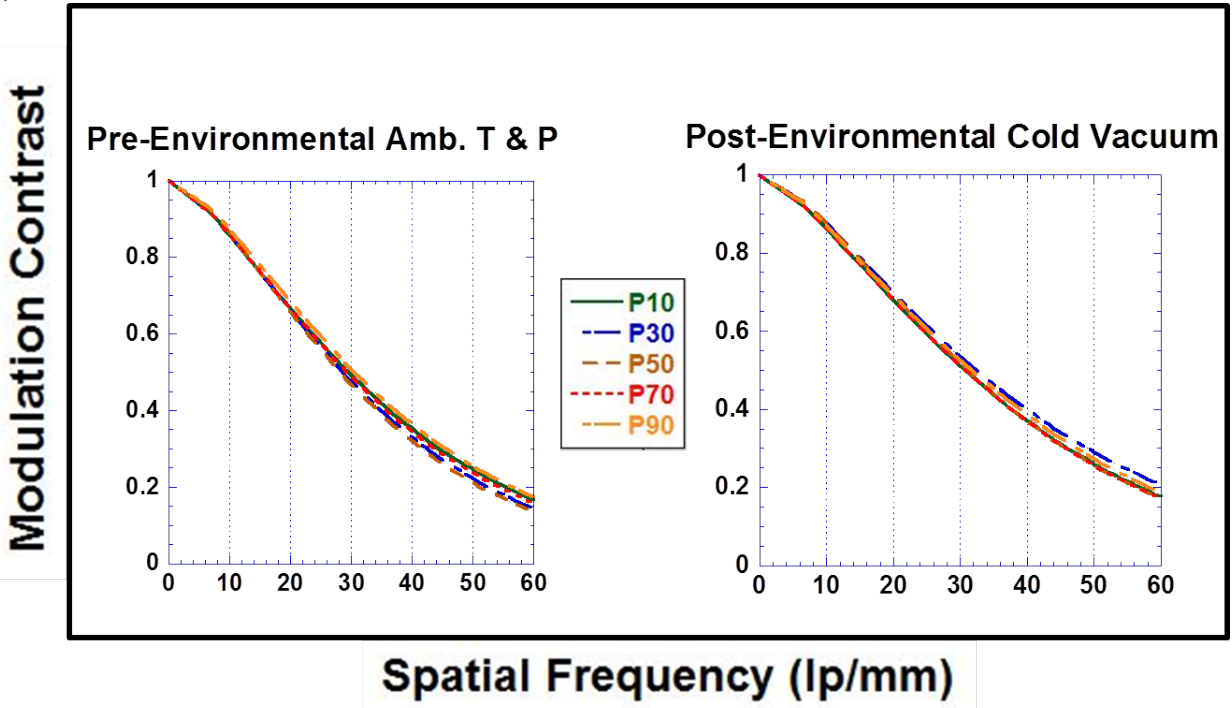

b)

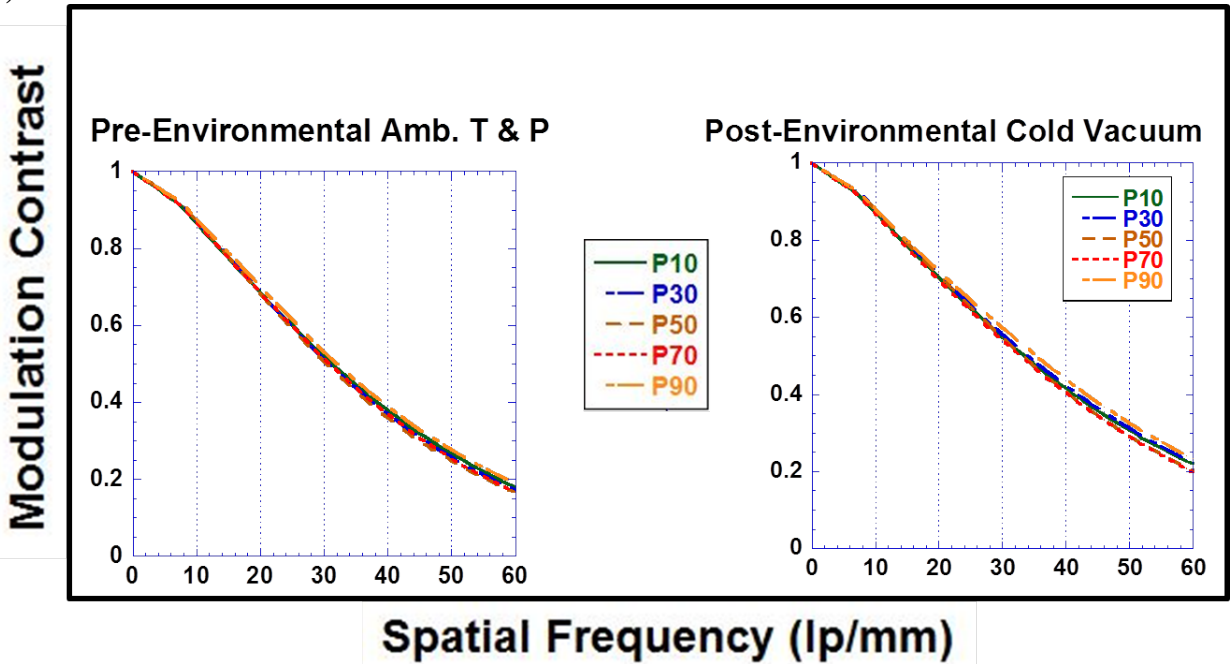

c)

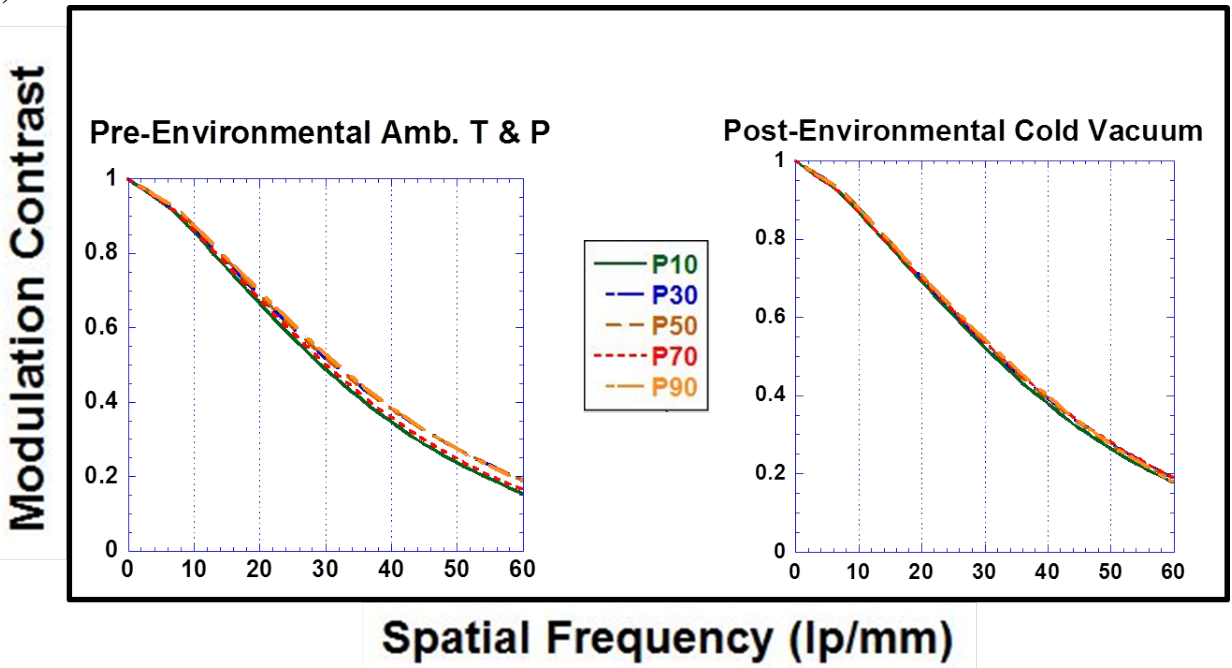

d)

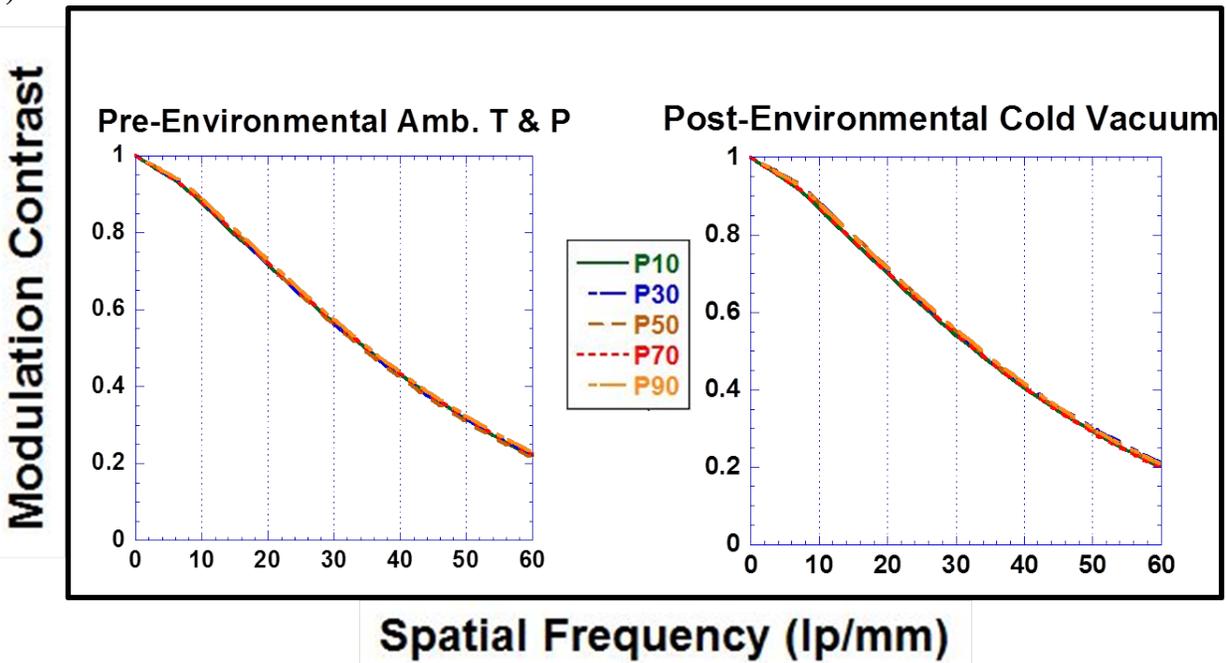

e)
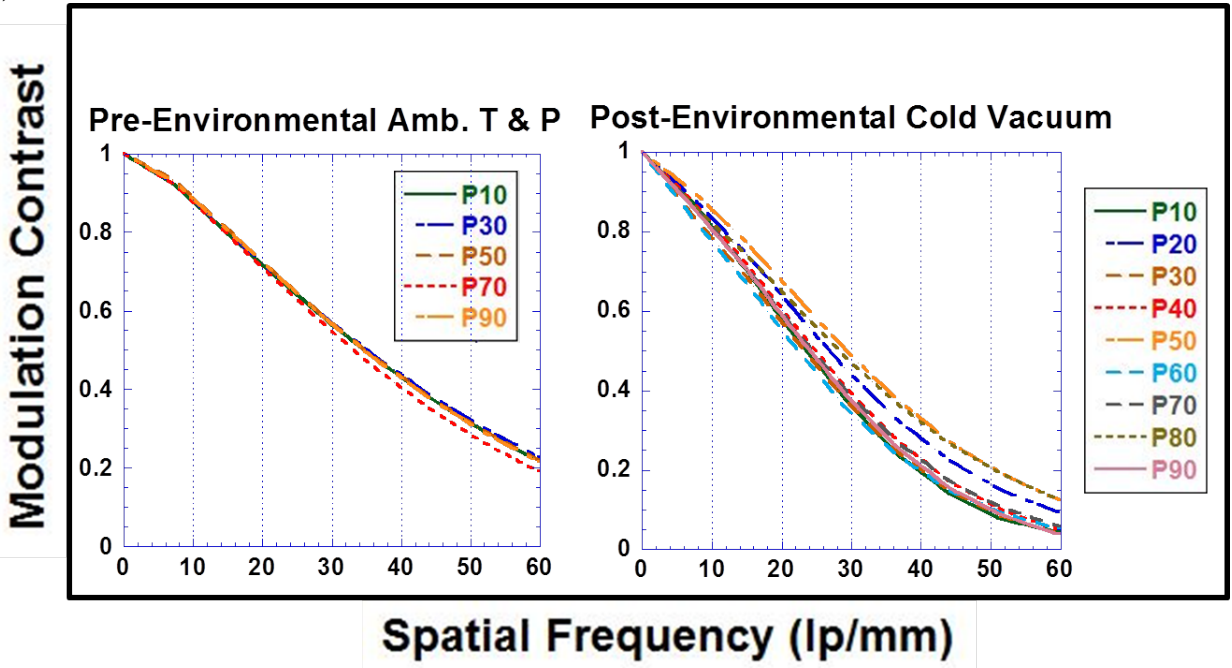

Figure 35

a)

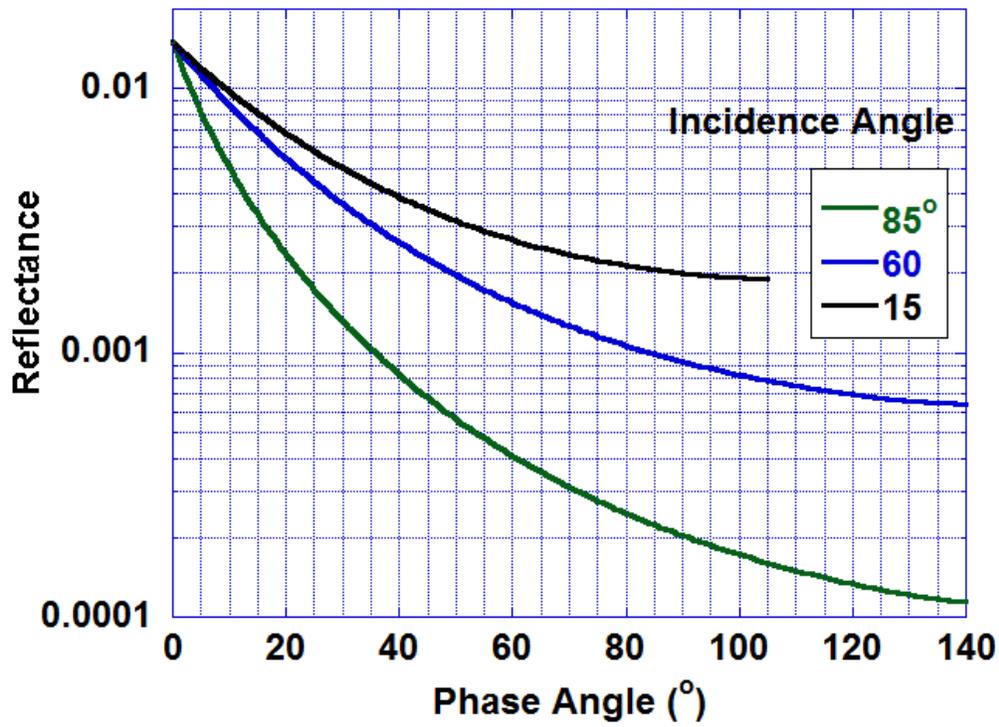

b)

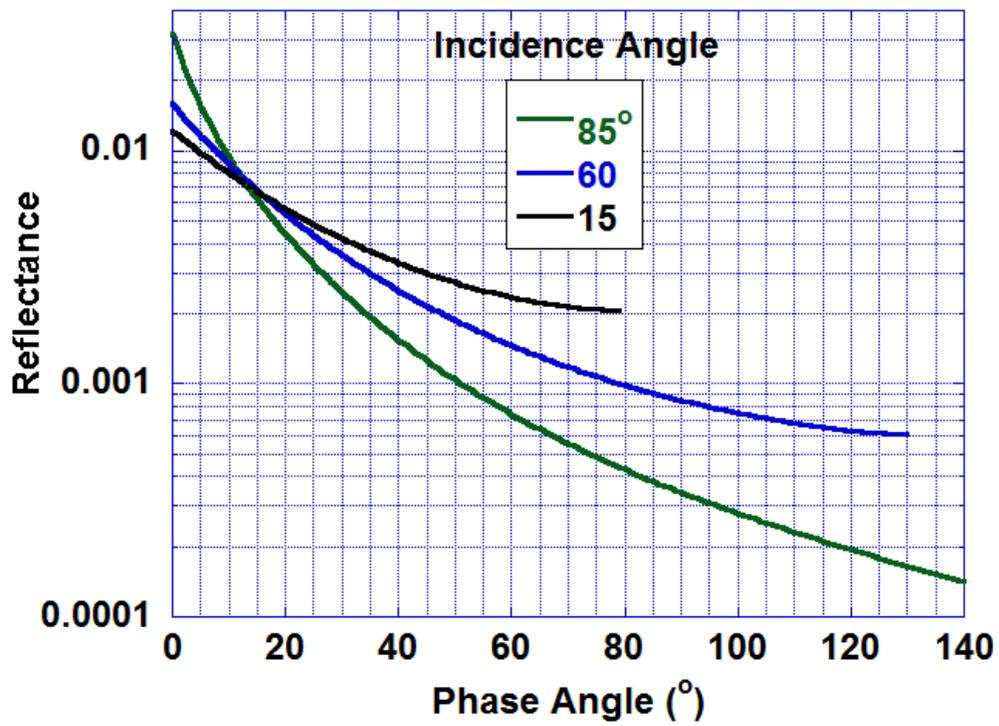

c)

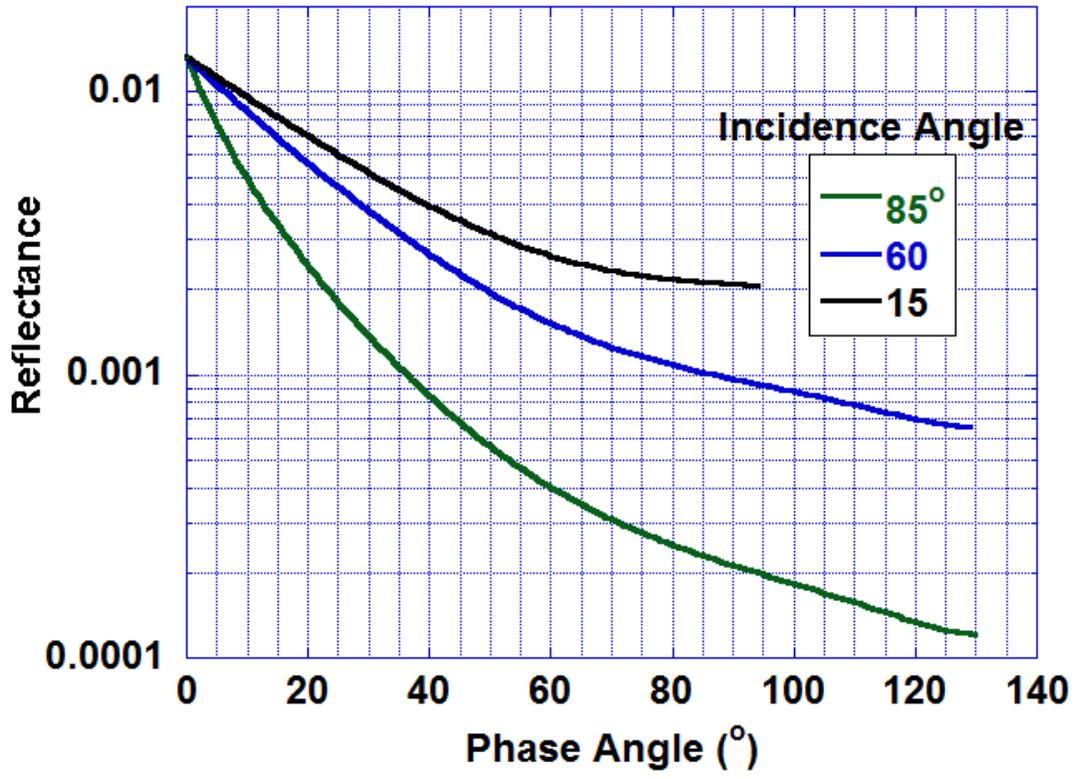

Figure 36

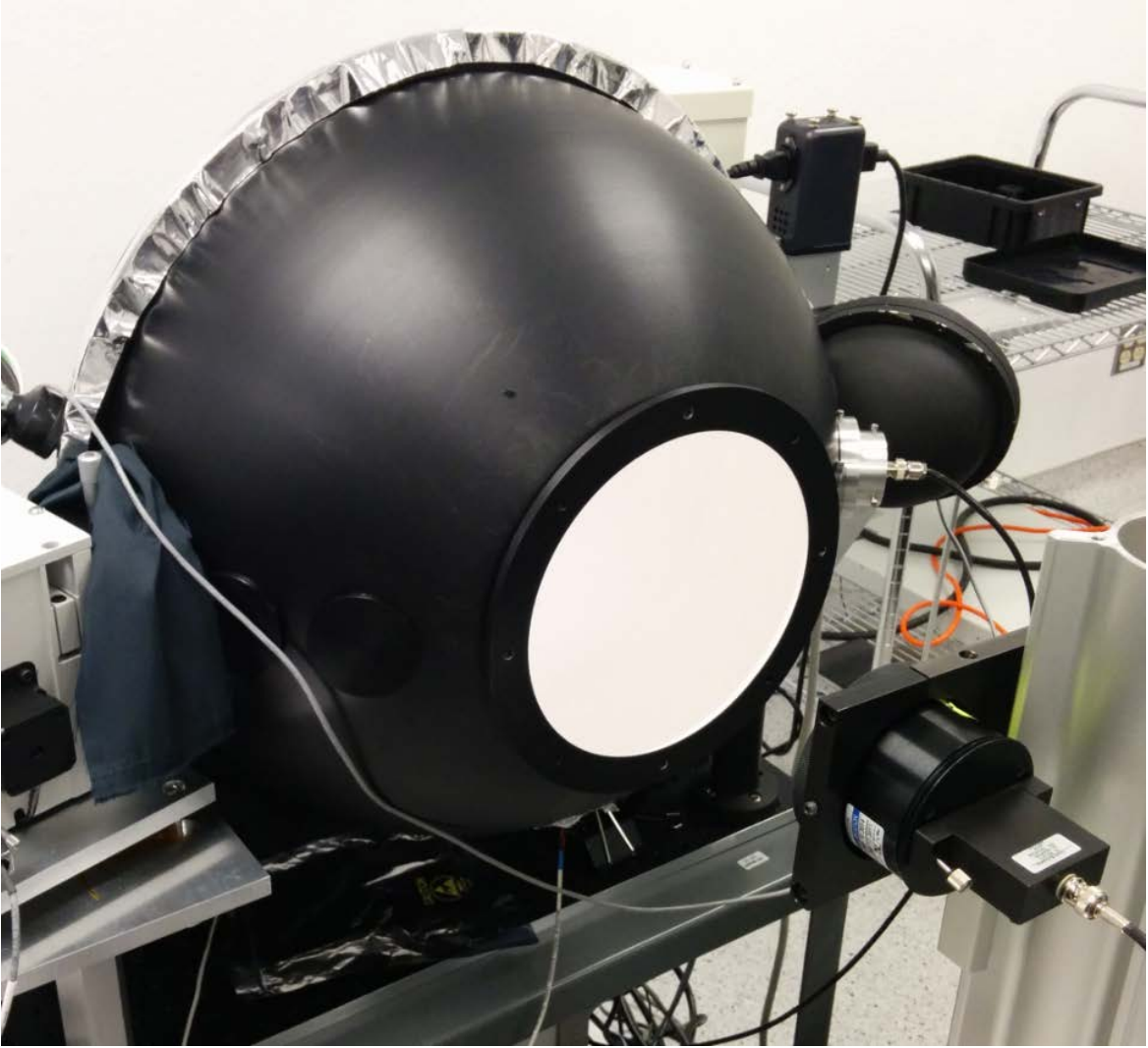

Figure 37

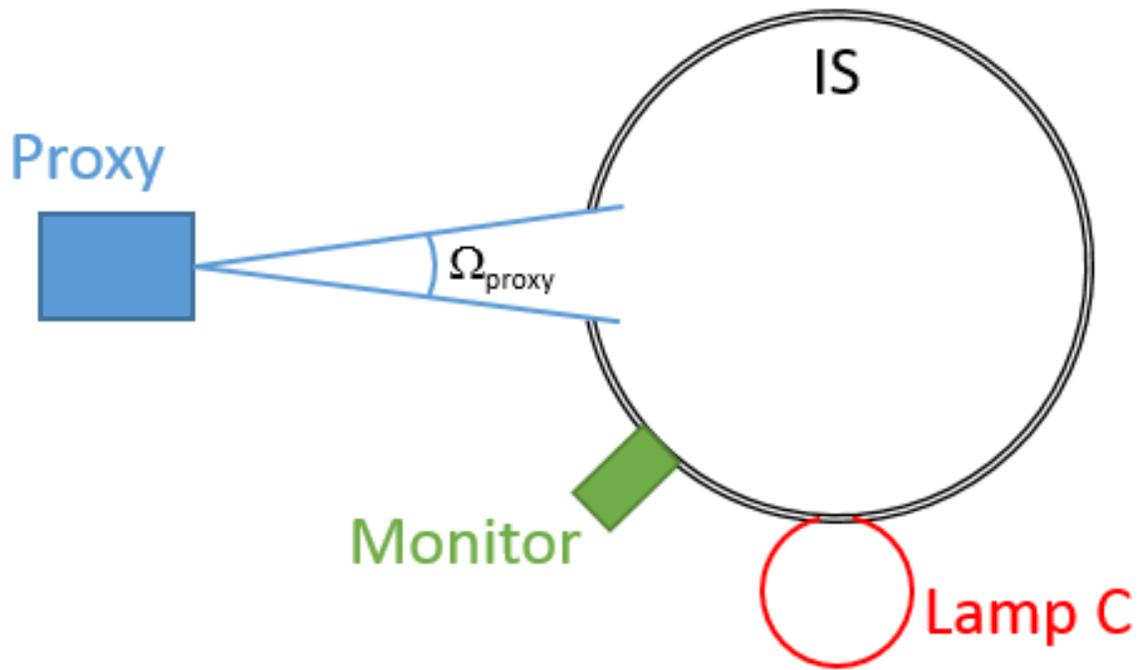

Figure 38

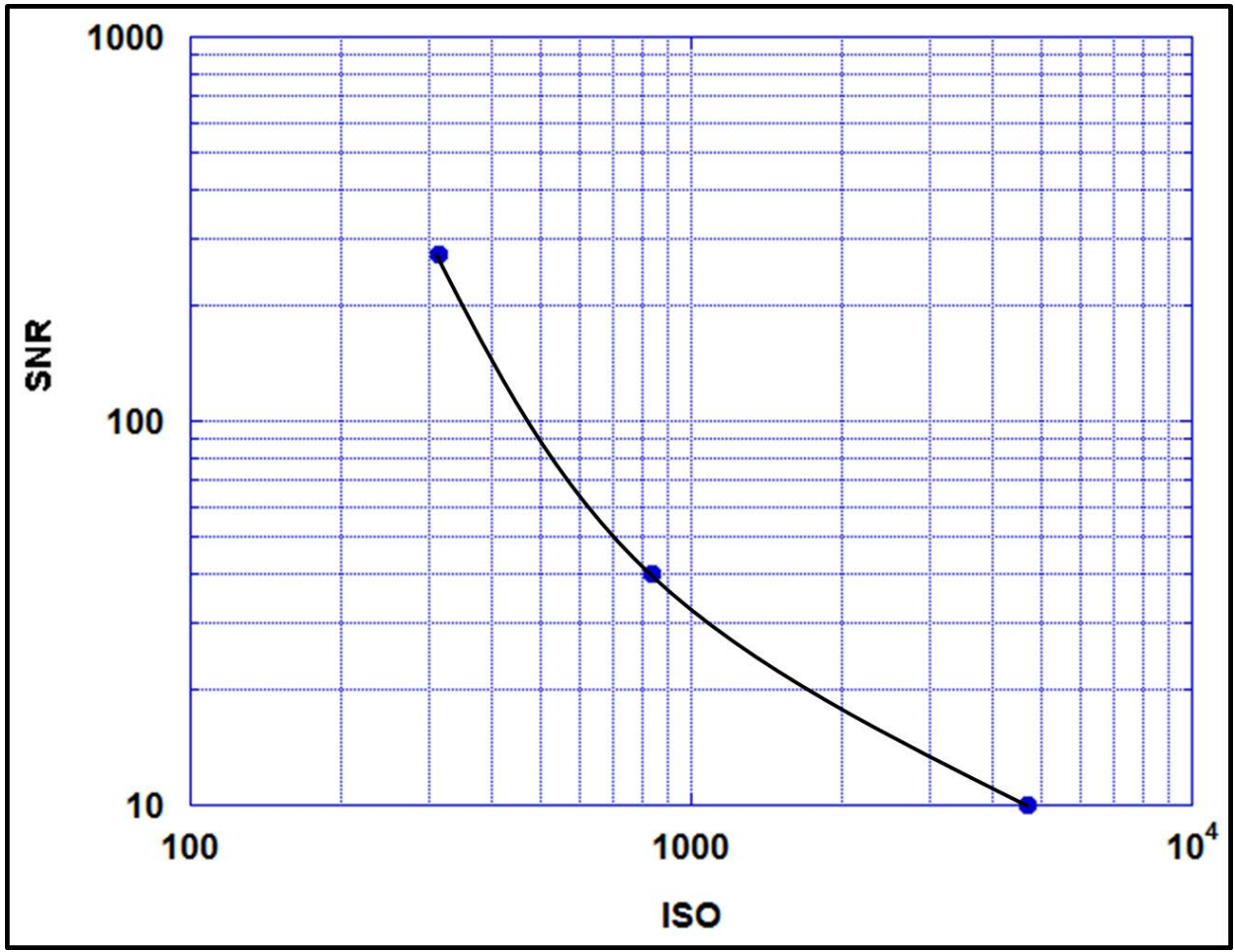

Figure 39

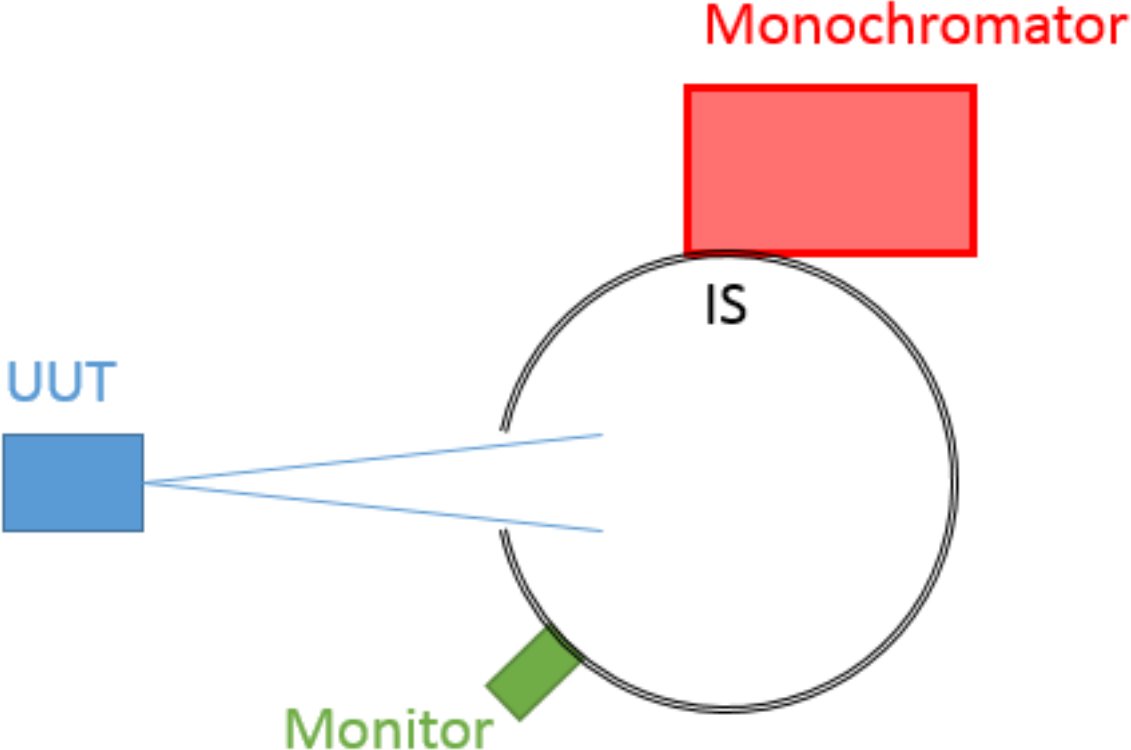

Figure 40

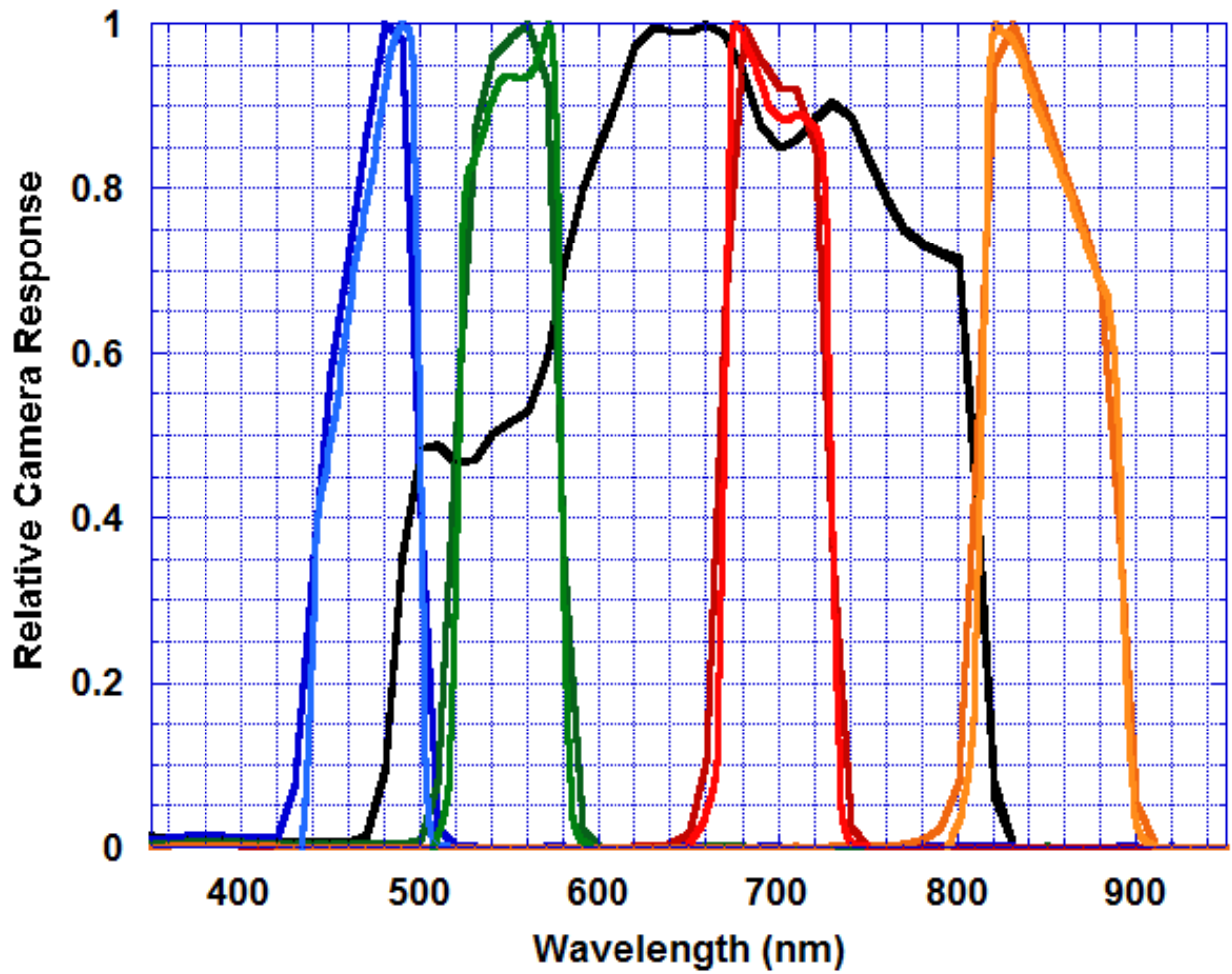

Figure 41

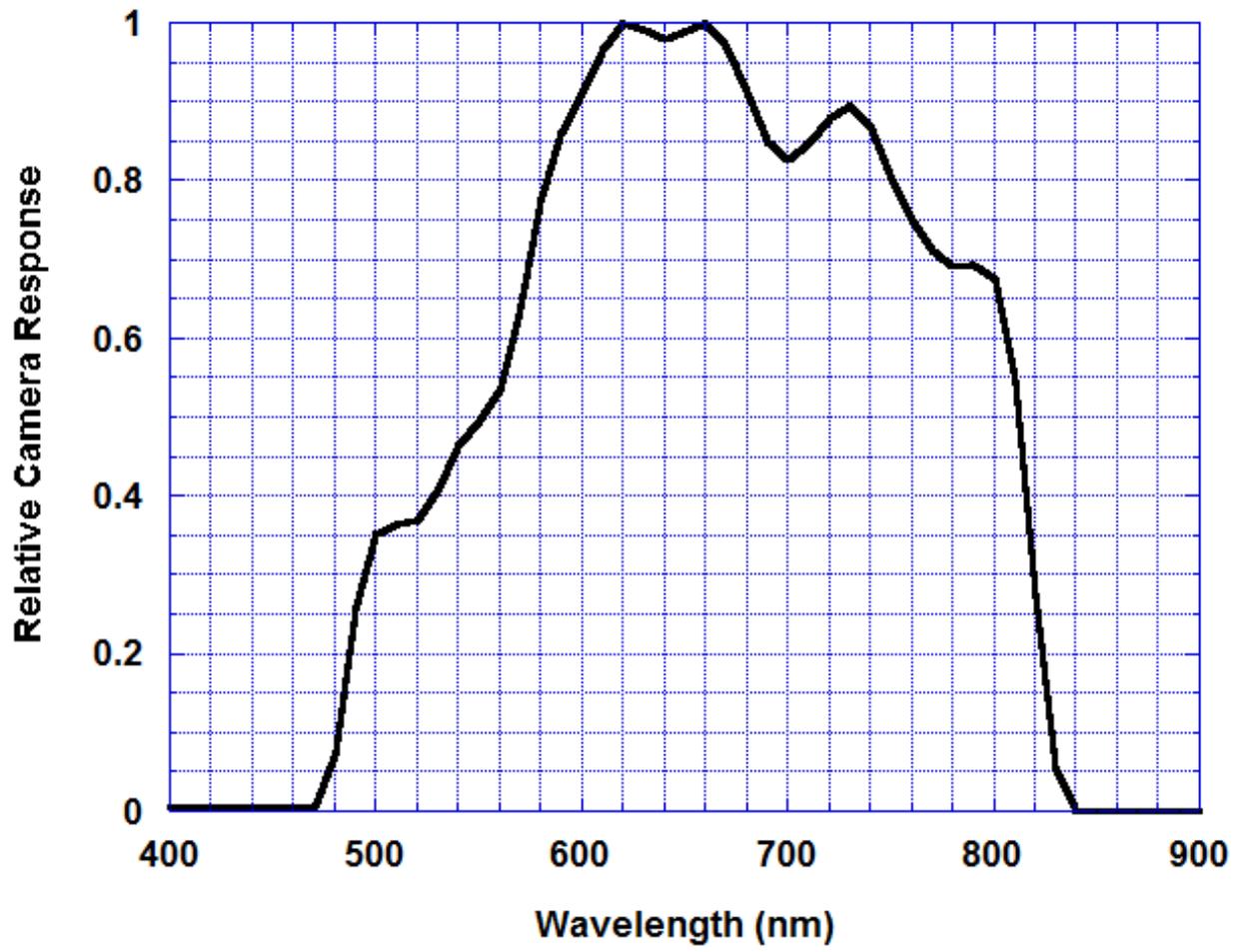

Figure 42

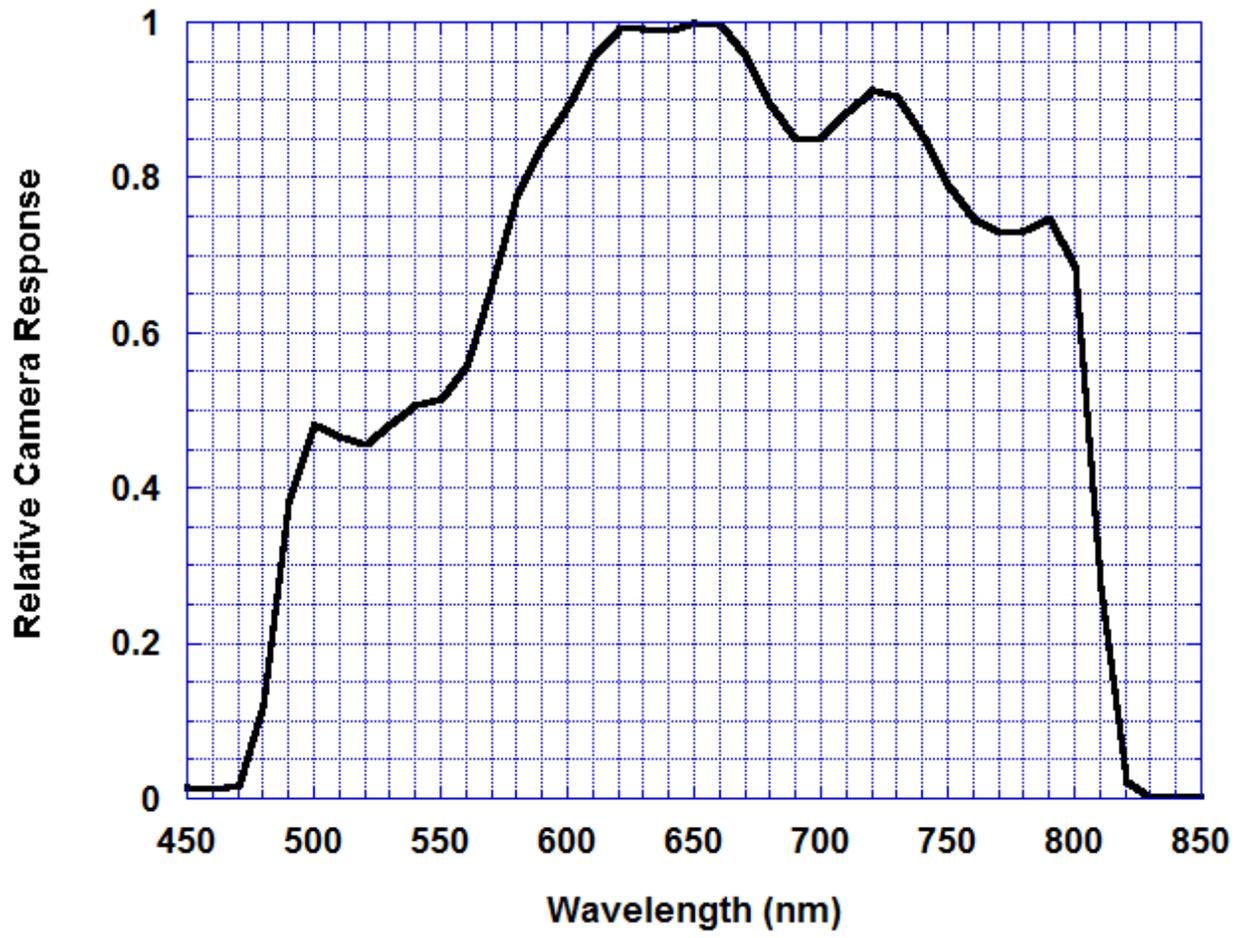

Figure 43

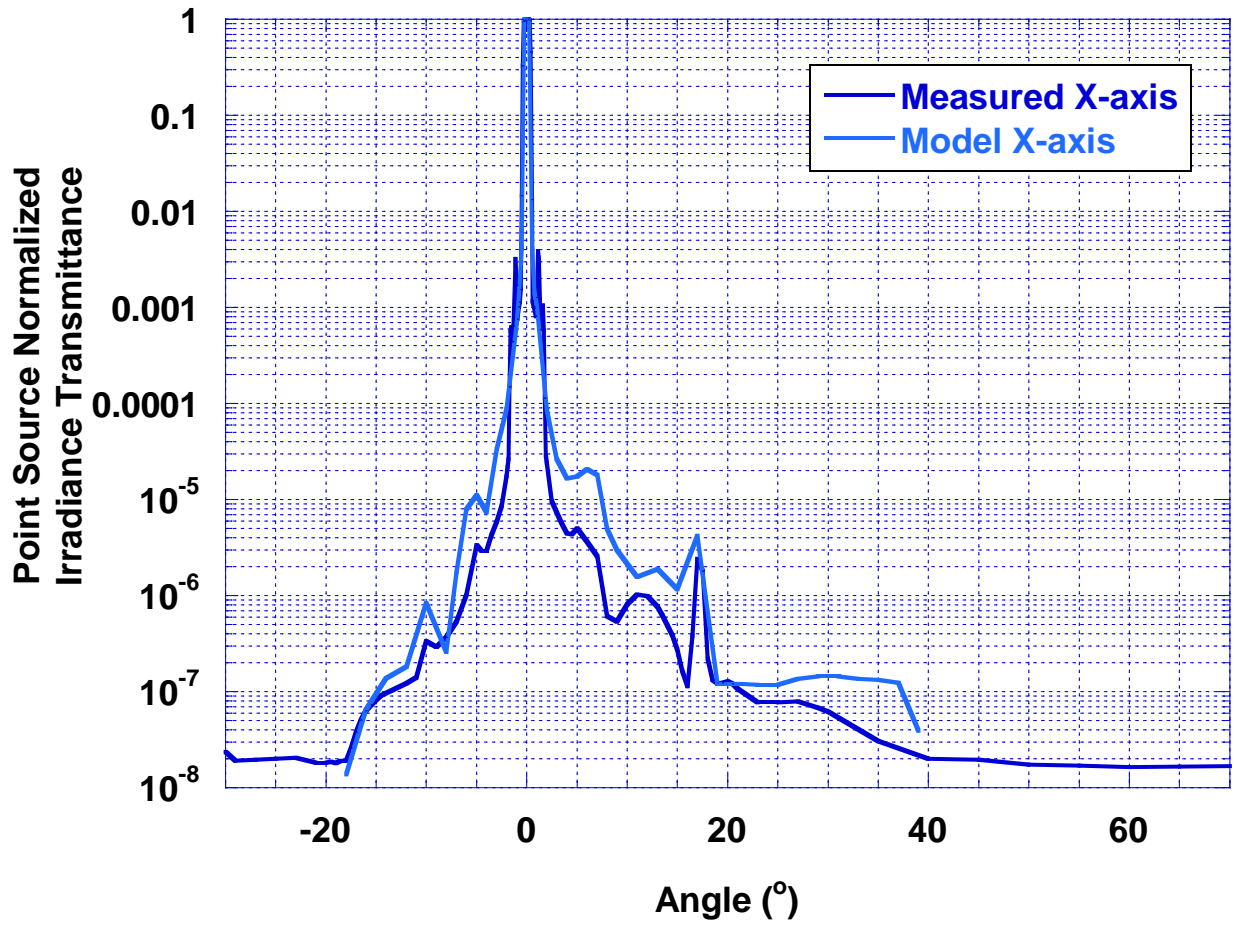

Figure 44

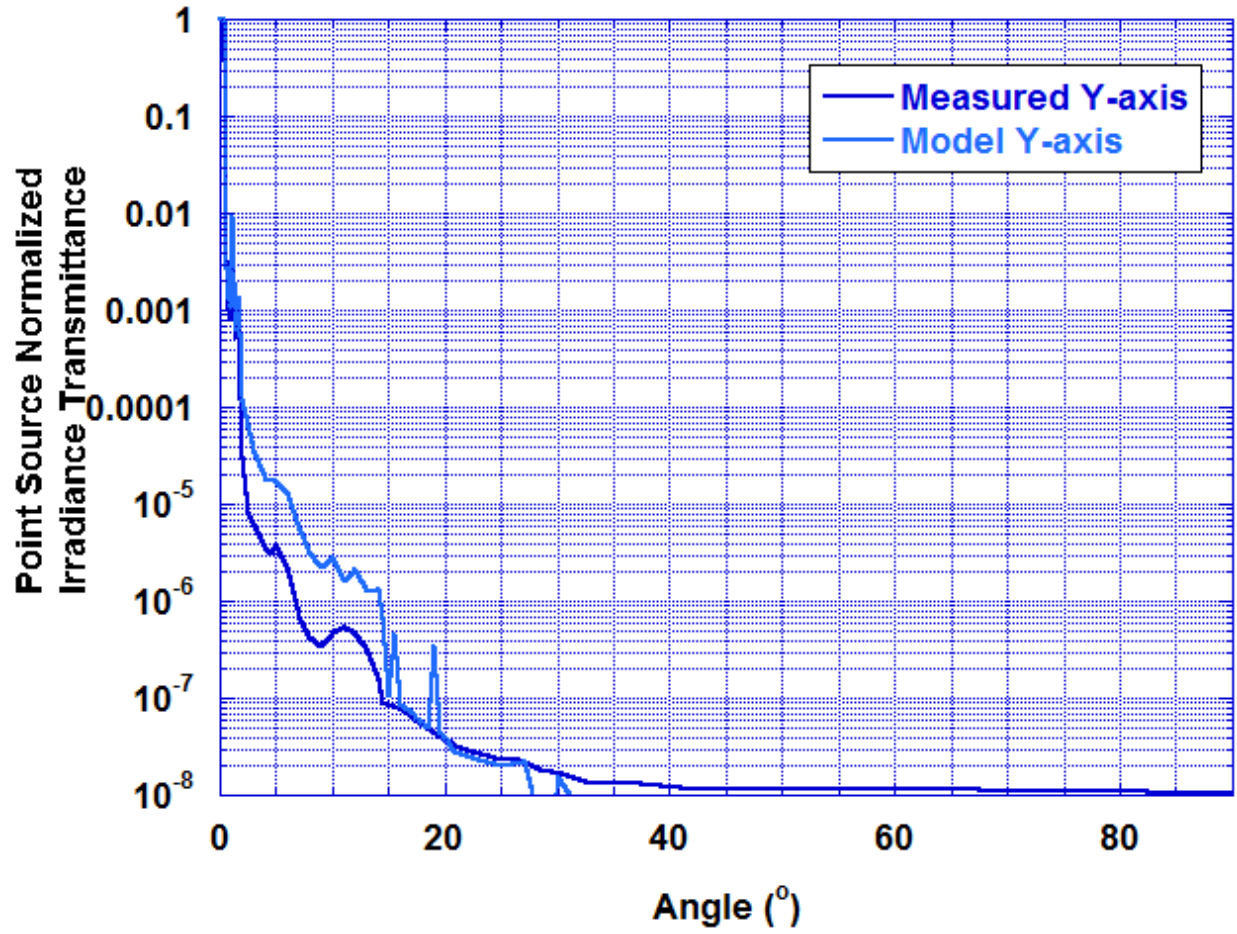

Figure 45

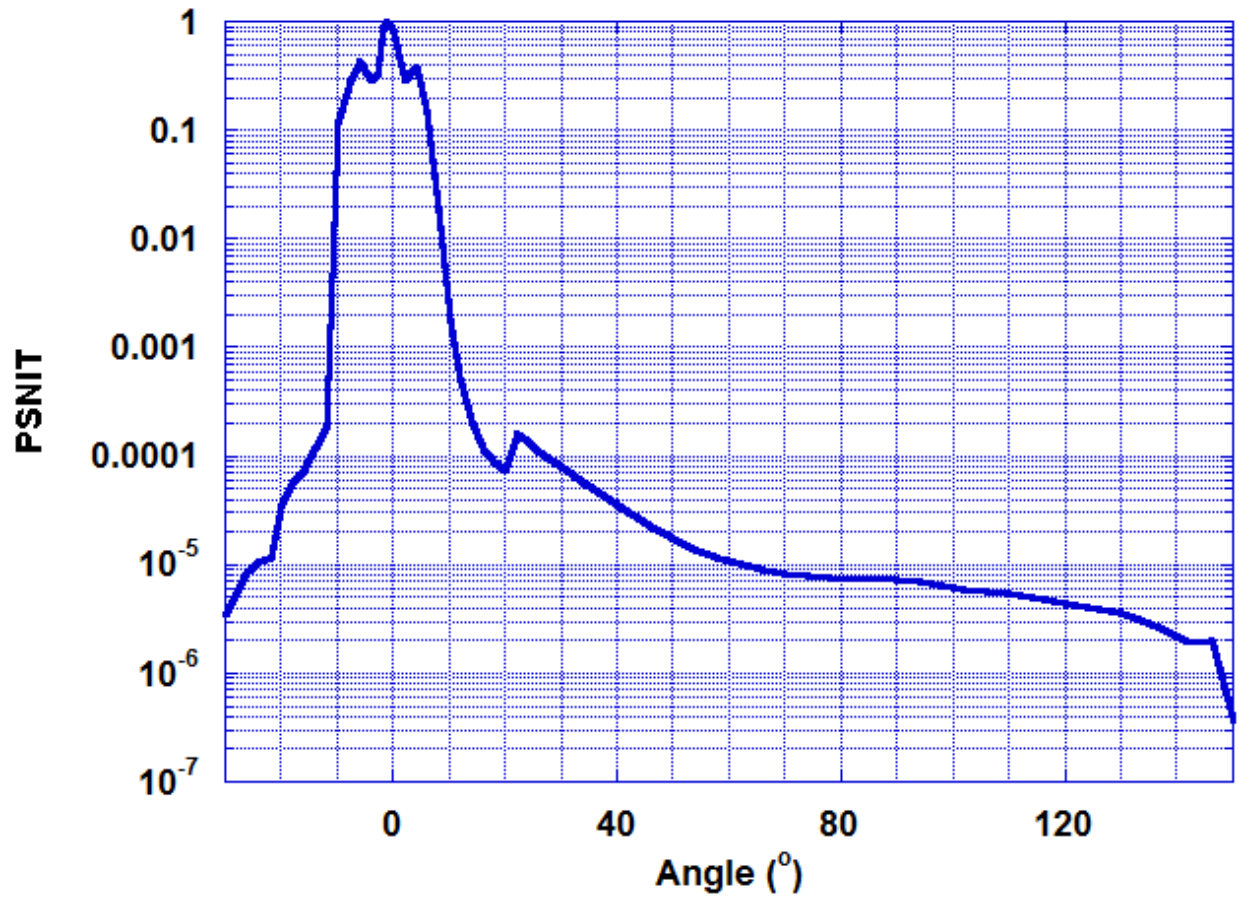

Figure 46

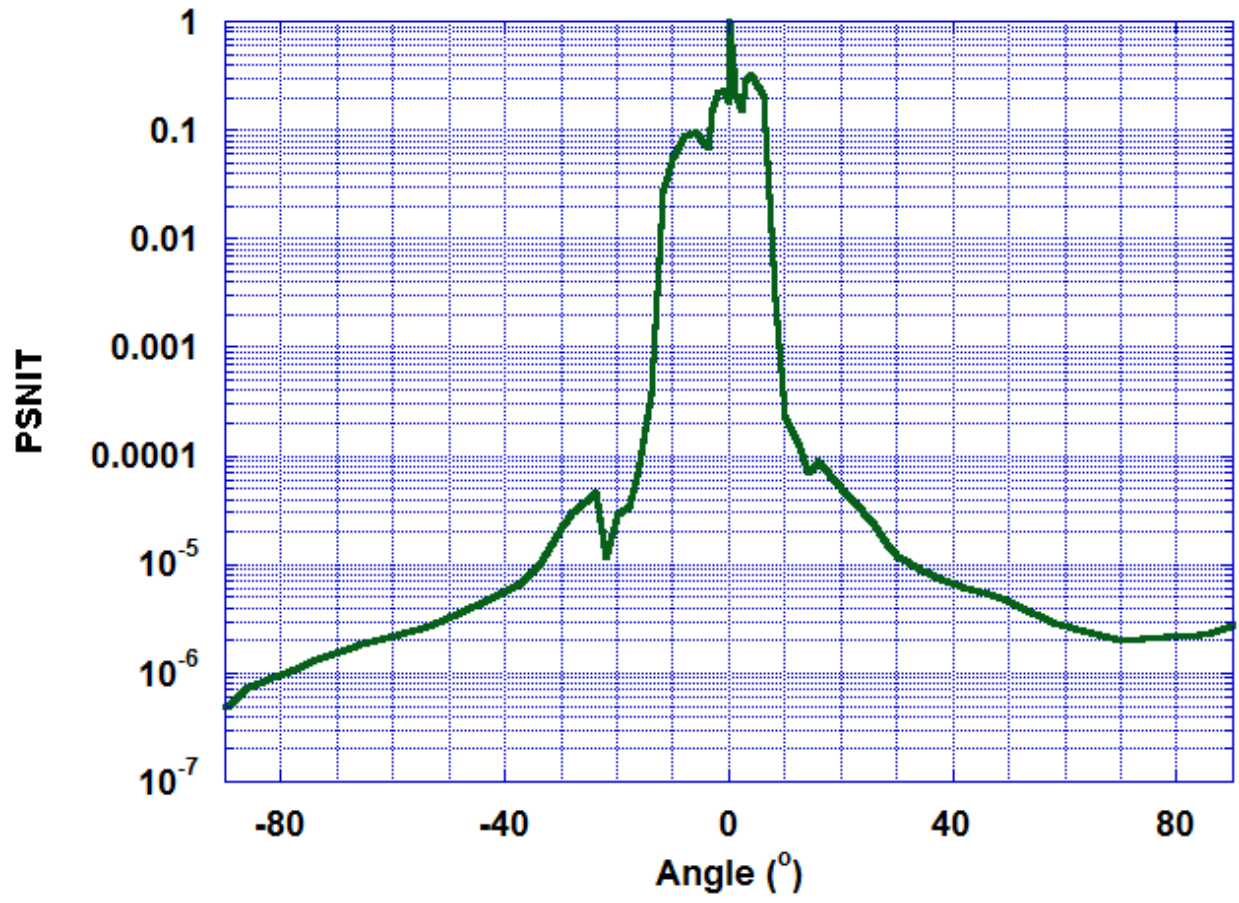

Figure 47

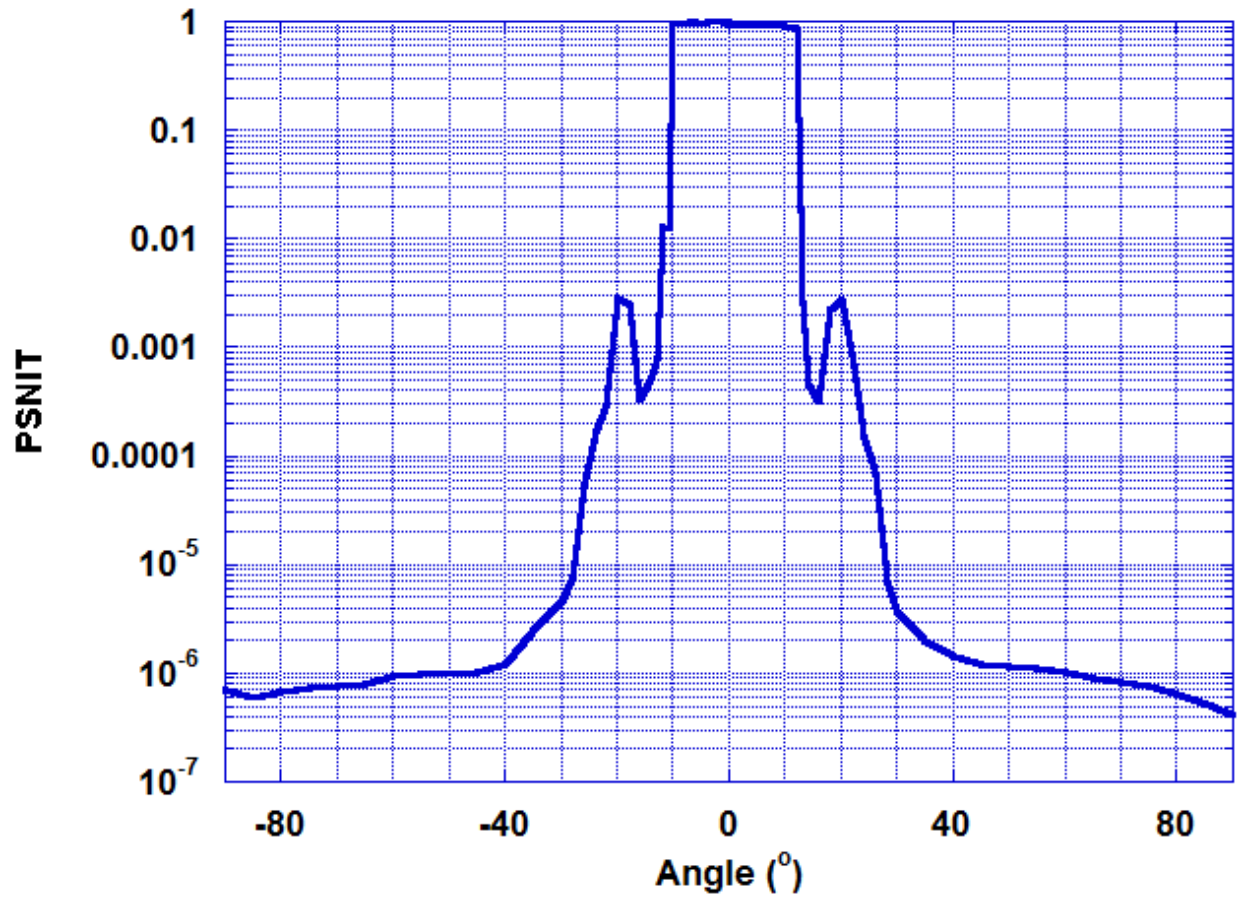

Figure 48

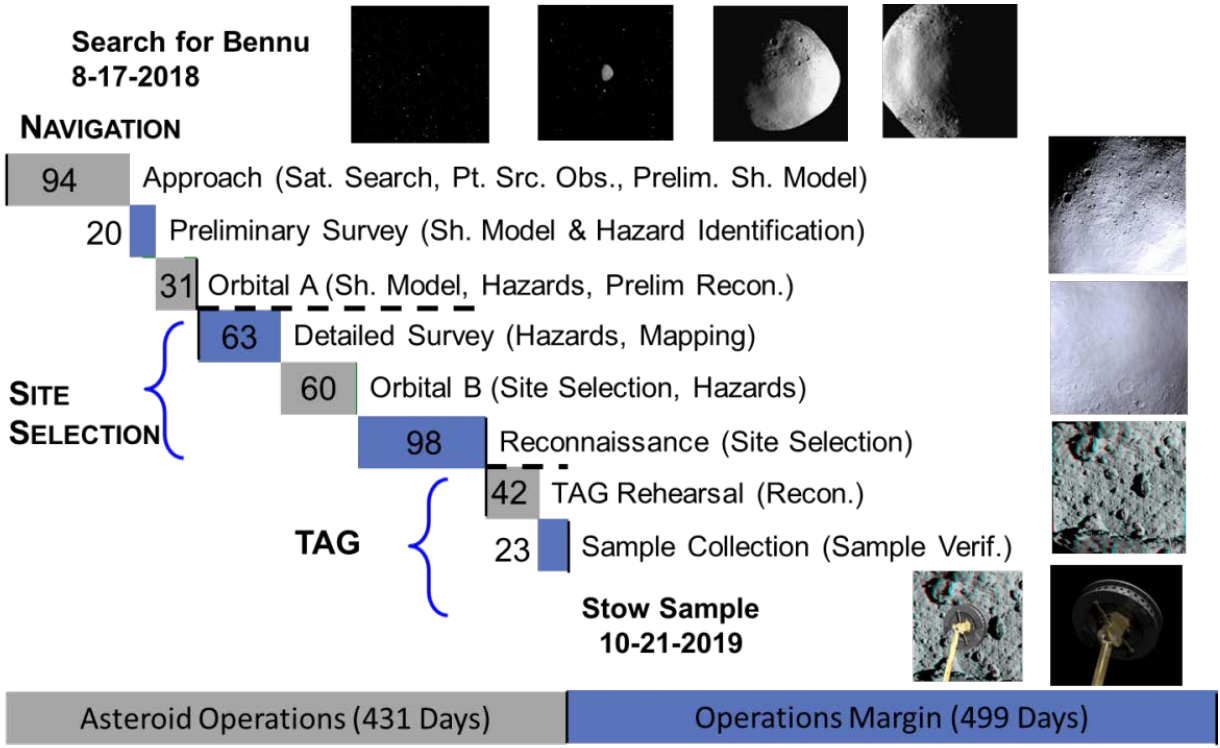

Figure 49

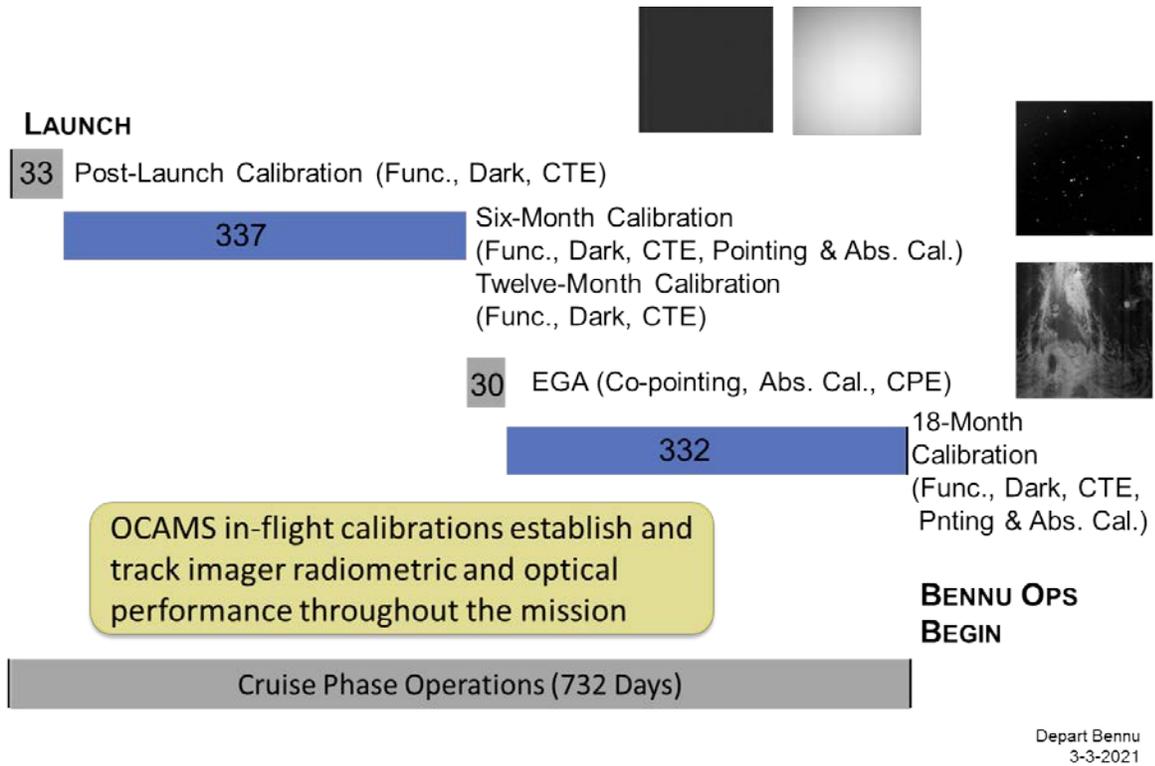

Figure 50

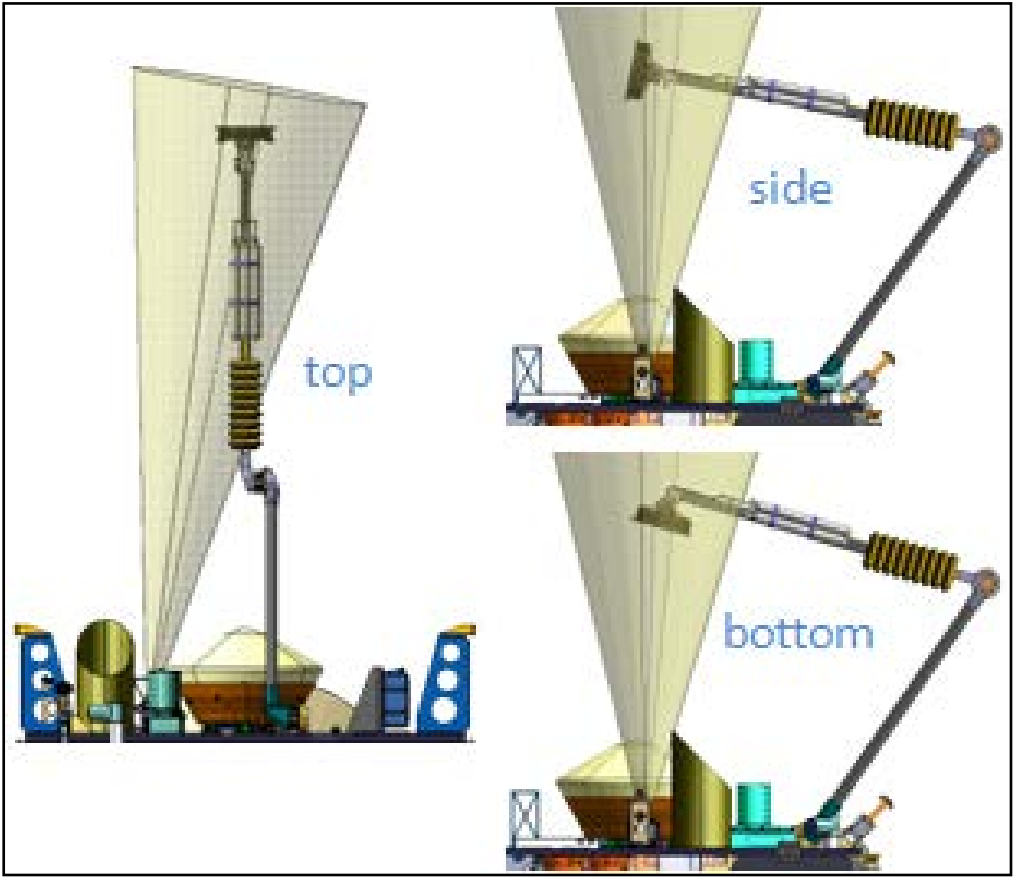